\documentclass[sigconf]{acmart}

\acmYear{2021}
\usepackage{xspace}
\usepackage{pifont}
\usepackage{makecell}
\usepackage{arydshln}
\usepackage{color, colortbl}
\usepackage[vlined,ruled,linesnumbered]{algorithm2e}
\usepackage{enumitem}
\usepackage{adjustbox}
\usepackage{booktabs}

\SetCommentSty{mycommfont}

\usepackage{subcaption}
\usepackage{tabularx}
\colorlet{lightgrey}{lightgray}
\usepackage{float}
\usepackage{multirow}

\newcommand{\sys}{\mbox{\textsc{DeepSemantic}}\xspace}
\newcommand{\syspre}{\mbox{\textsc{DS-Pre}}\xspace}
\newcommand{\sysbs}{\mbox{\textsc{DS-BinSim}}\xspace}
\newcommand{\syscopt}{\mbox{\textsc{DS-Toolchain}}\xspace}
\newcommand{\systask}{\mbox{\textsc{DS-Task}}\xspace}

\newcommand{\cc}[1]{\mbox{\smaller[0.5]\texttt{#1}}}

\newcommand{\ie}{\textit{i}.\textit{e}.}
\newcommand{\eg}{\textit{e}.\textit{g}.}

\input{code/fmt}

\def\Snospace~{\S{}}

\newcommand{\x}{$\times$\xspace}

\if 0

\fi

\newif\ifdraft\drafttrue
\newif\ifnotes\notestrue
\ifdraft\else\notesfalse\fi

\input{glyphtounicode}
\newcolumntype{R}[1]{>{\raggedleft\let\newline\\\arraybackslash\hspace{0pt}}p{#1}}

\newcommand{\includepdf}[1]{
  \includegraphics[width=\columnwidth]{#1}
}

\newcommand{\squishlist}{
\begin{itemize}[noitemsep,nolistsep]
  \setlength{\itemsep}{-0pt}
}
\newcommand{\squishend}{
  \end{itemize}
}

\usepackage{tikz}

\newcommand*\BC[1]{%
\begin{tikzpicture}[baseline=(C.base)]
\node[draw,circle,fill=black,inner sep=0.2pt](C) {\textcolor{white}{#1}};
\end{tikzpicture}}

\usepackage{xstring}
\newcommand{\PP}[1]{
\vspace{2px}
\noindent{\bf \IfEndWith{#1}{.}{#1}{#1.}}
}

\newcommand{\boxbeg}{
\vspace{2px}
\noindent\begin{tabular}{|l|}\hline
\begin{minipage}{3.2in}
\vspace{2px}
\noindent
}

\newcommand{\boxend}{
\vspace{2px}
\end{minipage}\\ \hline
\end{tabular}
\vspace{-10pt}
}

\begin{document}

%\fancyhead{}
%\copyrightyear{2021}
%\acmYear{2021}
%%\setcopyright{rightsretained}
%\acmConference[CCS '21]{Proceedings of the 2021 ACM SIGSAC Conference on Computer and Communications Security}{November 14--19, 2021}{Seoul, South Korea}
%\acmBooktitle{Proceedings of the 2021 ACM SIGSAC Conference on Computer and Communications Security (CCS '21), November 14--19, 2021, Seoul, South Korea}
%\acmPrice{15.00}
%\acmDOI{10.1145/3372297.3417866}
%\acmISBN{978-1-4503-7089-9/20/11}

% TODO: replace this section with code generated by the tool at https://dl.acm.org/ccs.cfm
%\begin{CCSXML}
%	<ccs2012>
%	<concept>
%	<concept_id>10002978.10003029.10011703</concept_id>
%	<concept_desc>Security and privacy~Usability in security and privacy</concept_desc>
%	<concept_significance>500</concept_significance>
%	</concept>
%	</ccs2012>
%\end{CCSXML}

%\ccsdesc{Security and privacy~System security; Browser security}
%% -- end of section to replace with generated code
%
%\keywords{Debloating; Browser; Program Analysis; Binary Rewriting}

\title{Semantic-aware Binary Code Representation with BERT}

% when 'make draft'
%\ifdefined\DRAFT
% \pagestyle{fancyplain}
% \lhead{Rev.~\therev}
% \rhead{\thedate}
% \cfoot{\thepage\ of \pageref{LastPage}}
%\fi

% blind
%\author{Anonymous Authors (Paper \#168)}

\author{
 Hyungjoon Koo$^\dagger$\; 
 Soyeon Park$^\ddagger$\;
 Daejin Choi$^\ast$\;
 Taesoo Kim$^\ddagger$\;
 \\
 \emph{Sungkyunkwan University$^\dagger$ }, 
 \emph{Georgia Institute of Technology$^\ddagger$}, 
 \emph{Incheon National University$^\ast$}
}

\begin{abstract}
A wide range of binary analysis applications,
such as bug discovery,
malware analysis and code clone detection,
require recovery of contextual meanings on a binary code.
Recently, binary analysis techniques based on machine learning
have been proposed
to automatically reconstruct
the code representation of a binary
instead of manually crafting
specifics of the analysis algorithm.
However,
the existing approaches utilizing machine learning
are still specialized
to solve one domain of problems,
rendering recreation of models
for different types of binary analysis.

In this paper,
we propose \sys utilizing BERT
in producing
the semantic-aware code representation of a binary code.
%that can be reused for other downstream tasks.
%
To this end, we introduce \emph{well-balanced instruction 
	normalization}
that holds rich information
for each of instructions
yet minimizing an out-of-vocabulary (OOV) problem.
\sys has been carefully designed 
based on our study with large swaths of binaries.
%
%The key idea of \sys is to adopt a two-stage training models:
%i) pre-training to generate a generic model, and
%ii) fine-tuning the model for each specific downstream task.
%%
%The essence of the design lies in re-purposing 
%a pre-trained model that is readily available as a one-time processing
%to quickly apply further applications.
%
Besides, \sys leverages the essence of 
the BERT architecture into \emph{re-purposing}
a pre-trained generic model that is readily available 
as a one-time processing, followed by
quickly applying specific downstream tasks
with a fine-tuning process.
We demonstrate \sys with two downstream tasks, namely,
binary similarity comparison and compiler provenance 
(\ie, compiler and optimization level) prediction.
Our experimental results show that the binary similarity model
outperforms two state-of-the-art binary similarity tools, 
DeepBinDiff and SAFE, 49.84\% and 15.83\% on average, respectively.
%comparison tools with an F1 of $96.4$, and
%the toolchain model with an F1 of $96.2$ and $91.0$ for 
%compiler and optimization classification, respectively.
\end{abstract}

\date{}
\maketitle

\sloppy

\section{Introduction}
\label{s:intro}
%\XXX{use well-balanced instead of balanced-grained.}
%
In a modern computing environment,
it is not uncommon
to encounter
binary-only software:
\eg, commodity or proprietary programs,
system software like firmware and device drivers,
etc.
Accordingly,
binary analysis plays a pivotal role
in implementing a wide range of popular use 
cases~\cite{binsim-survey, ml-based-binanalysis-survey, 
	dl-based-binanalysis-survey, ml-dm-binanalysis-survey}:
\eg,
code clone or software plagiarism detection
to protect against intellectual property infringement%
~\cite{binsimfse, asm2vec, deepbindiff},
vulnerability discovery on distributed software~\cite{
	bug-search-acsac14,
	xarch-bug, discovre, esh-pldi16, acfg, bingo, esh2,
	vulseeker, struct2vec, binarm, alphadiff},
malware detection~\cite{polymorphic-worm,
	self-mutating-malware, cfg-malware-detection} and
classification~\cite{mutantx, simcalc19},
program repair or patch analysis~\cite{exe-comp-dimva04,
	code-comp-saner16, spain-icse17},
and toolchain provenance~\cite{oglassesx, toolchain-recovery}
for the digital forensics purpose.

% why is it challenging?
However, analyzing a binary code
is fundamentally more challenging than 
the source code-based analysis
because it has to infer underlying contextual meanings
from the machine-interpretable binary code alone.
Unlike the representation of human-readable source code,
a binary code is a final product
out of a complicated compilation process
that involves massive transformations (\eg, optimizations),
such as control flow graph alteration,
function inlining, 
instruction replacement and
dead code elimination,
which eventually discards
a majority of high-level semantic information
useful for analysis.
Besides, there are other major factors that impact
code generation, such as
an architecture,
compiler,
compiler version or option
and code obfuscation.

% Summary of prior approaches
Recently,
machine learning-based techniques%
~\cite{
	binsim-survey, ml-based-binanalysis-survey, 
	dl-based-binanalysis-survey, ml-dm-binanalysis-survey,
	bingo, bindnn, gemini, asm2vec, alphadiff,
	deepbindiff, innereye, safe, ordermatters}
have been proposed
as a promising direction
to address this code semantic problem in binaries.
Although
traditional approaches like
static analysis (\eg, graph isomorphism on call graph~\cite{discovre, esh2})
or dynamic analysis (\eg, taint analysis~\cite{blanket, xarch-bug})
have shown a high accuracy in specific tasks,
machine learning-based approaches
are often much favorable
in rapidly changing computing environments:
as far as training data is provided,
one model can be reused for multiple platforms and architectures,
as well as
it can be constantly improved
with the increasing number of new inputs.
Indeed, the recent state-of-the-art
tools~\cite{deepbindiff, innereye, asm2vec, safe, ordermatters}
successfully generate code embedding (vector) for
semantic clone detection across
architecture~\cite{innereye, ordermatters},
optimization~\cite{asm2vec, deepbindiff, safe},
and even obfuscation~\cite{asm2vec}.
%
%Priors works have shed light on
%%repeatedly demonstrated that 
%borrowing the concept of cutting-edge deep neural networks
%is a viable approach.
%

However, we question both the means and quality of 
code embedding to infer code semantics
%particularly across both compilers and optimizations,
in terms of applicability in practice.
\begin{figure}[t!]
	\centering
    \includegraphics[width=\columnwidth]{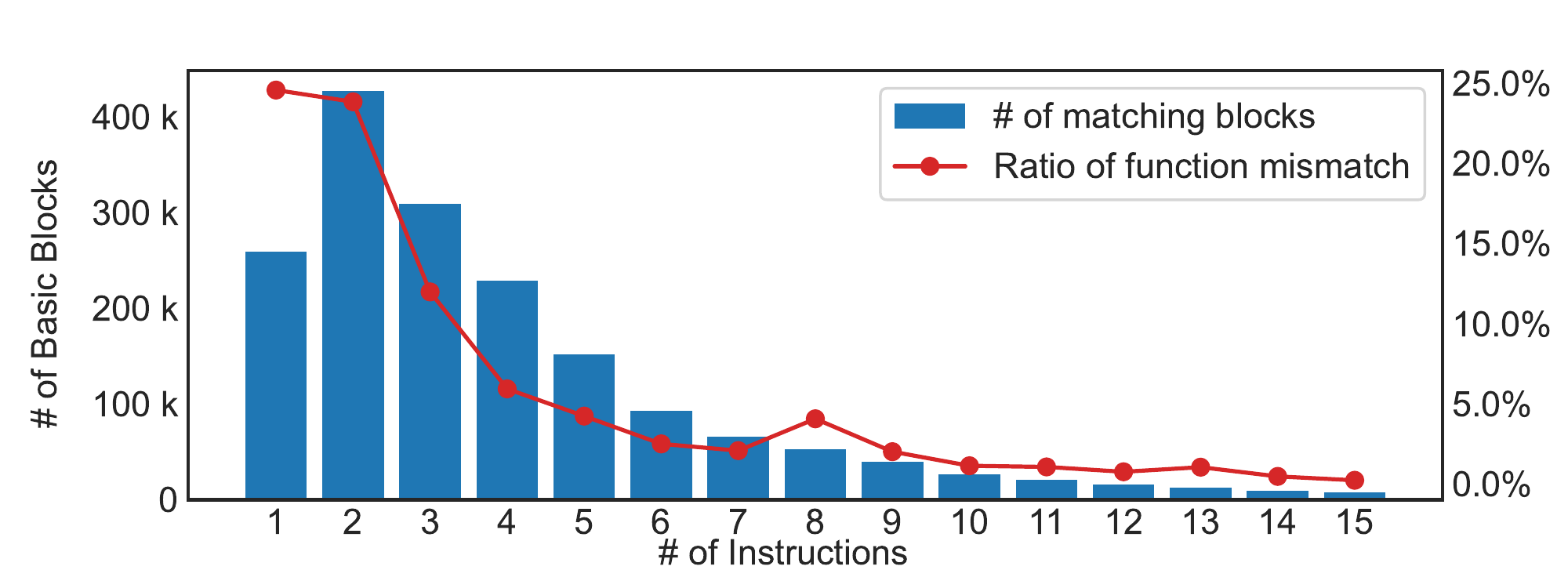}
	\caption{
        %Distributions of basic blocks by the number of instructions.
        %
        Histogram for the number of matching block 
        pairs (left label)
        and the ratio of function mismatch (right label)
        by the number of instructions per basic block
        with DeepBinDiff~\cite{deepbindiff}
        from our evaluation
        dataset (\autoref{s:eval}).
        The function mismatch represents 
        basic block pairs with a sequence of identical 
        instructions that do not belong to the same function.
        More than eight out of $10$ from
        the whole matching basic blocks
        incorporate $5$ instructions or less.
	}
	\label{fig:deepbindiff}
\end{figure}
For instance, \autoref{fig:deepbindiff} illustrates 
the matching basic block pairs and 
function mismatch cases with our
DeepBinDiff~\cite{deepbindiff} evaluation 
dataset (\autoref{s:eval}).
According to our experiment,
$83\%$ of the whole matching block pairs consist of 
%not large enough to infer code semantics
%(\ie, the number of instructions is 15 or less)
$5$ instructions or less.
%while \XXX{xx}\% of functions have the 
%same number of instructions.
%and
%quite a few pairs belong to wrong functions.
%
Besides, we examine function mismatch cases
(\ie, matching block pairs that consist of identical
instructions, which belong to different functions):
$14.1\%$ of such block pairs contain five instructions or less
whereas $1.5\%$ contain six or more, which
implies that a block granularity for
a binary similarity task is often insufficient
to deduce code semantics. 
In particular, most basic block pairs with one or two
instructions consist of \cc{nop}, \cc{jmp} or \cc{call}.

%A mismatched function case implies the weakness of CFG analysis
%with basic blocks in terms of infering binary semantics, 
%as it represents 
%basic blocks that have a sequence of identical normalized instructions
%in different context, \ie, function.
%%
%The output for basic blocks with less than five instructions 
%presents 14.12\% of mismatched function case,
%while 
%basic blocks with greather than five instructions
%generate 1.53\% of mismatched function case.
%%
%This result presents that basic block granularity 
%and getting contextual information from CFG are 
%not enough to infer code semantics to judge binary similarity.
%
In this regard, we investigate
large swaths of binary code that include approximately 
108 million machine instructions or 1.7 million
binary functions, summarizing our insights as follow.
First, the distribution of the instructions 
follows Zipf's law~\cite{zipfslaw} analogous to
a natural language. 
Second, oftentimes a function conveys 
contextually meaningful information.
Third, word2vec~\cite{word2vec} lacks diverse 
representations for the identical instruction
in a different position.
%necessitating a better representation reflecting a context
%(e.g., a single instruction may be represented 
%with multiple embeddings), 
%
Fourth, graph information (\eg, control flow graph)
may not play a pivotal role to determine code semantics,
which aligns with the recent findings~\cite{struct2vec}.

% Our approach
To this end, we present \sys, an architecture that is capable of
deeply inferring underlying code semantics based on the
cutting-edge BERT (Bi-directional Encoder Representations 
from Transformers) architecture~\cite{bert}.
%
% design choice
Based on the above insights, \sys has been carefully
designed so that it can leverage 
BERT to achieve our goals including
i)~\emph{function-level granularity}; \eg, the unit
of an embedding is a binary function,
ii)~\emph{function embedding as a whole}; \eg,
each instruction may have multiple representations
depending on the location of a function,
iii)~\emph{well-balanced instruction normalization}
that strikes a balance between too-coarse-grained
and too-fine-grained normalization, and
iv)~a \emph{two-phase training model}
to support a wide range of other downstream tasks
based on a pre-trained model.
%
% How did we model it?
\sys mainly consists of two separate training stages:
It creates a one-time \emph{generic code representation}
(\ie, pre-trained model or \syspre) 
applicable to \emph{any} downstream task 
that requires the inference of code semantics 
at a pre-training stage,
followed by generating a \emph{special-purpose code representation}
(\ie, fine-tuned model or \systask) for a given specific task 
based on the pre-trained model
at a fine-tuning stage.
Like the original BERT, the former stage employs
a general dataset 
with unsupervised learning,
whereas the latter stage makes use of
a task-oriented dataset 
with supervised learning.
%
% Strength
The key advantage of adopting the two-stage model
in \sys is to support potential applications
that allow for \emph{re-purposing}
a pre-trained model to quickly apply other
downstream tasks
using less expensive computational resources.
%

% Brief showcase of our result
%
We have applied \sys 
to both binary similarity comparison (\sysbs) and
toolchain provenance prediction (\syscopt) tasks.
The empirical results show that
\sysbs by far outperforms two
state-of-the-art binary similarity
comparison tools, obtaining 
a $49.8\%$ higher F1 than DeepBinDiff~\cite{deepbindiff} 
(up to 69\%) and
$15.8\%$ than SAFE~\cite{safe} (up to 28\%)
on average.
For \syscopt,
we obtain an F1 of $0.96$ and $0.91$ for 
compiler and optimization level (compilation provenance) 
prediction, respectively.
%
% Contribution
% \SP{I think it's better to move to related work 
% as we don't need to emphasize our rival}
% Note that this is not the first attempt
% (Zeping et al.~\cite{ordermatters})
% that leverages BERT to deduce code semantics; 
% however,
% to the best of our knowledge,
% \sys is the first approach that contributes
% the following:
%
In summary, we make the following contributions:
%this paper shares
%our experience and insights
%of binary analysis through a language representation model
%with BERT. 
%
\begin{itemize}[leftmargin=*]
	\setlength{\itemsep}{0pt}
	\setlength{\parskip}{0pt}
	\item 
		To the best of our knowledge, our work 
		is the first study to investigate binaries
		on a large scale to \emph{choose appropriate 
		design} for inferring code semantics.
		%We thoroughly investigate binary code on a large scale
		%to obtain fruitful insights for inferring code semantics.
	\item 
		We devise \emph{well-balanced instruction normalization}
		that can preserve as much contextual information
		as possible while maintaining efficient computation.
	\item
		We implement a simplified BERT architecture 
		atop our observations and insights on binaries, 
		which can generate
		semantic-aware code representations.
	\item
		We experimentally demonstrate both the effectiveness 
		% and efficiency of \sys with two useful
		% applications (binary similarity and toolchain prediction), 
		and efficiency of \sys with 
		binary similarity (\sysbs) and 
		compiler provenance prediction (\syscopt).
		% and, finally, shed light on myriad further applications.
		%
		In particular, \sysbs surpasses the two
		state-of-the-art binary similarity tools
		(\ie, DeepBinDiff~\cite{deepbindiff} and SAFE~\cite{safe}).
		%
%		Our pre-trained model (\syspre) can lead further applications
%		by utilizing code representations (\ie, function embeddings).

	%
\end{itemize}
The source code of \sys will be publicly available
including our pre-trained model
to foster further binary research in the near future.

\section{Background}
\label{s:bg}
%
%In this section, we describe a handful of 
%cutting-edge concepts to better understand 
%why our model has been designed to adopt them.
%
This section describes 
how we leverage a handful of cutting-edge concepts in the 
literature of natural language processing (NLP) into building \sys.
%
%Vulnerability detection
%Binary similarity (semantic distance)
%Code clone detection

%\XXX{use this section to implicitly highlight \sys's approach.}

%\subsection{Advanced Neural Network Architectures 
%	to Handle Sequential Data}
%\label{ss:transformer}
%
% RNN and Transformer
%\XXX{similarly, focus is not just RNN/Transformer,
%  but more about RNN/Transformer in the context
%  of ``binary'' analysis.}
%

\PP{Binary Code Representation}
\label{ss:coderepr}
Binary code represents machine instructions 
% with two numeric values (e.g., the digits of 0 and 1)
with two digits (\ie, 0 and 1)
as the final product after 
an extremely complicated compilation process. 
%
%A disassembler translates such machine instructions
%% into an assembly language (e.g., inverse process
%% of an assembler) to have better human-readability
%into an assembly language 
%to have better human-readability
%of binary code%
%~\footnote{The accuracy of the disassembly process is out of this
%	paper's scope, which has been widely discussed by
%	Pang \etal~\cite{sok-x86}.}.
%%
%Each instruction consists of a single opcode,
%and zero or more operands.
%
As a binary code holds a very concise representation
in that a majority of high-level concepts
(\eg, variable name, structure, type, class hierarchy)
have been lost due to a wide range of transformations at compilation,
it is quite challenging to deduce underlying contextual
meanings.
%
%A set of instructions may vary depending on different architectures,
%thus this paper focuses on \cc{x84\_64}.
%

\PP{Recurrent Neural Network.}
A recurrent neural network (RNN) is 
a specialized type of neural network designed
to process sequential data (\eg, text, audio, video
and even \emph{code}).
The RNN has shown great performance~\cite{linux-kernel-rnn}
on a sequence prediction task with
in-network memory that stores fruitful information
(\eg, state changes).
However, a na\"ive RNN struggles with capturing 
useful information from long sequences
because of the vanishing gradient problem~\cite{vanishing}.
%because the output at time $t$
%just relies on the previous input at time $t-1$.
%
A gating model, such as LSTM (Long Short-Term Memory)~\cite{lstm} 
and GRU (Gated Recurrent Unit)~\cite{gru}, has been proposed to
mitigate such a short memory issue by devising a special cell
for long-range error propagation.
%which becomes popular with an encoder-decoder 
%architecture~\cite{seq2seq} (known as a seq2seq model).
%
However, there are still several downsides: 
i)~limited capability of tracking long-term dependencies; 
simply put, a single vector from an encoder that implies 
all previous words may lose partial information, and
ii)~prohibiting parallelizable computation due to sequentiality.
A binary function often consists of 
a number of instructions, necessitating
a better architecture than either a strawman RNN or its variants.
%
%\XXX{for example, you can describe
%  why the normalized instruction sequences used
%  is too lengthy (so missing semantics) in binary analysis.}
%

\PP{Attention and Transformer}
%
%A recent advancement introduces an 
The main idea of the Attention~\cite{attention} mechanism 
is to consider all input words (at each time step)
when predicting an output word, particularly
paying attention to a specific word that is
associated with the output word for prediction.
This helps to capture a contextual relationship 
between words in a sentence
without worrying about a gradient vanishing problem,
which is now widely used in a machine translation domain.
%
%remember a long sequence 
%with an emphasis on the relevant
%word of an input, enhancing
%the quality of a neural machine translation task.
%
%
\begin{figure}[t!]
	\centering
	\includegraphics[width=0.99\linewidth]{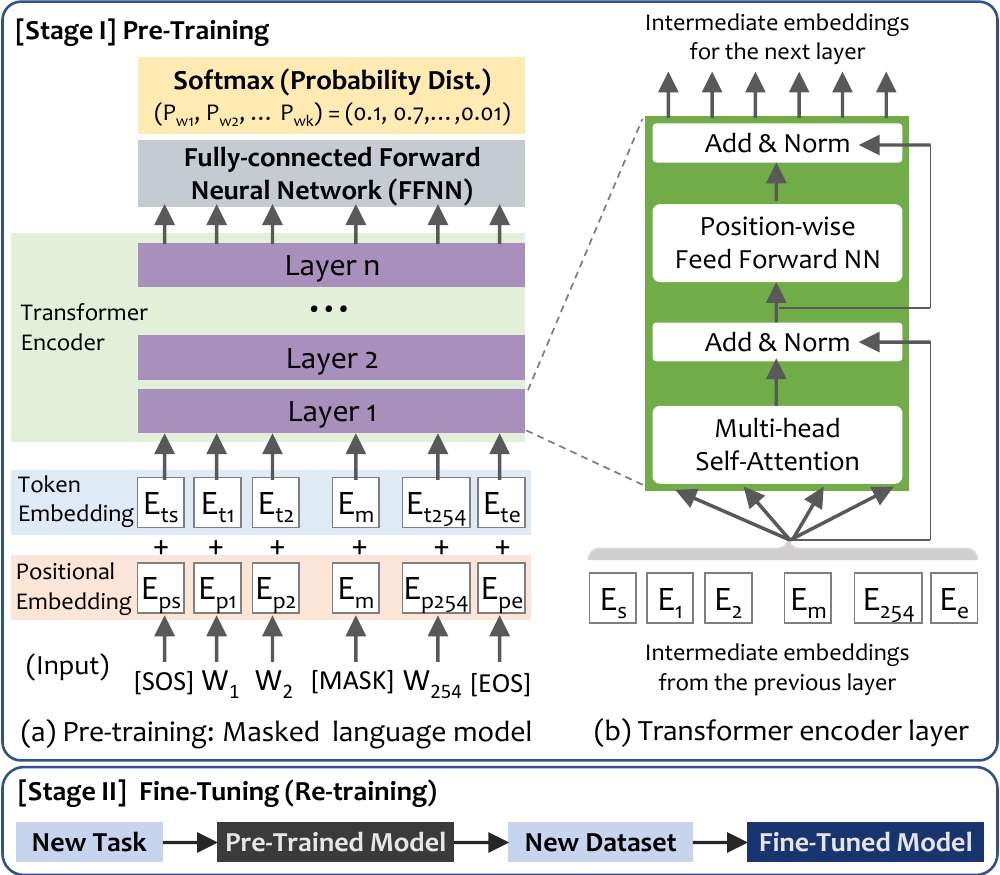}
	\caption{%\XXX{add all legends used in the figure (on the figure)}
		The simplified BERT structure.
		The key aspects of BERT are i)~taking a pre-training stage
		that generates a generic model and then reuses the model
		at a fine-tuning stage to create
		a specific model tailored to a downstream task, and
		ii)~adopting the Transformer's encoder alone.
		%
		%		Pre-training is a one-time processing, and
		%		fine-tuning is computationally less expensive.
		%
		\cc{E} denotes an embedding per word \cc{W}, 
		and the subscripts
		\cc{p}, \cc{t}, \cc{m}, \cc{s}, \cc{e}, and 
		a number represent a position, token, \cc{[MASK]}, 
		\cc{[SOS]}, \cc{[EOS}], and a word identifier,
		respectively.
		%
		%\XXX{emphasize one key aspect of BERT
		%	that readers should spot in the figure.}
	}
	\label{fig:bert-transformer}
\end{figure}
Transformer~\cite{transformer} proposes 
a \emph{multi-head self-attention} technique
for highly inferring the context of a sentence
(a binary function for our purpose)
atop the Attention's encoder-decoder architecture.
Self-attention focuses on the inner relationship 
between input words, and multi-head 
conceptually considers multiple Attention vectors
(\eg, multiple words with positional information
to predict the next word).
\autoref{fig:bert-transformer}~(b) illustrates
a single encoder layer: i)~taking input vectors
from the previous layer % as intermediate word embeddings
for computing an Attention matrix,
ii)~feeding the resulting vectors into
a feed forward neural network (FFNN) in turn,
iii)~applying both a batch (\eg, across training examples) and 
layer (\eg, across feature dimensions) normalization
between the self-attention and FFNN layers.
The original Transformer~\cite{transformer} 
has six encoders (\ie, six layers as shown
in~\autoref{fig:bert-transformer}~(a)) and six decoders.
%
%Such design of the Transformer well fits our goal
%to effectively capture the contextual relation 
%between instructions, eventually generating semantic-aware
%code representation.
%

\PP{Language Model and BERT}
BERT~\cite{bert}, Bi-directional Encoder Representations 
from Transformers,
is one of the state-of-the-art architectures to provide 
the rich vector representation of a natural language by 
capturing the contextual 
meanings of words and sentences
(instructions and functions for our model)
%corresponding instructions and functions, respectively, in our model
by adopting the encoder layer of Transformer~\cite{transformer}.
%with a few modifications.
%
It encompasses a number of advanced concepts
such as ELMo~\cite{elmo},
semi-supervised sequence learning~\cite{semi-supervised} and
Transformer.
BERT consists of two training phases 
as in \autoref{fig:bert-transformer}: 
a \emph{pre-training} process builds a generic model
with a large amount of corpus 
%to produce generic representations, 
and a \emph{fine-tuning} process updates the pre-trained model 
that is applicable to a specific downstream task.
The former takes two strategies: 
masked language model (MLM) and 
next sentence prediction (NSP) for
building a language model that considers
context and the orders of words and sentences, which
%\XXX{Fix}
can be achieved by unsupervised learning
with an unlabeled dataset.
%
%It is noteworthy that our scheme modifies the original BERT
%by removing NSP because,
%%a function logically associates with each other in a call-invocation manner.
%unlike two subsequent sentences that
%have a contextual relationship in NLP,
%two consecutive functions in a binary
%do not have any meaningful relationship
%(a caller and callee relationship instead).
%
In \autoref{fig:bert-transformer} (a), 
the \cc{[MASK]} token represents an input word
that has been masked, and \cc{[SOS]} and \cc{[EOS]}
are tokens for the start and end of a sentence,
respectively~\footnote{\cc{[CLS]}
	and \cc{[SEP]} tokens in the original BERT 
	correspond to our
	\cc{[SOS]} and \cc{[EOS]}.}.
The \cc{[UNK]} token is used for unknown words.
This example has a 256 fixed-length input 
(254 words with masked ones excluding
two special tokens: start/end of a sentence
at both ends) at a time.
Once the pre-training is complete,
the pre-trained model can be recycled for
varying user-defined downstream tasks
% with a labeled dataset (supervised learning).
with supervised learning.
We adopt BERT because the structure 
can fit into our objective seamlessly:
creating a pre-trained model that 
contains a generic binary code representation, and 
re-training that model
for a wide range of different classification
tasks with relatively lower computational resources
(See \autoref{ss:pre-training} in detail).
%
%\XXX{\sys just benefits from BERT; nothing but an application?
%maybe we may need to carry out additional experiment; e.g.,
%applying all instructions with the original BERT!}
%
%In particular, we will demonstrate the following two applications:
%i)~binary similarity comparison and 
%ii)~compiler and optimization level prediction
%(See \autoref{ss:fine-tuning}).
%
%
%\XXX{last remark should be: ``attention'' is not yet applied
%  nor known its effectiveness in the
%  context of binary representation,
%  and our work makes a first step toward demonstrating that
%  attention can effectively capture the contextual relation b/w instructions.}

%\XXX{perhaps in the next subsection,
%  to avoid the impression that we just applied BERT,
%  you might hint on some challenges of adopting BERT
%  in terms of binary representation.}

%\subsection{Language Model and BERT Structure}
%\label{ss:bert}
%

%
%\XXX{clarify. ``general representation'' and ``typically''
%  do not make sense in this paragraph.
%  It means, general representation is typically expensive
%  but by specializing into three applications,
%  you could avoid the problem of expensive? computation?}
%Typically creating general representations is fairly
%expensive.
%

\section{Binary Code Semantics}
\label{s:overview}
\begin{figure}[t]
	\centering
	\includegraphics[width=0.8\linewidth]{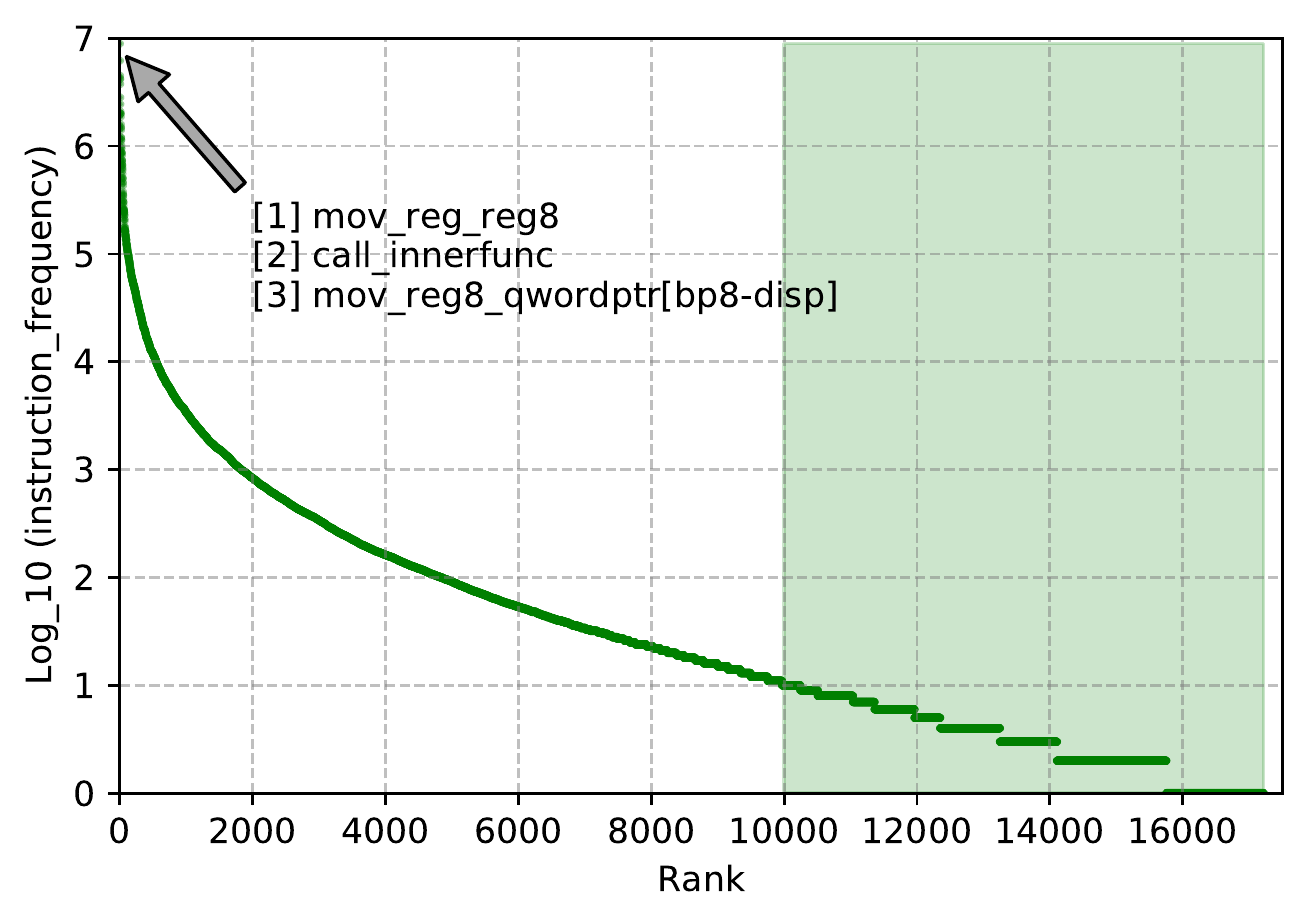}
	\caption{
		%\XXX{could you put say, three well-know instructions
		%    marked with arrows on the graph?}
		Log scale of instruction frequencies
		based on each rank in our vocabulary
		corpus.
		The curve closely follows Zipf's law, akin to
		a natural language.
		The green area illustrates 
		a long tail that has been rarely seen 
		(\eg, less than 10 times).
		Interested readers refer to~\autoref{t:freqrank}
		in Appendix.
	}
	\label{fig:rankfreq}
\end{figure}
In this section, we discuss the definition of code semantics,
followed by highlighting a few insights of binary
codes with common compilation toolchains and optimizations.
\subsection{Definition of Code Semantics}
%
%\XXX{We define the code semantics in binary representation
%  differently from the ones in source code. In specific,
%  we define the code semantics as XXX (described below).}
%
We view code semantics in a binary representation
differently from the ones in source code.
Specifically, the \emph{equivalent semantics of a binary code}
can be defined as a sequence of instructions 
that carries out an identical task
from a logical function in the original source.
A binary function differs from a programmer-written function
due to varying transformations by a compiler toolchain.
%
%\XXX{check the logic here!}
%For instance, two binary functions perform 
%the same sorting task; one with a Bubble Sort algorithm and
%another with QuickSort, then these two functions
%are considered different according to our definition.
%
We use a cosine similarity score ([-1,1] range) that 
represents the relationship between the two binary functions,
meaning that the higher value, the closer in code semantics.
%
%\XXX{say, two binary functions with varying compilation optimizations
%  that are generated from the same source
%  are ideally 1.
%  %
%  And note that two binary functions,
%  say two sorting algorithms,
%  that are from different, original code,
%  are considered different (-1)
%  according to our definition.}.
%
%\XXX{This sentence doesn't buy you much from reviewers.
%  It meant to \emph{imply} that we didn't overlook
%  by saying that others are careless.
%  We better just say it we first to highlight our insights derived from the
%  dataset or binaries}
%It is commonplace that researchers and practitioners overlook
%deep understanding of a dataset itself 
%albeit their awareness of its importance
%when generating a model with machine learning techniques.
%
%\PP{Motivating Example}
%%
%\KK{An example with two disassembly from different optlevels
%but the same source code; show how much it looks differently}
%
\subsection{Observations and Insights}
\label{ss:insights}
A binary code is a sequence of machine instructions 
analogous to a natural language.
Indeed, InnerEye~\cite{innereye} borrows the ideas of
Neural Machine Translation (NMT) to a binary
function similarity comparison task by regarding
instructions as words and basic blocks as sentences.
For successful binary code representation
with deep neural networks, it is essential to
carefully understand its properties.
Here are several insights based on our observation of
machine instructions.
\begin{figure}[t]
	\centering
	\begin{subfigure}{0.48\textwidth}
		\includegraphics[width=\columnwidth]{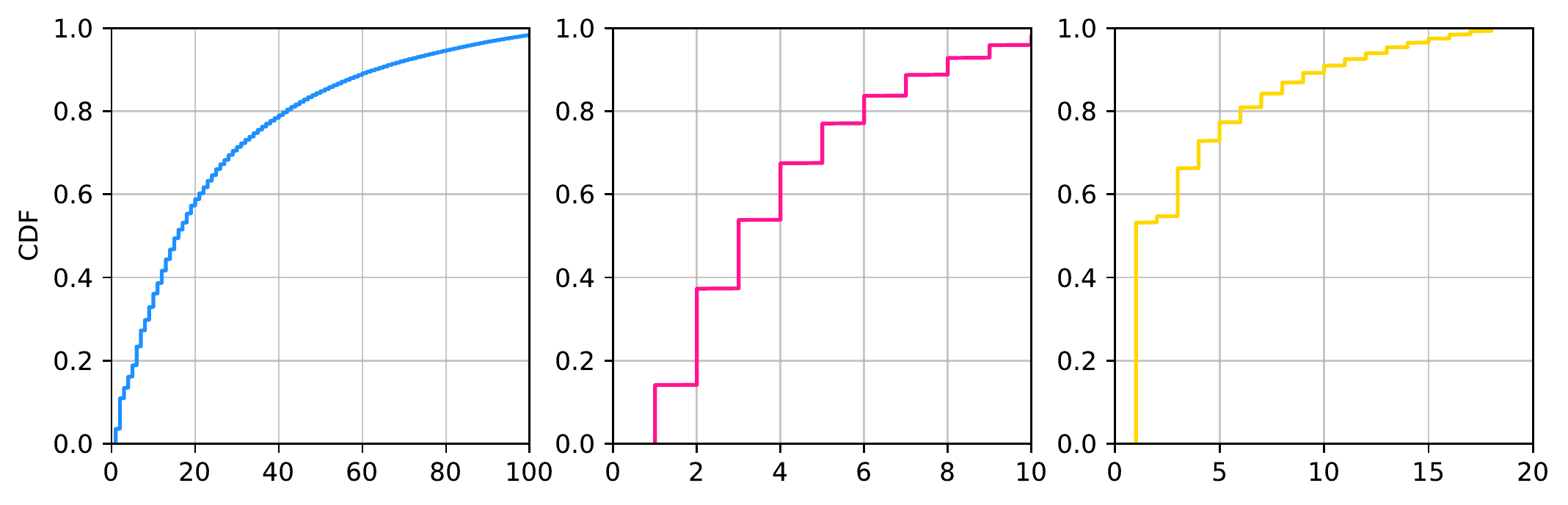}
		%\caption{CDF of occurrence rates. }
		\label{figsub:fbicdf}
	\end{subfigure}

	\hfill
	
	\begin{subfigure}{0.48\textwidth}
		\includegraphics[width=\columnwidth]{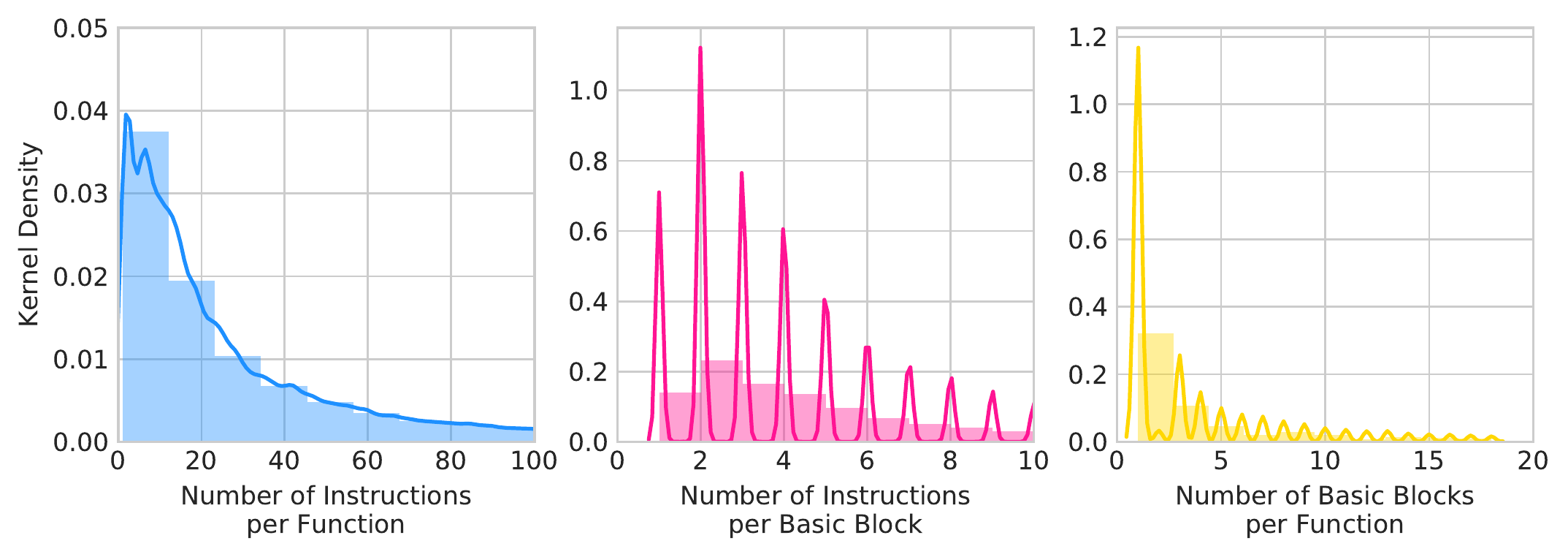}
		%\caption{Distribution of occurrence rates for 
		%	each statistics.}
		\label{figsub:fbidist}
	\end{subfigure}
	\caption{
%			\XXX{general comments.
%			for most figures you have, you describe what they are,
%			so users should interpret what you provided
%			and infer your intention of showing the graph.
%			answer ``so what?''.
%			in figure3, you did a good job by concluding that it's zipf-like
%			distribution that follows natural language.}
		CDFs and histograms with kernel
		density estimates ($10$ bins) for the number of
		instructions per function (left) and basic block (middle), 
		and the number of basic blocks per function (right)
		after removing outliers.
        A majority of basic blocks (around 80\%) contain 
        merely five instructions or less, indicating that 
		a larger granularity is needed
		to imply contextually fruitful information. 
	}
	\label{fig:fbi}
\end{figure}
\begin{figure}[t]
	\centering
	\includegraphics[width=0.99\linewidth]{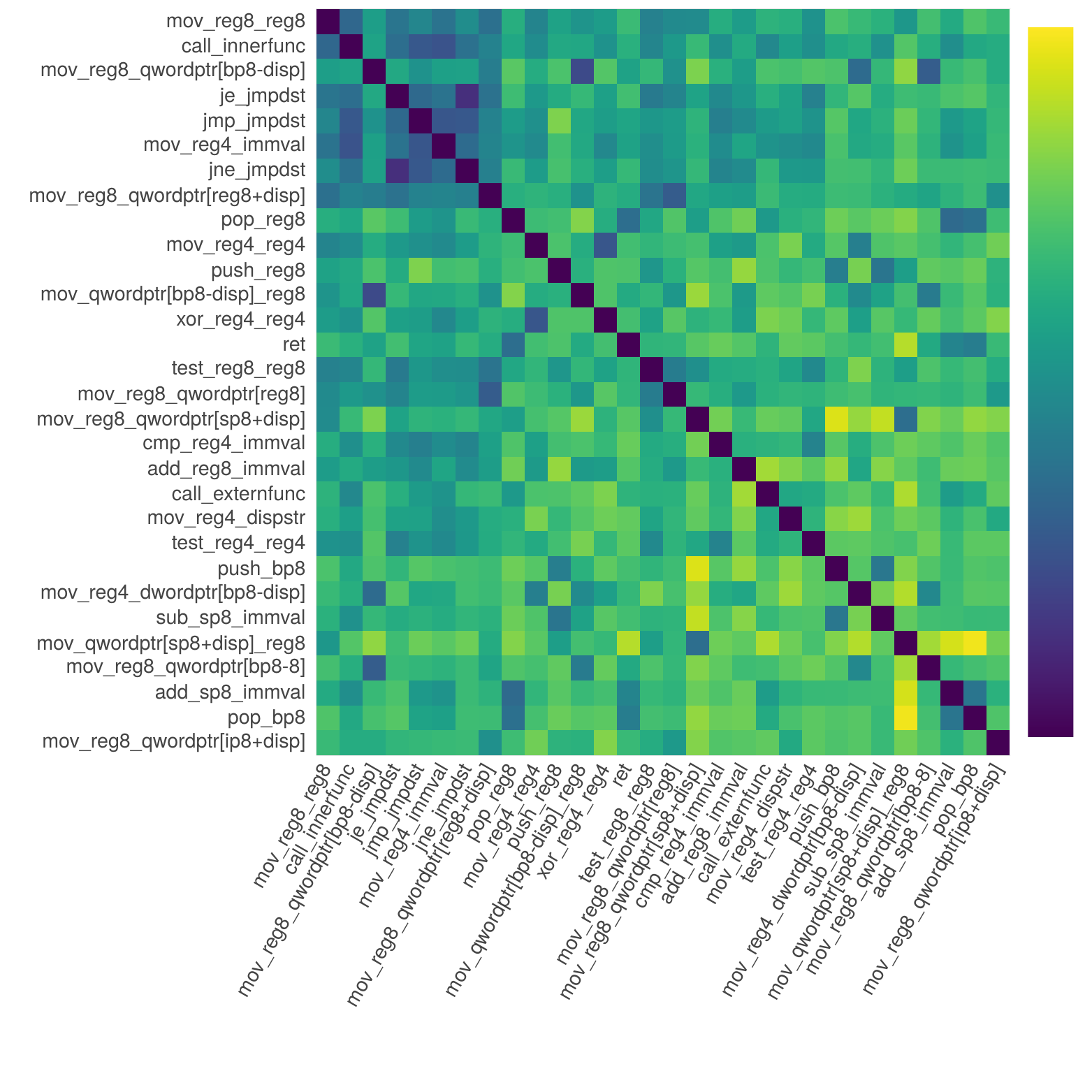}
	\caption{Visualization of a similarity matrix 
		for the most common 30 instructions
		with a Word2Vec model.
		The dark color represents
		that the relationships of two instructions
		are close to each other.
		For example, oftentimes \cc{ret} (middle-left) 
		comes with a series of function epilogue instructions
		(\eg, \cc{pop bp8} and \cc{add sp8 immval}
		at the right-bottom corner).
	}
	\label{fig:w2vmodel}
\end{figure}
\begin{itemize}[leftmargin=*]
	\setlength{\itemsep}{0pt}
	\setlength{\parskip}{0pt}
	\item \textbf{Machine instructions follow Zipf's Law.}
	\autoref{fig:rankfreq} depicts the relationship
	between the rank of instructions
	and the log scale of their frequencies.
	Our finding shows that the curve of the instruction
	distribution closely follows Zipf's law~\cite{zipfslaw}
    like a natural language, which implies that utilizing  
    effective techniques in an NLP domain such as BERT 
    works for a binary task.
%	%
%	\KK{I added this part to support that
%	using BERT from an NLP domain makes sense;
%    It looks interesting to see as no one
%    seems to explore it before.}
	%
	\item \textbf{A function often conveys a
		meaningful context.}
	We analyze $1,681,467$ functions ($18,751,933$ basic blocks
	or $108,466,150$ instructions) excluding linker-inserted ones
	in our corpus (\autoref{t:dataset}).
	We measure several statistics:
	i)~the number of instructions per function on average (I/F)
	  is $64.5$ (median=$19$, std=$374.7$),
	ii)~the number of basic blocks per function on average (B/F)
	  is $5.8$ (median=$4$, std=$16.4$), and
	iii)~the number of instructions per basic block on average (I/B)
	  is $11.2$ (median=$3$, std=$95.8$).
	As the standard deviation is quite large, we remove outliers
	by cutting off the values that are bigger than 
	the standard deviation 
	times three, which is around $12\%$, finally obtaining
	a mean of (I/F, B/F, I/B) = $(25.1, 3.9, 3.7)$.
	\autoref{fig:fbi} illustrates CDFs (upper) and histograms 
	 (below) without outliers, counterintuitively showing that 
    approximately 70\% of basic blocks
    include five instructions or less where
    a binary function contains
	around four basic blocks with 25 instructions 
	on average.
	%    
%    \autoref{fig:fbi} implies that a majority of functions can contain more 
%    contextual information than a majority of basic blocks can.
	%
	We choose a granularity
	as a single function that is large enough
	to be able to convey
	contextually meaningful information.
	Indeed, quite a few matching blocks
	from our experimental results (\autoref{fig:deepbindiff})
	using DeepBinDiff~\cite{deepbindiff} contain
	a couple of instructions (\eg, \cc{jmp}, \cc{call}),
	which are highly likely to miss surrounding contexts
	~\footnote{Instead, DeepBinDiff considers 
	CFGs to read underlying context.}.
	We discuss other cases that a fine-grained granularity
	becomes beneficial (See~\autoref{s:discussion}). 
	\item \textbf{Word2vec lacks diverse representations for
		the same instructions in a different position.}
	A majority of prior works~\cite{innereye, asm2vec, deepbindiff}
	adopt a Word2vec~\cite{word2vec} algorithm to
	represent a binary code.
	Word2vec is an embedding technique
	that aims to learn word relationships from a large 
	corpus text, representing each distinct word with
	a vector.
	\autoref{fig:w2vmodel} illustrates
	%\XXX{here suddenly normalized pops up, confusing}
	the Top 30 most common instructions
	that are well associated with each other.
%    \SP{ I think the Figure 4 is fancy but too hard to 
%         read data. For example, I spent almost 5 mins to find "pop reg8"
%         and "add sp8 imm" in the Figure to understand the following example.
%         You can consider to drop this figure and explain the relationship of the 
%         two instructions (in the example) with embedding data (numbers).
%     }
	% 
	%\XXX{rephrase this part - may be confusing}
	%
	Word2vec itself cannot differently represent
	the identical instruction in a distinct context
	due to the absence of position information.
	For instance, a behavior of popping a register
	for a function epilogue differs from the one
	for other computation in the middle of the function.
	However, Word2vec represents the identical
	representation (embedding) for the same word 
	regardless of their contextual differences, 
	necessitating 
	a better embedding means
	with the \emph{restricted} number of vocabularies.
	\item \textbf{CFG within a function 
		may not be fruitful.}
	Previous works 
	often employ a control flow graph~\cite{deepbindiff, gemini} 
	($G(V,E)$; basic block as a vertex
	and flow as an edge) or graph isomorphism~\cite{discovre}
	as a feature to compare a binary code.
	However, our finding shows that
	both the numbers of vertices and edges
	do not often stay identical ($|V|\neq|V'|,\ |E|\neq|E|$)
	across different optimization levels,
	to which the isomorphism cannot be applicable.
	The empirical results of a recent study~\cite{struct2vec} 
	align with our insight.
	Instead of having CFGs, we feed such fruitful 
	flow information with well-balanced 
	normalization (\autoref{ss:balanced-grained-normalization})
	to deep neural networks
	(\eg, by defining new words for calling \cc{libc} functions).
	%
%	\XXX{Reviewers seem to be hard to accept this part
%	maybe because of a bold claim? Maybe a little
%	different way to explain.}
	%
\end{itemize}
%
%\PP{Necessity of Better Instruction Normalization 
%	and Neural Network Architecture 
%	for Code Semantics.}
%
Unlike a natural language, the number of
possible instructions is countless when
mapping each instruction into a single word 
(token or vocabulary); an immediate value in a 64-bit operand 
of the instruction may produce $2^{64}$ different words,
prohibiting further computation.
%
%\XXX{put the topic sentence first in this par.}
In this regard, most of prior approaches that harness
deep learning techniques~\cite{asm2vec, 
	deepbindiff, ordermatters, innereye}
adopt \emph{normalization} before feeding 
a sequence of instructions
as an input into a training process.
In particular, striking a balance is crucial 
so that instruction normalization
is neither too generic nor too specific because 
each token holds rich information yet minimizing an OOV problem,
necessitating a better instruction normalization technique
to capture code semantics for neural networks.
%
%For example, if the destination of a \cc{call} opcode
%has been converted into a single target (e.g., \cc{call\_dst}),
%it would lose contextual semantics.
%
\autoref{ss:deepsemantic-overview} justifies
the decision of \sys design to comply with
our insights.
%
%We have carefully observe the distribution of x86 instructions.
%Extreme carefulness must be taken when normalizing instructions.
%
%\XXX{perhaps, visualizing a normalization process (w/ a list of instructions)
%  where common normalization approaches fail to capture
%  some semantics but \sys can.}
%

\section{\sys Design}
\label{s:design}

In this section, we provide an outline of \sys, and
portray the design of \sys in detail.
%\XXX{RD: clarify the concept of pre-training and fine-tuning.}

\subsection{\sys Overview}
\label{ss:deepsemantic-overview}
\begin{figure}[t!]
	\centering
	\includegraphics[width=0.99\linewidth]{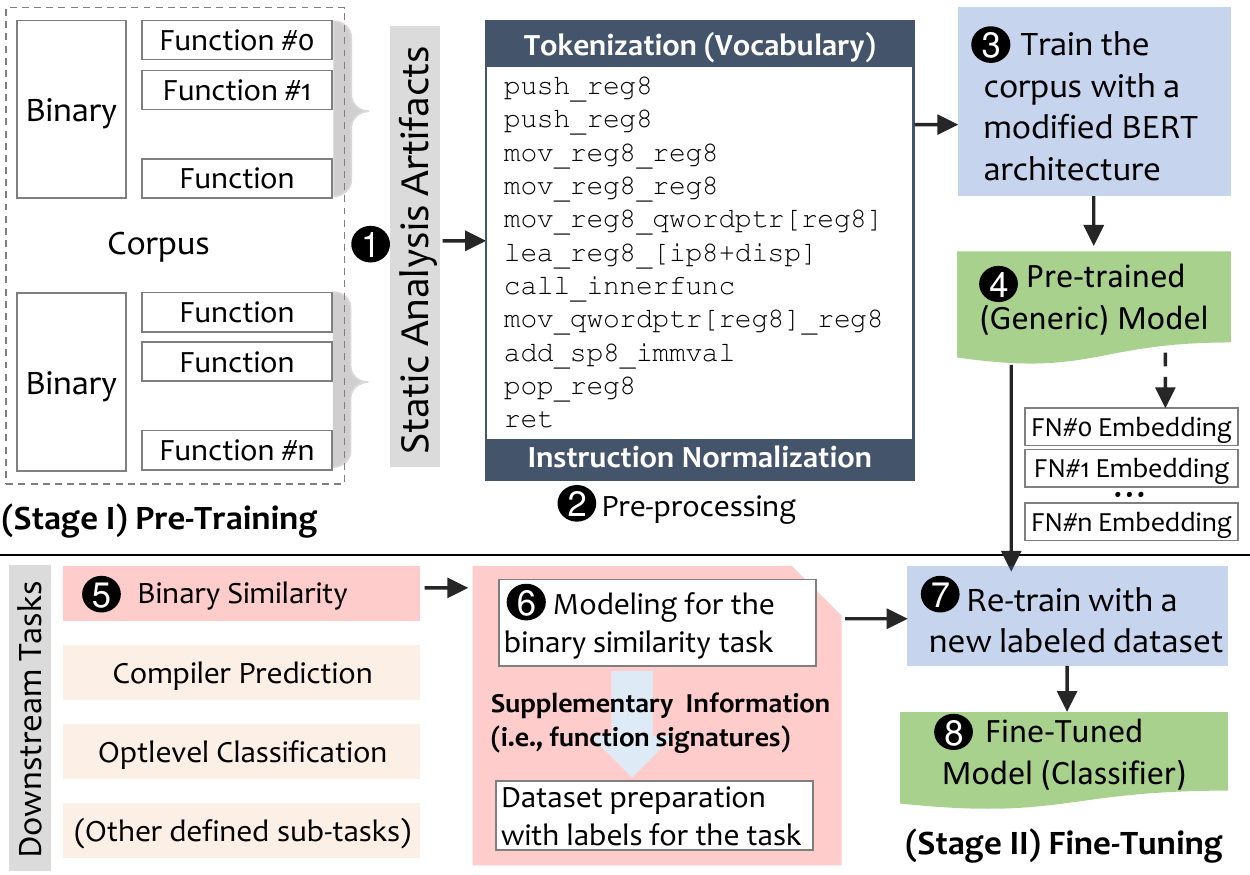}
	\caption{\sys Overview. 
		We collect artifacts via static analysis
		with a pre-defined binary corpus [1].
		A pre-processing step tokenizes all instructions 
		with normalization [2]
		(\autoref{ss:balanced-grained-normalization}).
		The entire corpus is pre-trained with BERT [3],
		generating a pre-trained model [4]
		(\autoref{ss:pre-training})
		We define a downstream task (\eg, binary similarity comparison)
		[5], and prepare another dataset with labels 
		accordingly [6].
		Lastly, a fine-tuned model is obtained [8]
		(\autoref{ss:fine-tuning})
		after re-training the dataset for the task [7].
		%
%		We adopt Masked Language Model from the original 
%		BERT architecture that allows for unsupervised learning
%		while initial embedding generation. A downstream task
%		contains a label with a new dataset for re-training.
%		This example illustrates a task of 
%		binary similarity prediction for fine-tuning.
	}
	\label{fig:overiew}
\end{figure}
\sys consists of two separate stages, 
as illustrated in~\autoref{fig:overiew}:
i) a pre-training stage that creates a general 
model applicable to a downstream task, and
ii) a fine-tuning stage that
generates another model for a specific task
on top of the pre-trained model.
%
%The design of \sys architecture adopts
%the original BERT (see, \autoref{ss:bert})
%without employing the strategy of NSP (Next Sentence Prediction)
%because a function logically associates
%with others in a call-invocation manner 
%regardless of the proximity of their locations or placements.
%
%In brief, we view each instruction as a word and
%a function as a sentence in NLP.
%
%However, unlike two subsequent sentences have a contextual relationship in NLP,
%two consecutive functions in a binary
%do not employ any meaningful relationship.
%A caller and callee relationship in a function call, instead,
%should be taken into account
%in representing the semantics in a binary.
%
The following describes four design decisions and their rationales
behind \sys:
\begin{itemize}[leftmargin=*]
	\setlength{\itemsep}{0pt}
	\setlength{\parskip}{0pt}
	\item \textbf{Function-level Granularity.}
	We determine a function as
	a minimum unit that can imply meaningful semantics
	from our insights in~\autoref{ss:insights}.
	%
%	As shown in~\autoref{fig:fbi}, a single basic block
%	holds approximately five instructions on average
%	(median is even lower) \KK{check!}, which often
%	suffers from a unique context representation.
	%
	Indeed, our experiment shows
	a large portion (\eg, $83.25\%$) of 
	basic block matching 
	results from previous results~\cite{deepbindiff} 
	come from very small basic blocks 
	as in \autoref{fig:deepbindiff}.% (\eg, five or less).
	\item \textbf{Function Embedding.}
	Along with the function-level granularity, 
	\sys generates an embedding per function as a whole,
	rather than per instruction (\eg, word2vec) 
	for code representation.
	This means even an identical instruction would have
	a different embedding upon its position
	and surrounding instructions (\autoref{fig:bert-transformer}).
%	which is part of input information (\eg, positional embedding)
%	that feeds into Transformer encoder.
	%
%	For example, the representation of \cc{ret} in
%	the middle of a function
%	may differ from that of another at the end.
	%
	\item \textbf{Well-balanced Normalization.}
	We leverage the existing static binary analysis
	to normalize instructions so that a pre-training model
	can naturally embrace important features in a deep neural network.
	%with supplementary information.
	%
	We intentionally attempt to remain as much information 
	(manually engineered features from previous studies~\cite{blanket})
	as possible (\autoref{ss:balanced-grained-normalization}).
	\item \textbf{Model Separation.}
	Unlike prior approaches, our model requires
	two trainings: one for pre-training
	and another for fine-tuning per user-defined task. 
	The flexible design of \sys suits 
	a wide range of domain-specific sub-tasks, which we will showcase 
	two applications: a binary similarity and
	toolchain prediction (\autoref{ss:fine-tuning}) task.
	%
%	Since we adopt a masked language model from BERT,
%	the pre-training stage can be done without labeling
%	(unsupervised learning) where the fine-tuning stage
%	requires a labeled dataset (supervised learning).
	%
\end{itemize}
As \sys inherently generates multiple models, 
the pre-training and fine-tuning models are
dubbed \syspre and \systask, respectively,
depending on the task (\eg, \sysbs for
binary similarity, \syscopt for compiler
provenance).
% , to avoid further confusion.
%
\begin{figure*}[!ht]
	\centering
	\includegraphics[width=0.99\linewidth]{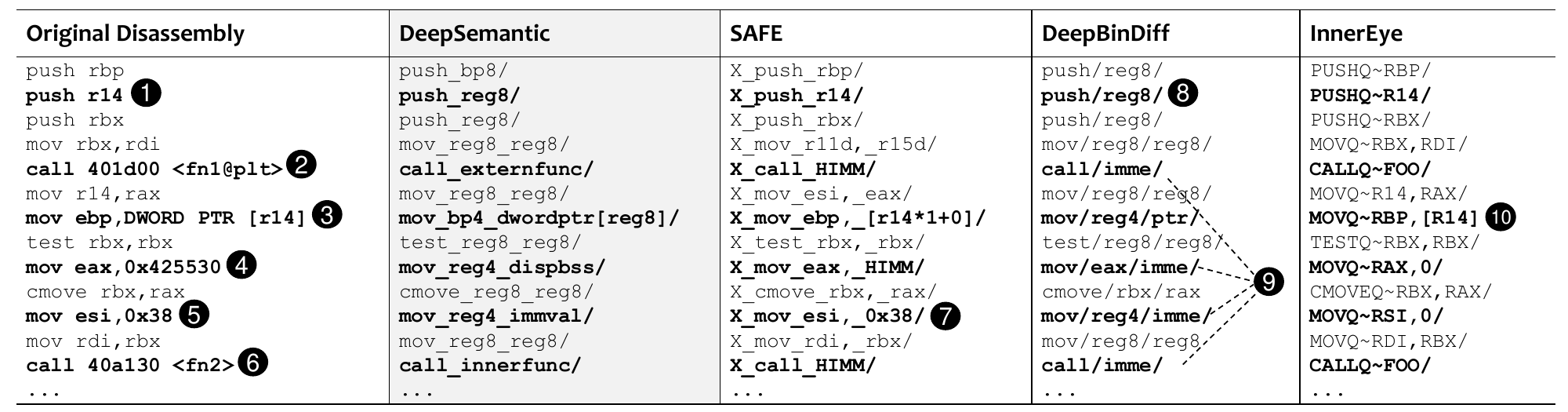}
	\caption{%\XXX{what are circled numbers for?}
		Varying strategies of normalizing (disassembled)
		instructions prior to code embedding generation. 
		A slash (/) represents a token separator; \eg, 
		DeepBinDiff~\cite{deepbindiff} separately tokenizes 
		opcode and corresponding
		operand(s) by splitting an instruction.
		Note that a strategy that is too coarse-grained 
		(\eg, DeepBinDiff~\cite{deepbindiff}, InnerEye~\cite{innereye})
		may lose contextual information; on the other hand,
		a strategy that is too fine-grained (\eg, SAFE~\cite{safe})
		may suffer from an OOV problem 
		(See \autoref{ss:balanced-grained-normalization}).
	}
	\label{fig:norm-ex}
\end{figure*}

\subsection{Well-balanced Normalization}
\label{ss:balanced-grained-normalization}
An instruction normalization process is essential
to prepare its vectorization form fed into a neural network
as adopted by many prior approaches~\cite{safe, 
	deepbindiff, innereye, asm2vec, ordermatters}.
However, a too coarse-grained normalization,
  such as stripping all immediate values~\cite{asm2vec, 
  		deepbindiff, innereye}
loses a considerable amount of contextual information
whereas a too fine-grained normalization
 close to instruction disassembly
 raises an OOV problem
 due to a massive number of unseen instructions (tokens).
% \eg, an arbitrary immediate value 
% ranges from $-2^{32}$ to $2^{31}-1$.
%
%Indeed, a number of prior approaches that adopt NLP techniques%
%~\cite{safe, deepbindiff, innereye, asm2vec, ordermatters} normalize
%instructions as a preprocessing step before vectorizing them.
%
%\XXX{put this topic sentence first. revise like this:
%  Coarse-grained normalization methods
%  such as stripping immediate values [cite], etc,
%  do not convey capture contextual information
%  of instructions when used as input for training.
%}
%
We observe that the previous approaches sorely
perform mechanical conversion of 
either an opcode or operand(s) 
without thorough consideration
on their contextual meanings.
\autoref{fig:norm-ex} shows
different normalization strategies
taken by several approaches for binary similarity detection.
DeepBinDiff~\cite{deepbindiff} considers a register size
to symbolize an n-byte register (\BC{8}); however,
it converts all immediates into \cc{imme} (\BC{9}).
Meanwhile, InnerEye~\cite{innereye}
discards the size information of registers (\BC{10})
%by converting all registers into a 64-bit register 
for a 64-bit machine instruction set.
SAFE~\cite{safe}
%originally designed for creating a binary
%function embedding, %\XXX{please clarify ``some''}
retains immediate values (\BC{7}).
%which possibly encounters an OOV problem 
%upon unseen instructions.
%
Besides, all three cases convert the destination of
a call invocation into a single notation 
(\eg, \cc{HIMM}, \cc{imme}, or \cc{FOO}),
rendering every call instruction identical.

To this end, we establish a
\emph{well-balanced normalization strategy} 
to strike a balance between 
expressing binary code semantics 
as precisely as possible and
%\XXX{need clarification here}
maintaining reasonable amount of tokens, that is,
a small number of tokens may lose original semantics
whereas a large number of tokens may suffer from 
an OOV problem.
%
%\XXX{clarify: in what sense?
%  \eg, acceptable for computation?}
%because a too coarse-grained approach
%may lose the original semantics (\eg, converting
%all call operands into a single target) whereas
%a too fine-grained approach may suffer from
%both an expensive computation (\eg, back propagation) 
%and %\XXX{you already used OOV before}
%OOV problem.
%
The quality of instruction normalization is of significance
because word embeddings eventually rely on 
an individual normalized instruction, 
holding final contextual information.
For instance, an immediate can imply one of 
the following: a target of library, call invocation within 
or out of the current binary, destination to jump to,
string reference, or statically-allocated variable;
however dropping such an implication makes
the embedding rarely distinguishable
from each other.
For instance, the two most frequent words
(\cc{mov\_reg8\_ptr} and \cc{mov\_reg8\_reg8})
out of two thousand vocabularies
when applying coarse-grained normalization 
account for more than 20\% of appearances, according to
our experiment,
which is unable to convey a valid context.
Note that the OOV problem can be minimized 
with our strategy. 

\autoref{t:normalization-rules} summarizes three
basic principles in mind to balance seemingly 
conflicting goals above:
i)~an immediate can fall into 
a jump or call destination
(\eg,
\BC{2} \cc{0x401d00} $\rightarrow$ \cc{externfunc},
\BC{6} \cc{0x425530} $\rightarrow$ \cc{innerfunc}),
a value itself
(\eg,
\BC{5} \cc{0x38} $\rightarrow$ \cc{immval})
or a reference
(\eg,
\BC{4} \cc{0x425530} $\rightarrow$ \cc{dispbss}),
according to a string literal, statically allocated variable
or other data;
ii)~a register can be
classified with a size by default
(\eg,
\BC{1} \cc{r14} $\rightarrow$ \cc{reg8},
\BC{4} \cc{eax} $\rightarrow$ \cc{reg4});
but the ones with a special purpose stay intact such as 
a stack pointer, instruction pointer, or base pointer
(\eg,
\BC{3} \cc{ebp} $\rightarrow$ \cc{bp4}); and
iii)~a pointer expression follows the original 
format, ``\cc{base+index*scale+displacement}''
(\eg,
\BC{3} \cc{DWORD PTR [r14]} $\rightarrow$ \cc{dwordptr[reg8]})
so that certain memory access information can be preserved;
moreover, the same rule applies if and only if
the displacement refers to a string reference 
(\eg, \cc{dispstr}).
Note that opcode is not part of
our normalization process.

\begin{table*}[t]
	\footnotesize
	\renewcommand{\arraystretch}{1.05}
	\scalebox{1.10} {
		\begin{tabular}{lllll}
			\toprule
			\textbf{Operand} & \textbf{Rule} & \textbf{Description or Expression} & \textbf{Notation} & \textbf{Note} \\
\hline
\textbf{Immediate} & call target & \cc{libc} library call & \cc{libc}[name] & points to a libc function\\
& & recursive call & self & points to a function itself \\
& & function call within a binary & innerfunc & points to a \cc{.text} section \\
& & function call out of a binary & externfunc & points to a \cc{.got} or \cc{.plt} section \\
\cline{2-5}
& jump family & destination to jump to & jmpdst & destination within a function \\
\cline{2-5}
& reference & string & dispstr & refers to a string\\
& & statically allocated variables & dispbss & points to a \cc{.bss} section \\
& & data & dispdata & refers to data other than a string \\
\cline{2-5}
& default & all other immediate values & immval & all cases other than the above\\

\hline
\textbf{Register} & size & [e|r]*[a|b|c|d|si|di][x|l|h]*, r[8-15][b|w|d]* & reg[1|2|4|8] & size information \\
& stack/base/instruction & [e|r]*[b|s|i]p[l]* & [s|b|i]p[1|2|4|8] & special registers (\eg, stack) \\
& special purpose & cr[0-15], dr[0-15], st([0-7]), [c|d|e|f|g|s]s & reg[cr|dr|st], reg[c|d|e|f|s]s & special registers (\eg, flags)\\
& special purpose (avx) & [x|y|z]*mm[0-7|0-31] & reg[x|y|z]*mm & Advanced vector extensions registers \\
\hline
\textbf{Pointer} & direct (small size) & byte,word,dword,qword,ptr & memptr[1|2|4|8] & pointer with a small size (<=8 bytes)\\
& direct (large size) & tbyte,xword,[x|y|z]mmword & memptr[10|16|32|64] &  pointer with a large size (>8 bytes) \\
& indirect (string) & [base+index*scale+displacement] & [base+index*scale+dispstr] & pointer that refers to a string\\
& indirect (others) & [base+index*scale+displacement] & [base+index*scale+disp] & pointer with a displacement \\

			\bottomrule
		\end{tabular}
	}
	\centering
	\caption{
		%\XXX{add your justification of each rules in the right most column}
          Well-balanced normalization rules
		for the representation of \cc{x86\_64}
		instruction operands: 
		immediate, register and pointer.
		%\XXX{capstone removed! clearly state innerfunc/externfunc etc.}
		%
		Our strategy aims to remain as much contextual information
		as possible for further embedding generation.
		%
%		Note that the following rules have been
%		applied to disassembly expressions from 
%		Capstone~\cite{capstone}.
	   }
	\label{t:normalization-rules}
\end{table*}
\subsection{Pre-training Model}
\label{ss:pre-training}
%
%\XXX{put the topic sentence first. revise.} 
It is necessary to train \sys
with a completely new dataset (\eg, machine 
instructions in our corpus)
with an entirely different vocabulary
(\eg, normalized instructions).
However, in general, it is possible to employ
a pre-trained model 
with a large corpus of human words (\eg, from Wikepedia)
%that is readily available 
in an NLP domain because
BERT training is not only computationally expensive 
but also excellent to apply to various downstream pipelines 
by design~\cite{squad, glue, multinli}.
%
%As with \autoref{eq:kqv-attention}, the heart of
%the pre-training model is computing probability of
%all normalized instruction distributions. \KK{check!}
%

We adopt the original BERT's masked language model
(MLM) that probabilistically masks a pre-defined portion of 
normalized instructions (\eg, 15\%), 
followed by predicting them in a given function
during pre-training.
%
%As explained in~\autoref{s:bg}, 
It is noteworthy that \sys does not employ 
NSP (\ie, prediction of next sentence)
because two consecutive functions often
do not connect semantically.
We visualize the internal structure
of multi-head self-attention with multiple layers
in Appendix (\autoref{s:appendix}) for the interested readers.
Besides, our model instinctively takes advantage
of Transformer (comparing to prior RNN models), which allows for 
direct connection between all instructions 
effectively, and 
highly parallelizable computation (\eg, GPU resource).

%\XXX{going to fill more?}

%BERT and transfer learning
%to train a model in one domain, 
%and then leverage the acquired knowledge 
%to improve the model’s performance in another domain. 
%%
%one unsupervised task: masked language modeling 
%(predicting a missing word in a sentence)
%NSP; does not adapt it
%%
%underlying mechanism: attention 
%(enabling BERT to learn a variety of rich lexical relationships) 
%and Transformer 
%how the words relate to each other in the context of the sentence. 
%%
%The attention mechanism enables the model to do this, 
%by forming composite representations that the model can reason about. 

%
\subsection{Fine-tuning Model}
\label{ss:fine-tuning}
\begin{figure}[t!]
	\centering
	\includegraphics[width=0.99\linewidth]{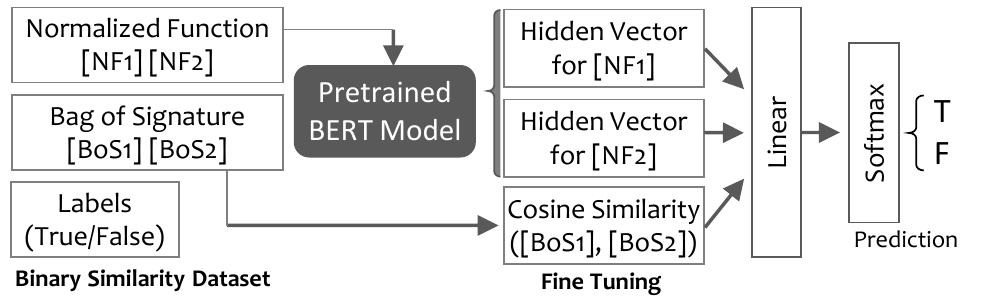}
	\caption{Binary similarity prediction model (\sysbs) 
		as a fine tuning task.}
	\label{fig:binsim-model}
\end{figure}
\begin{figure}[t!]
	\centering
	\includegraphics[width=0.99\linewidth]{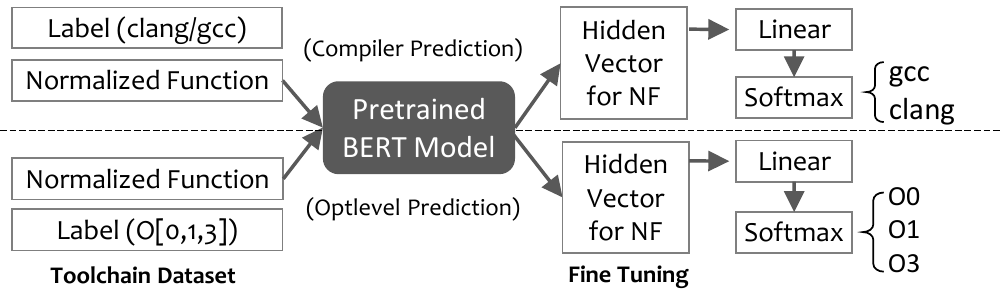}
	\caption{Toolchain (compiler and optimization level)
		classification model (\syscopt) 
		as a fine tuning task.}
	\label{fig:copt-model}
\end{figure}
\sys aims to support specific downstream tasks that
need to infer contextual information of a binary code
with a relatively quick re-training based on 
a pre-trained model for generic code representation.
%
%Unlike a pre-training stage (\autoref{ss:pre-training}),
A fine-tuning stage requires
another dataset preparation with a label (supervised learning)
to learn a specific task.
In this paper, we define two downstream tasks
to demonstrate the effectiveness of our model:
a binary similarity (\sysbs) task
that predicts whether two functions are similar and 
a toolchain prediction task (\syscopt) 
that classifies either a compiler or optimization level.

\subsubsection{\sysbs Model}
\label{ss:deepsemantic-binsim}
%
%\XXX{either front or end,
%  add what useful applications can be built
%  by using binary similarity.
%similarly, for Copt, add its applications too.}
%
%In general, binary code similarity has a broad range of
%real-world applications, including code clone 
%detection~\cite{binsimfse, asm2vec, deepbindiff},
%malware classification~\cite{mutantx, simcalc19}, 
%and bug discovery~\cite{bug-search-acsac14, 
%	xarch-bug, discovre, esh-pldi16, acfg, bingo, esh2, 
%	vulseeker, struct2vec, binarm, alphadiff}.
%
%including but not limited to:
%i)~code clone detection~\cite{binsimfse, asm2vec, deepbindiff}, 
%ii)~malware family classification~\cite{mutantx, simcalc19} 
%and detection~\cite{polymorphic-worm, 
%	self-mutating-malware, cfg-malware-detection},
%iii)~authorship identification~\cite{malware-author-acsac12, 
%	lineage-inference}, 
%iv)~known bug discovery~\cite{bug-search-acsac14, 
%	xarch-bug, discovre, esh-pldi16, acfg, bingo, esh2, 
%	vulseeker, struct2vec, binarm, alphadiff}, and
%v)~patching analysis~\cite{exe-comp-dimva04,
%	code-comp-saner16, spain-icse17}.
%
We define our first downstream task as
estimating the similarity of two binary functions
that originate from the same source code.
%
%\XXX{about why, w/o fine-training, is this task is difficult.
%below is general description of why this task is challenging.
%}
%
%Deciding the similarity of the two binary functions
%%(\eg, a different set of machine instructions
%%are compiled from the same source code or not)
%is a very challenging task in case that
%varying optimizations have largely transformed 
%the original code at compilation, including
%register allocation, instruction replacement,
%basic block splitting and combination
%%\XXX{please recheck all use cases of \eg, and \eg,}
%(\eg, cmov) and/or
%dead code elimination.
%
We define a new dataset that comprises
two normalized functions (NFs) %with additional information, 
with a label (whether an identical function pair).
\autoref{fig:binsim-model} illustrates
our binary similarity model as a downstream task.
We load \syspre as a basis
to obtain two hidden vectors (size=$h$)
from each NF.
Note that we also include supplementary information,
dubbed \emph{Bag of Signature (BoS)}~\footnote{
	Feeding additional information indeed improves
    our model, however, the difference is marginal
    (0.09\% as in \autoref{fig:binsim-model}).
    Interested readers refer to Appendix (\autoref{s:appendix})}, 
to enhance the binary similarity task.
Next, we concatenate three vectors (\ie,
two hidden vectors and cosine similarity of 
two BoSes) and then 
pass them through a linear layer
where the number of inputs and outputs are
$2 * h + 1$ and $2$, respectively.
\subsubsection{\syscopt Model}
\label{ss:deepsemantic-toolchain}
We define another downstream task
to predict either a compiler or optimization level 
because recovering the toolchain provenance of 
binary code is an important task
in the literature of digital forensics
~\cite{oglassesx, toolchain-recovery}.
Note that
our model experimentally shows poor performance 
in categorizing both compiler and optimization together
(See \autoref{ss:eval-application} in detail);
hence, we separate \syscopt into two sub-tasks:
compiler and optimization level classification.
%
%\XXX{why it's either? not and?}
%
\autoref{fig:copt-model} depicts our model,
which is simpler than \sysbs.
Similarly, we extract a hidden vector from each NF and
then train the classifier with a linear and softmax layer.
\subsubsection{Model and Loss Function}
Both \autoref{ss:deepsemantic-binsim} and 
\autoref{ss:deepsemantic-toolchain} tasks are
%We consider each task as
a multi-class (including binary) classification problem. 
In particular,
the logits of both downstream tasks can be calculated as:
\begin{equation}
\label{eq:logits}
%\resizebox{.85\linewidth}{!}{$
%	\displaystyle
  \hat{y} = \text{Softmax}(\mathcal{F}(h))
%	$}
\end{equation}
where $\mathcal{F(\cdot)}$ and \textit{h} are a fully-connected layer 
and the hidden vector of the given function returned 
from \syspre, respectively.
To obtain the optimal network parameters 
in the fine-tuning layers, we use
cross-entropy as a loss function. In other words, we find 
the network parameters $\theta$ that satisfy:
\begin{equation}
\label{eq:loss}
%\resizebox{.85\linewidth}{!}{$
%	\displaystyle
  \theta = \underset{\theta}{\arg\min} \sum_{c \in C} p(c|y) log(p(c|\hat{y}))
%	$}
\end{equation}
where $C$, $p(c|y)$, and $p(c|\hat{y})$ denote a set of classes 
(\eg, decision of function similarity in \sysbs or 
toolchain prediction in \syscopt), 
the ground-truth class distribution, 
and the estimated probability
for the class $c$ by the logits $\hat{y}$ 
calculated from \autoref{eq:logits}.

\section{Implementation}
\label{s:impl}

This section briefly describes 
the artifacts from static binary analysis and
\sys implementation.

\paragraph{Binary Analysis Artifacts.}
With our corpus (\autoref{t:dataset}),
we extract essential artifacts from one of the state-of-the-art
static binary analysis tools, IDA Pro 7.2~\cite{idapro}.
%
%As part of pre-processing, 
We leverage \cc{IDAPython}~\cite{idapython} 
(a built-in IDA Pro~\cite{idapro} plugin)
to build an initial database of binary analysis artifacts 
including function names,
\cc{libc} library calls, 
cross references (\eg, string literals, numeric constants), 
section names and call invocations 
(\eg, internal calls, external calls),
which further assists achieving 
a well-balanced normalization process
as described in \autoref{t:normalization-rules}.
%
%including a function and basic block boundary, 
%function name, numeric constant and string literal,
%\cc{glibc} function invocation, and a section name that a 
%reference points to, facilitating further 
%fine-grained instruction normalization.
%
%Each artifact contains instruction disassembly
%with basic analysis results including
%
Although we provide 
a binary with debugging symbols available
to confirm the ground truth (\eg, function boundaries)
during static analysis, 
\sys is agnostic to the availability of
such debugging information.
Any binary analysis tool would suffice
to recognize binary functions
such as angr~\cite{angr}, 
Ghidra~\cite{ghidra}, or radare2~\cite{radare2}.
\paragraph{BERT Model without NSP}
We develop \sys with 
Tensorflow~\cite{tensorflow} and PyTorch~\cite{pytorch}
on top of a few existing BERT
implementations~\cite{transformers, bertimpl, google-bertimpl}.
\begin{table}[!t]
	\centering
	\footnotesize
	\renewcommand{\arraystretch}{1.0}
	%\scalebox{0.87} {
		\begin{tabular}{lrlr}
			\toprule
			%\multicolumn{2}{l}{ \textbf{BERT Language Model Parameters} } & 
%\multicolumn{2}{l}{ \textbf{Optimizer Parameters }} \\
(B) Dimension of embeddings & 256 & (O) Loss rate & 0.0005 \\
(B) Number of hidden layers & 128 & (O) Adam beta1 & 0.9 \\
(B) Number of attention layers & 8 & (O) Adam beta2 & 0.999 \\
(B) Number of attention heads & 8 & (O) Adam weight decay & 0.01 \\
(B) Maxium length of encoding & 250 & (O) Adam epsilon & 1.00E-06 \\
(B) Position dropout & 0.1 & (O) Warmup & Linear \\
(B) Conv1d dropout & 0.2 & (T) Epochs & 5 \\
(B) Number of conv1d layers & 3 & (T) Batch size & 96 \\
(B) Size of conv1d kernel & 5 & (T) Train dataset ratio & 0.9 \\
(B) Feed forward network dropout & 0.1 & (T) Valid dataset ratio & 0.05 \\
(B) Self-attention dropout & 0.1 & (T) Test dataset ratio & 0.05 \\
			\bottomrule
		\end{tabular}
	%}
	\caption{Hyperparameters for 
		(B)ERT, (O)ptimizer, and (T)rainer
		during a training phase. % (unless otherwise stated).
	}
	\label{t:hyperparam}
\end{table}
As described in~\autoref{s:design}, \sys does not
compute NSP when building a language model
unlike the original BERT architecture because 
the semantics of a function is \emph{location-independent}
from that of its adjacent functions.
We pre-train with a batch size of 96 sequences,
where each sequence contains 256 tokens
(\eg, 256 * 96 = 24,576 tokens/batch
including special tokens) using
%for 50,000 steps \XXX{check this out},
five epochs over the 1.3M binary functions.
We use the Adam optimizer with a learning rate of 
$0.0005$, $\beta_1 = 0.9$, $\beta_2 = 0.999$,
L2 weight decay rate of $0.01$, and
liner decay of the learning rate.
We use a dropout rate of 0.1 on all layers,
and \cc{ReLU} activation function.
\autoref{t:hyperparam} encapsulates all hyperparameters 
when we build our models for the BERT language 
model, optimizer, and trainer.
The number of trainable parameters is $8,723,914$
for \syspre. 

\section{Evaluation}
\label{s:eval}
%
%\XXX{lack of security findings; I think we can ignore this
%	as code semantics itself implies many applications.}
%
As the approach of \sys inherently involves
a training process twice -- \syspre and \systask --
accordingly, we set up various experiments 
for accurate and fair evaluation.
%thorough assessment.
%
We first build a \syspre model
(\autoref{ss:pre-training model}),
%\XXX{don't abuse footnote. simply say in here or abstract/conclusion}
followed by answering the following five research questions
from three aspects:
i)~effectiveness (RQ1-3) that focuses on \sysbs,
ii)~applicability (RQ4) with \syscopt, and
iii)~efficiency (RQ5) of \sys in practice.
%
%We will release a pre-computed model (\syspre)
%in the near future
%to foster further binary analysis research.
%\XXX{revise the questions to be ``how'' not yes/no question.}

\begin{itemize}[leftmargin=*]
	\setlength{\itemsep}{0pt}
	\setlength{\parskip}{0pt}
	\item \textbf{RQ1.}
	How much does \sysbs outperform
	existing cutting-edge approaches 
	(\eg, DeepBinDiff~\cite{deepbindiff}, SAFE~\cite{safe})
	for a binary similarity detection task
	that requires the inference 
	of underlying binary semantics
	(\autoref{ss:eval-binsimtask})?
	\item \textbf{RQ2.}
	How well does a fine-tuning task enhance
	binary code representation?
	We assess how \sysbs updates the original
	function embedding vectors from \syspre
	(\autoref{ss:eval-eprenstation}).
	\item \textbf{RQ3.}
	%
	%How much improvement does supplementary information 
	%contribute to \sysbs model?
	How much improvement does well-balanced normalization
	offer for \sysbs model?
	%
%	In particular, we evaluate two techniques 
%	to improve the quality of
%	code representation: 
%	well-balanced instruction normalization 
%	and Bag of Signature
%	(\autoref{ss:eval-justification}).
	%
	%	Does the means of well-balanced instruction normalization
	%	in \sys improve model performance?
	%	Is BoS (Bag of Signature) information fruitful to enhance 
	%	a fine-tuning model for a binary similarity task?
	%	
	\item \textbf{RQ4.}
	Can \sys be applicable to
	other downstream tasks (\eg, compiler 
	and optimization level prediction)
	(\autoref{ss:eval-application})?
	\item \textbf{RQ5.}
	How efficient is \sys 
	in practice (\autoref{ss:eval-performance})?
\end{itemize}

\PP{Environment}
We evaluate \sys on a 64-bit Ubuntu 18.04 system
equipped with Intel(R) i9-10900X CPU
(with 20 3.70 GHz cores), 128 GB RAM and
NVIDIA Quadro RTX 8000 GPU.

\PP{Dataset} %\XXX{Move corpus in impl to here.}
\begin{table*}[!t]
	\centering
	\footnotesize
	\renewcommand{\arraystretch}{1.3}
	\scalebox{1.0} {
		\begin{tabular}{l|rrrrr|rr|rrr|rrr}
			\toprule
			\textbf{ } & \multicolumn{5}{c|}{ \textbf{Number of Occurrence} } & \multicolumn{2}{c|}{ \textbf{Rate} } & \multicolumn{3}{c|}{ \textbf{Metrics} } & \multicolumn{3}{c}{ \textbf{Average Occurrence per FN} } \\
\hline
\textbf{Testsuite} & \textbf{Binaries} & \textbf{FNs} & \textbf{BBs} & \textbf{INs} & \textbf{Vocas} & \textbf{Small FNs} & \textbf{Large FNs} & \textbf{BBs/F}N & \textbf{INs/FN} & \textbf{INs/BB} & \textbf{Immediate} & \textbf{String} & \textbf{Libc} \\
\hline
GNUtils & 1,000 & 446,014 & 3,983,384 & 22,216,363 & 5,328 & 13.70\% & 3.14\% & 8.93 & 49.81 & 5.58 & 7.99 & 1.01 & 0.88 \\
Spec2006 & 176  & 408,496 & 4,664,973 & 28,321,431 & 9,632 & 14.96\% & 5.01\% & 11.42 & 69.33 & 6.07 & 9.72 & 0.81 & 0.41 \\
Spec2017 & 120 & 755,673 & 9,339,087 & 52,952,315 & 11,933 & 13.90\% & 4.83\% & 12.36 & 70.07 & 5.67 & 10.37 & 0.68 & 0.28 \\
Utils & 32 & 80,532 & 788,861 & 5,083,417 & 4,963 & 18.89\% & 4.94\% & 9.80 & 63.12 & 6.44 & 11.95 & 0.97 & 0.25 \\
\hline
Total & 1,328 & 1,690,715 & 18,776,305 & 108,573,526 & 17,220 & 15.36\% & 4.48\% & 10.63 & 63.08 & 5.94 & 10.01 & 0.87 & 0.46 \\
			\bottomrule
		\end{tabular}
	}
	\caption{Summary of our whole corpus. 
		%
%		Each binary has been
%		compiled with two different compilers 
%		(\eg, \cc{gcc}, \cc{clang}) and four optimization levels
%		(\eg, \cc{O[0-3]}), respectively. 
%		The number in a parenthesis
%		is the number of unique source code.
%		(*) denotes that some binaries were
%		not successfully compiled
%		with a default configuration.
		%
		FN, BB and IN represent a function, 
		basic block and instruction, respectively.
		A small function indicates the number of
		instructions per function (I/F) is less than
		or equal to 5, whereas a large function 
		indicates I/F is greater than 250.
		We also investigate the number of immediate
		operands, string references and libc call
		invocations per function to devise 
		well-balanced normalization.
	}
	\label{t:dataset}
\end{table*}
\autoref{t:dataset} defines the whole corpus for our dataset.
We generate the $1,328$ binaries compiled with two
compilers (\eg, \cc{gcc} 5.4 and \cc{clang} 6.0.1) and four
optimization levels (\eg, \cc{O[0-3]})
from three different testsuites (\eg, GNUtils, SPEC2017,
the utilities including \cc{openssl}~\cite{openssl}, 
\cc{nginx}~\cite{nginx}, and \cc{vsftpd}~\cite{vsftpd}. 
The $176$ SPEC2006 binaries are
borrowed from the official release~\cite{sok-x86}.
%
%Although, in general, it is quite challenging to define
%a well-representative-and-generalized set 
%with a limited number of binaries due to the nature
%of software diversity,
%we choose our corpus considering a size, language (C/C++),
%and purpose (\eg, utility, server, benchmark).

\subsection{\syspre Generation}
\label{ss:pre-training model}
With the artifacts corresponding to each binary from~\autoref{s:impl},
we normalize all instructions, define tokens,
and create a dataset for BERT pre-training.
We obtain $17,225$ tokens in total from all $107.88$ million
instructions in our corpus including
i)~each instruction token
(every instruction represents 
an individual token in our model) 
from a normalized instruction
and ii)~five special tokens 
(\cc{[SOS]}: start of a sentence,
\cc{[EOS]}: end of a sentence, 
\cc{[UNK]}: unknown token,
\cc{[MASK]}: mask to predict a word, 
\cc{[PAD]}: padding symbol to fill out an input length)
for \syspre generation,
which is a modified BERT model
described in \autoref{fig:bert-transformer}.
The total number of all normalized binary functions
is $1,690,715$ in our corpus; however, we solely include 
around $1.3$ million functions after filtering out
either functions that are too small (the number of instructions
is less than or equal to five; $15.4\%$) or 
too large (greater than $250$ instructions; $4.5\%$) ones.

This is because i)~the BERT structure can hardly learn
functions that are too small in a mask language model;
\eg, we use a masking rate of 15\% like the
original BERT model; thus at least six instructions
(as a single sentence) can play a role of
a meaningful semantic chunk for training, which
generates an appropriate number
of masks for prediction during training, and
ii)~functions that are too large may hinder performance
of pre-training model generation; and thus
they have been excluded (4.5\%).
%
%One might concern quite a few small functions
%that have been ruled out, however, we believe
%such a small instruction set cannot distinguish  
%underlying code semantics from each other.
%\KK{example here!}
%
It is noteworthy that we have not excluded
the identical NFs during pre-training model generation
because they can be regarded as a different training
dataset considering the nature of the probabilistic masking
mechanism in MLM.
Additionally, the ratio of OOV in the test set
is around $0.93\%$, that is,
merely $82$ vocabularies are unknown 
out of $8,783$ tokens (the number of vocabularies
in the training set is $17,024$).
\begin{figure}[t!]
	\centering
	\includegraphics[width=0.9\linewidth]{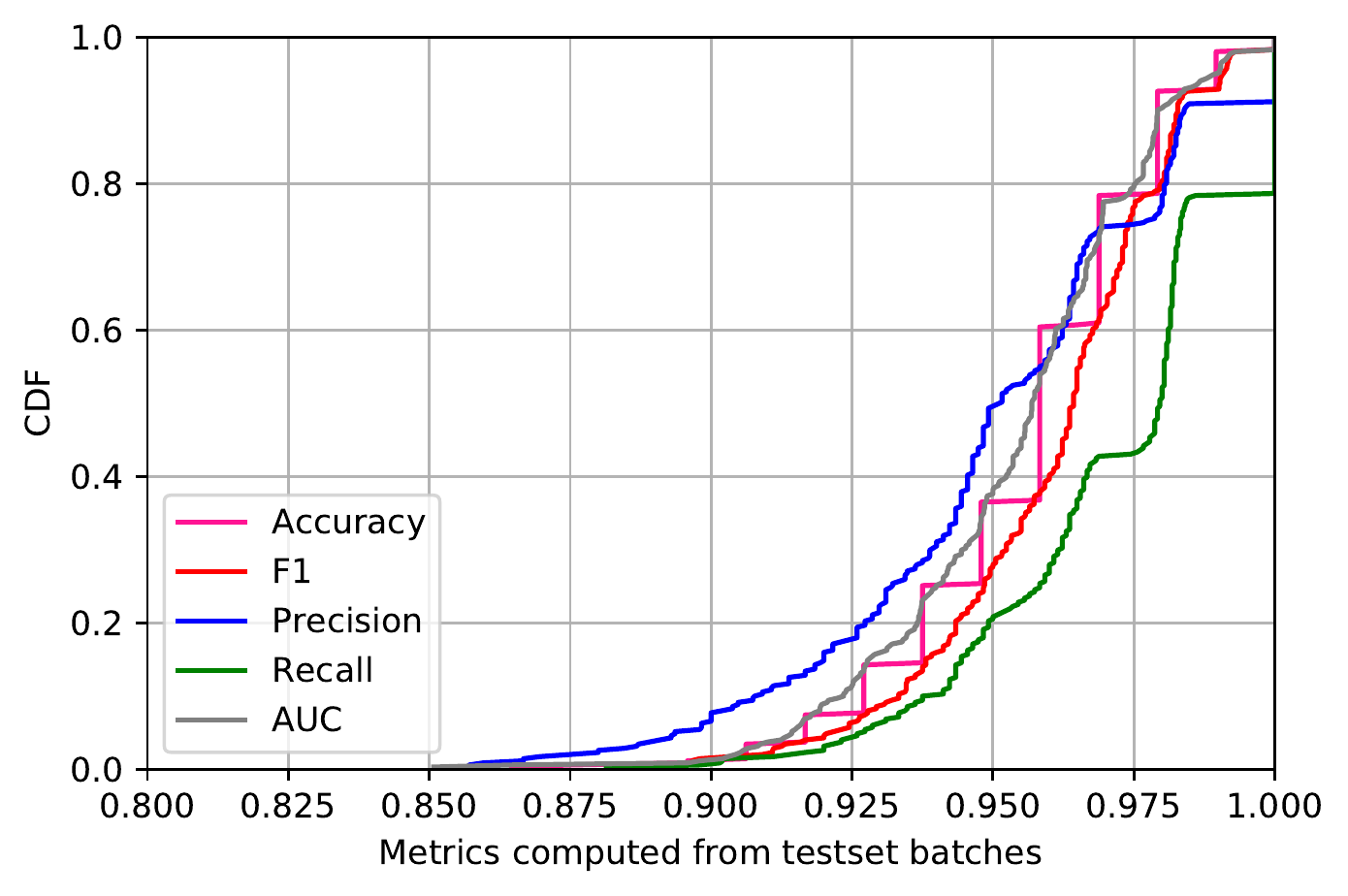}
	\caption{
		CDF of varying metrics to show the performance
		of \sysbs (See \autoref{t:fgn-bos-comp}).
	}
	\label{fig:binsim-cdf}
\end{figure}
%
%\begin{table*}[h]
%	\centering
%	\footnotesize
%	\renewcommand{\arraystretch}{1.05}
%	\scalebox{0.99} {
%		\begin{tabular}{l|rrr|rrr|rrrrr||rrr|rrrr}
%			\toprule
%			\input{tbls/deepbindiff-comparison}
%			\bottomrule
%		\end{tabular}
%	}
%	\caption{Precision, Recall, and F1 results of binary similarity comparison
%		across different (Compiler, Optlevel) pairs 
%		between \sysbs and DeepBinDiff~\cite{deepbindiff},
%		%	SAFE~\cite{safe} and \sysbs. 
%		%
%		We conduct two separate experiments with a limited testset A (left)
%		due to the constraint of DeepBinDiff and a full testset B (right).
%		%
%		\sysbs significantly outperforms DeepBinDiff (\eg, around 50\%) 
%		across all pair combinations, and slightly exceeds SAFE.
%		%
%		However, the experiment with the full testset, \sysbs surpasses
%		SAFE (\eg, 16\% on average).
%		%
%		\XXX{This is too much; maybe separate it?}
%	 	%
%%		\XXX{so what? state the conclusion
%%			(summarize your analysis instead of letting people to infer or
%%			lookup other places).}
%	}
%	\label{t:deepbindiff-comp}
%\end{table*}

\begin{table}[ht]
	\centering
	\footnotesize
	\renewcommand{\arraystretch}{1.05}
	\scalebox{0.99} {
		\begin{tabular}{l|rrr|rrr|r}
			\toprule
			\textbf{Experiment} & \multicolumn{7}{c}{\textbf{Limited Testset}} \\
\hline
\multirow{2}{*}{ \textbf{Pair} } & 
\multicolumn{3}{c}{\textbf{DeepBinDiff}} & \multicolumn{4}{c}{\textbf{DeepSemantic}} \\
& \textbf{P} & \textbf{R} & \textbf{F1} & \textbf{P} & \textbf{R} & \textbf{F1} & \textbf{Diff} \\
\hline
\textbf{(CO0,CO1)} & 0.574 & 0.197 & 0.293 & 0.907 & 0.907 & 0.907 & 61.40\% \\
\textbf{(CO0,CO2)} & 0.528 & 0.176 & 0.264 & 0.924 & 0.910 & 0.917 & 65.29\% \\
\textbf{(CO0,CO3)} & 0.501 & 0.178 & 0.263 & 0.910 & 0.924 & 0.917 & 65.43\% \\
\textbf{(CO0,GO0)} & 0.693 & 0.279 & 0.398 & 0.921 & 0.930 & 0.925 & 52.72\% \\
\textbf{(CO0,GO1)} & 0.529 & 0.173 & 0.261 & 0.934 & 0.915 & 0.925 & 66.40\% \\
\textbf{(CO0,GO2)} & 0.506 & 0.151 & 0.233 & 0.930 & 0.920 & 0.925 & 69.22\% \\
\textbf{(CO0,GO3)} & 0.488 & 0.151 & 0.231 & 0.932 & 0.905 & 0.918 & 68.74\% \\
\textbf{(CO1,CO2)} & 0.727 & 0.643 & 0.682 & 0.928 & 0.921 & 0.924 & 24.20\% \\
\textbf{(CO1,CO3)} & 0.705 & 0.612 & 0.655 & 0.930 & 0.911 & 0.921 & 26.55\% \\
\textbf{(CO1,GO0)} & 0.599 & 0.202 & 0.302 & 0.931 & 0.917 & 0.924 & 62.15\% \\
\textbf{(CO1,GO1)} & 0.613 & 0.330 & 0.429 & 0.923 & 0.909 & 0.916 & 48.67\% \\
\textbf{(CO1,GO2)} & 0.592 & 0.375 & 0.459 & 0.914 & 0.911 & 0.913 & 45.33\% \\
\textbf{(CO1,GO3)} & 0.546 & 0.346 & 0.424 & 0.921 & 0.915 & 0.918 & 49.41\% \\
\textbf{(CO2,CO3)} & 0.947 & 0.863 & 0.903 & 0.925 & 0.917 & 0.921 & 1.79\% \\
\textbf{(CO2,GO0)} & 0.509 & 0.187 & 0.274 & 0.933 & 0.918 & 0.926 & 65.21\% \\
\textbf{(CO2,GO1)} & 0.600 & 0.302 & 0.402 & 0.919 & 0.903 & 0.911 & 50.88\% \\
\textbf{(CO2,GO2)} & 0.606 & 0.343 & 0.438 & 0.918 & 0.918 & 0.918 & 47.99\% \\
\textbf{(CO2,GO3)} & 0.598 & 0.344 & 0.437 & 0.934 & 0.913 & 0.923 & 48.65\% \\
\textbf{(CO3,GO0)} & 0.494 & 0.179 & 0.263 & 0.910 & 0.910 & 0.910 & 64.73\% \\
\textbf{(CO3,GO1)} & 0.577 & 0.300 & 0.395 & 0.937 & 0.930 & 0.934 & 53.87\% \\
\textbf{(CO3,GO2)} & 0.605 & 0.349 & 0.443 & 0.928 & 0.929 & 0.929 & 48.60\% \\
\textbf{(CO3,GO3)} & 0.579 & 0.342 & 0.430 & 0.920 & 0.920 & 0.920 & 49.01\% \\
\textbf{(GO0,GO1)} & 0.582 & 0.221 & 0.320 & 0.916 & 0.910 & 0.913 & 59.24\% \\
\textbf{(GO0,GO2)} & 0.555 & 0.190 & 0.283 & 0.943 & 0.916 & 0.929 & 64.59\% \\
\textbf{(GO0,GO3)} & 0.515 & 0.187 & 0.274 & 0.920 & 0.916 & 0.918 & 64.35\% \\
\textbf{(GO1,GO2)} & 0.784 & 0.556 & 0.651 & 0.930 & 0.927 & 0.929 & 27.80\% \\
\textbf{(GO1,GO3)} & 0.751 & 0.536 & 0.626 & 0.924 & 0.900 & 0.912 & 28.62\% \\
\textbf{(GO2,GO3)} & 0.848 & 0.737 & 0.789 & 0.942 & 0.929 & 0.935 & 14.65\% \\
\hline
\textbf{Average} & 0.613 & 0.337 & \textbf{0.422} & 0.925 & 0.916 & \textbf{0.921} & \textbf{49.84\%} \\

			\bottomrule
		\end{tabular}
	}
	\caption{Precision, Recall, and F1 results of 
		binary similarity comparison
		%across different (Compiler, Optlevel) pairs 
		between \sysbs and DeepBinDiff~\cite{deepbindiff}.
		\sysbs significantly outperforms DeepBinDiff (\eg, around 49.8\%) 
		across all pair combinations.
	}
	\label{t:deepbindiff-comp}
\end{table}

\begin{table}[ht]
	\centering
	\footnotesize
	\renewcommand{\arraystretch}{1.05}
	\scalebox{0.99} {
		\begin{tabular}{l|rrr|rrr|r}
			\toprule
			\textbf{Experiment} & \multicolumn{7}{c}{\textbf{Full Testset}} \\
\hline
\multirow{2}{*}{ \textbf{Pair} } & 
\multicolumn{3}{c}{\textbf{SAFE}} & \multicolumn{4}{c}{\textbf{DeepSemantic}} \\
& \textbf{P} & \textbf{R} & \textbf{F1} & \textbf{P} & \textbf{R} & \textbf{F1} & \textbf{Diff} \\
\hline
\textbf{(CO0,CO1)} & 0.879 & 0.728 & 0.796 & 0.958 & 0.975 & 0.967 & 17.03\% \\
\textbf{(CO0,CO2)} & 0.774 & 0.676 & 0.722 & 0.973 & 0.967 & 0.970 & 24.79\% \\
\textbf{(CO0,CO3)} & 0.830 & 0.627 & 0.715 & 0.958 & 0.969 & 0.963 & 24.87\% \\
\textbf{(CO0,GO0)} & 0.788 & 0.996 & 0.880 & 0.955 & 0.979 & 0.967 & 8.68\% \\
\textbf{(CO0,GO1)} & 0.836 & 0.704 & 0.764 & 0.943 & 0.968 & 0.956 & 16.05\% \\
\textbf{(CO0,GO2)} & 0.698 & 0.662 & 0.679 & 0.940 & 0.975 & 0.957 & 27.79\% \\
\textbf{(CO0,GO3)} & 0.713 & 0.648 & 0.679 & 0.945 & 0.983 & 0.963 & 28.44\% \\
\textbf{(CO1,CO2)} & 0.770 & 0.824 & 0.796 & 0.954 & 0.973 & 0.963 & 16.70\% \\
\textbf{(CO1,CO3)} & 0.685 & 0.824 & 0.748 & 0.963 & 0.989 & 0.976 & 22.78\% \\
\textbf{(CO1,GO0)} & 0.796 & 0.794 & 0.795 & 0.954 & 0.981 & 0.967 & 17.22\% \\
\textbf{(CO1,GO1)} & 0.871 & 0.955 & 0.911 & 0.942 & 0.974 & 0.958 & 4.69\% \\
\textbf{(CO1,GO2)} & 0.737 & 0.882 & 0.803 & 0.954 & 0.984 & 0.969 & 16.55\% \\
\textbf{(CO1,GO3)} & 0.732 & 0.909 & 0.811 & 0.949 & 0.960 & 0.955 & 14.37\% \\
\textbf{(CO2,CO3)} & 0.662 & 0.978 & 0.789 & 0.959 & 0.971 & 0.965 & 17.55\% \\
\textbf{(CO2,GO0)} & 0.679 & 0.689 & 0.684 & 0.951 & 0.981 & 0.965 & 28.14\% \\
\textbf{(CO2,GO1)} & 0.813 & 0.952 & 0.877 & 0.968 & 0.982 & 0.975 & 9.75\% \\
\textbf{(CO2,GO2)} & 0.825 & 0.942 & 0.880 & 0.965 & 0.986 & 0.975 & 9.54\% \\
\textbf{(CO2,GO3)} & 0.830 & 0.949 & 0.885 & 0.939 & 0.978 & 0.958 & 7.24\% \\
\textbf{(CO3,GO0)} & 0.731 & 0.746 & 0.738 & 0.945 & 0.990 & 0.967 & 22.85\% \\
\textbf{(CO3,GO1)} & 0.864 & 0.941 & 0.901 & 0.950 & 0.961 & 0.955 & 5.42\% \\
\textbf{(CO3,GO2)} & 0.909 & 0.961 & 0.934 & 0.948 & 0.977 & 0.962 & 2.81\% \\
\textbf{(CO3,GO3)} & 0.791 & 0.966 & 0.870 & 0.947 & 0.962 & 0.954 & 8.45\% \\
\textbf{(GO0,GO1)} & 0.757 & 0.802 & 0.779 & 0.964 & 0.967 & 0.966 & 18.66\% \\
\textbf{(GO0,GO2)} & 0.815 & 0.687 & 0.745 & 0.953 & 0.962 & 0.958 & 21.20\% \\
\textbf{(GO0,GO3)} & 0.796 & 0.651 & 0.716 & 0.955 & 0.982 & 0.968 & 25.21\% \\
\textbf{(GO1,GO2)} & 0.815 & 0.993 & 0.895 & 0.961 & 0.967 & 0.964 & 6.92\% \\
\textbf{(GO1,GO3)} & 0.854 & 0.944 & 0.897 & 0.961 & 0.969 & 0.965 & 6.80\% \\
\textbf{(GO2,GO3)} & 0.755 & 0.938 & 0.837 & 0.949 & 0.982 & 0.965 & 12.85\% \\
\hline
\textbf{Average} & 0.786 & 0.835 & \textbf{0.805} & 0.954 & 0.975 & \textbf{0.964} & \textbf{15.83\%} \\
			\bottomrule
		\end{tabular}
	}
	\caption{Precision, Recall, and F1 results of 
		binary similarity comparison
		%across different (Compiler, Optlevel) pairs 
		between \sysbs and SAFE~\cite{safe}.
		\sysbs surpasses SAFE (\eg, around 15.8\%)
		across all pair combinations.
	}
	\label{t:safe-comp}
\end{table}

\subsection{Effectiveness of \sysbs}
\label{ss:eval-binsimtask}
As one of downstream tasks with \sys,
this section demonstrates the effectiveness of \sysbs
for a binary similarity task by comparing with
the two state-of-the-art tools using deep learning.
According to the model
illustrated in \autoref{fig:binsim-model},
we prepare a dataset that consists of pairs of
two normalized functions (NFs), and their labels. 
%
%Note that we carefully chooses the functions under the same condition
%with \sysbs (\eg, the number of instructions is between $5$ and $250$).
%
If two binary functions are from the same source
code, the label indicates $1$ (true) or $0$ (false) otherwise.
With the same hyperparameters in \autoref{t:hyperparam},
we generate a fine-tuning model (\sysbs) 
that allows for predicting binary similarity.
\autoref{fig:binsim-cdf} illustrates the CDF of
varying metrics, including an F1 of $0.965$ and
AUC of $0.959$ on average (See other metrics in the 
first column in \autoref{t:fgn-bos-comp}).
This means our classifier can accurately make
a prediction if two binary functions are
compiled from the same source
\emph{regardless of} a wide range of code transformations
from arbitrary combination of compilers and optimization levels.

%\XXX{Argues unfairness the comparison with DeepBinDiff}
%
\PP{Comparison with DeepBinDiff}
We conduct an experiment to compare \sys with
open-sourced DeepBinDiff~\cite{deepbindiff-tool}.
%
%at the basic block granularity,
%binary similarity tool at function granularity.
%
Unfortunately, the officially released DeepBinDiff
did not handle even medium-size binaries,
% due to the memory limit (\eg, 128GB) and
%performance issues (\eg, larger binaries 
%than \cc{hmmer})
we had no choice but to define a limited dataset 
(96 executables in total)
including part of the SPEC2006 and findutils
~\footnote{\cc{astar}, \cc{bzip2}, \cc{hmmer}, \cc{lbm}, \cc{libquantum}, 
	\cc{mcf}, \cc{milc}, \cc{namd}, \cc{sphinx\_livepretend},
	and findutils (\cc{find}, \cc{xargs}, \cc{locate})}.
By design, DeepBinDiff performs basic block matching
by taking two binary inputs, whereas \sysbs aims for
comparisons at the function granularity.
In this respect, we classify each case into a true positive
for a fair comparison with DeepBinDiff
when it discovers \emph{any pair of matching basic blocks}
that belong to the same function (relaxed judgment).
%train \sysbs with the above limited dataset
%
%It is noteworthy mentioning that we exclude all functions in
%the limited dataset during pre-training and fine-tuning.
%
It is noteworthy mentioning that 
we extend the evaluation of DeepBinDiff 
across different compilers (\ie, gcc VS clang)
as well as optimization levels
(\ie, 28 different combinations) to demonstrate
the effectiveness of \sysbs.
\autoref{t:deepbindiff-comp} shows that 
our approach considerably outperforms DeepBinDiff ($49.8\%$)
at all combinations of compiler and optimization levels.
Our experiment aligns with the results across optimizations
from DeepBinDiff~\cite{deepbindiff}. 

\PP{Comparison with SAFE}
We conduct another experiment with a full test dataset 
in our corpus to compare with SAFE~\cite{safe}.
By design, SAFE merely computes a cosine similarity value
rather than making a decision on (True/False).
Hence, we found a threshold of $0.5$ (\ie, if the value
is larger than the threshold, the decision is true)
where SAFE gives us the best performance.
To this end, we create a database for our whole 
dataset(\autoref{t:dataset})
to query a function embedding
with the open-sourced version of SAFE~\cite{safe-tool}.
%
%In addition, we test \sysbs and SAFE on full dataset that
%includes every binary in \autoref{t:dataset} to compare their
%performance on more general binaries. 
%
\autoref{t:safe-comp} demonstrates that 
our approach surpasses SAFE ($15.8\%$) on the full testset.
In particular, we observe a substantial difference
when code semantics is more difficult to infer
(\eg, high optimization level difference).

\boxbeg
\textbf{Answer to RQ1.}
\sysbs by far outperforms 
DeepBinDiff
%the existing
%cutting-edge binary similarity detection solution,
(up to 69\%) and SAFE (up to 28\%) across all 28 
combinations of different
compilers and optimization levels.
\boxend

\subsection{Code Representation with Fine-Tuning}
\label{ss:eval-eprenstation}
We assess whether a $256$-dimensional
embedding (\autoref{t:hyperparam}) that represents
a function can be updated after fine-tuning.
It is worth noting that there are several ways
to extract a contextualized embedding
~\cite{illustrated-bert}; here we use a mean
of all hidden states.

%and ii)~fine-tuning
%does not usually alter the pre-trained model
%(\eg, \syspre); we have done it 
%for the experimental purposes.
%
%Embedding results: [\# layers, \# batches, \# tokens, \# features]
%
To this end, we extract 179,163 unique function pairs
from our corpus
including approximately half for similar ones
and another half for dissimilar ones, and then
compute cosine similarity scores.
Note that we exclude all obvious cases in which
two NFs are exactly identical because the score becomes
$1$ at all times.
For example, the similarity score of the
similar function embedding pair,
\cc{ngx\_resolver\_resend\_handler}
between the clang O1 (70 instructions) 
and gcc O2 (86 instructions) is $0.805$
with the \syspre model.
%where a value in a parenthesis denotes the 
%number of instructions.
%
After fine-tuning with the \sysbs model,
we obtain an improved similarity value 
of $0.826$ for the above example,
indicating the vectorized values for
the similar pair is close to $1$.
\begin{table}[!t]
	\centering
	\footnotesize
	\renewcommand{\arraystretch}{1.1}
	\scalebox{1.0} {
		\begin{tabular}{lrrr}
			\toprule
			\textbf{Category} & \textbf{\syspre} & \textbf{\sysbs} \\
\hline
Similar & 0.647 & 0.751  \\
Dissimilar & 0.273 & 0.309  \\
			\bottomrule
		\end{tabular}
	}
	\caption{Cosine similarity of 
		similar/dissimilar function pairs
		on average between \syspre and \sysbs.
	\label{t:safe}
	}
\end{table}
%
%Besides, we compare the binary function representations from
%\sys with the ones from SAFE~\cite{safe} that
%produces a $100$-dimensional embedding
%when a binary function is given.
%
\autoref{t:safe} briefly shows cosine similarity
values on average.
We observe that the binary
function representation have been enhanced;
\ie, the difference between the similar pairs  
in a vector space become broader on average. 
%and ii) SAFE has comparable results.
%

\boxbeg
\textbf{Answer to RQ2.}
Our empirical results indicate that
a fine-tuning process 
successfully updates code representations
for a specific task (\eg, \sysbs).
%
%Besides, a pre-training model (\eg, \syspre)
%itself holds a fairly good embedding as well.
%
\boxend
\subsection{Effectiveness of Well-balanced Normalization}
\label{ss:eval-justification}
%
%Similar to the original BERT architecture,
%\sys divides into two stages by design
%to generate a better model for a certain task.
%context-aware embedding.
%
%In this section, we demonstrate that the quality 
%of code representation can be boosted
%by enclosing supplementary information
%at both a pre-training and fine-tuning stage.
%
\begin{table}[!t]
	\centering
	\footnotesize
	\renewcommand{\arraystretch}{1.00}
	\scalebox{0.99} {
		\begin{tabular}{l|rrr|rrr}
			\toprule
			\multirow{2}{*}{ \textbf{Metric} } & \multicolumn{3}{c|}{ \textbf{Normalization Granularity} } & \multicolumn{3}{c}{ \textbf{Bag of Signature (BoS)} } \\
& \makecell{\textbf{Balanced}} & \makecell{\textbf{Coarse}} & \textbf{(Diff)}
& \makecell{\textbf{With}} & \makecell{\textbf{Without}} & \textbf{(Diff)} \\
\hline
FPR & 0.057 & 0.110 & -5.250\% & 0.058 & 0.059 & -0.109\% \\
TPR & 0.975 & 0.962 & 1.347\% & 0.974 & 0.973 & 0.075\% \\
Accuracy & 0.961 & 0.934 & 2.712\% & 0.960 & 0.959 & 0.101\% \\
Precision & 0.955 & 0.932 & 2.323\% & 0.954 & 0.953 & 0.093\% \\
Recall & 0.975 & 0.962 & 1.347\% & 0.974 & 0.973 & 0.075\% \\
F1 & 0.965 & 0.946 & 1.869\% & 0.964 & 0.963 & 0.088\% \\
AUC & 0.959 & 0.926 & 3.299\% & 0.958 & 0.957 & 0.091\% \\
			\bottomrule
		\end{tabular}
	}
	\caption{Metric comparison
		to show the effectiveness
		of well-balanced normalization
		and BoS for model generation.
		Supplementary information
		clearly enhances the final model.
	}
	\label{t:fgn-bos-comp}
\end{table}

%
%\subsubsection{Well-balanced Normalization}
%\label{sss:balance-grained}
%
This section depicts 
how well-balanced normalization with pre-defined 
rules (\autoref{t:normalization-rules}) enhances
\sysbs.
To assess its effectiveness, we attempt to 
compare \sys with a fine-grained model 
without normalization (note that the only rule applied is
replacing every immediate operand with \cc{immval}).
With our corpus, the number of tokens is $4,917,904$
in a training set, which is
$207.4$ times bigger than well-balanced normalization,
requiring $283.6$ GB GPU memory to update $1.38$ billion
trainable parameters. 
The problem has been exacerbated as well as a resource constraint
considering that most of vocabularies ($3,716,287$ or $75.6\%$)
appear only once, failing to update 
corresponding instruction vectors while training.
Moreover, $97.8\%$ ($127,805$ out of $130,736$) of all tokens 
in the test set are not shown in the training set, rendering
further learning pointless due to a severe OOV problem
(\ie, most of vocabularies would be regarded as \cc{UNK}).

Next, we set up 
our experiment by re-defining a
relatively coarse-grained normalization rule;
simply put, all immediates and pointers have
been replaced with \cc{immval} and \cc{ptr}
without taking any information into account
(we merely include a normalization rule for registers),
followed by generating a new dataset 
($1,174,060$ functions).
This conversion shrinks the number of
tokens up to $2,022$ including five special ones
from $17,225$ of well-balanced normalization
(around $88.3\%$ reduction).
We generate another \syspre model for coarse-grained
normalization with $2,870,759$ trainable parameters.
Then, we prepare another \sysbs dataset with labels,
taking 100K pairs (the ratio of similar
and dissimilar labels are close to 1:1) from our corpus.
%followed by
%splitting it into the (train/valid/test) dataset
%with the ratio of $(0.90:0.05:0.05)$.
%
It is worth noting that we exclude all 
identical NF pairs for similar pairs
(\eg, all pairs should have at least 
one or more discrepancies) to compute a robust model.
We use negative sampling for dissimilar pairs.
\begin{figure}[t!]
	\centering
	\includegraphics[width=0.99\linewidth]{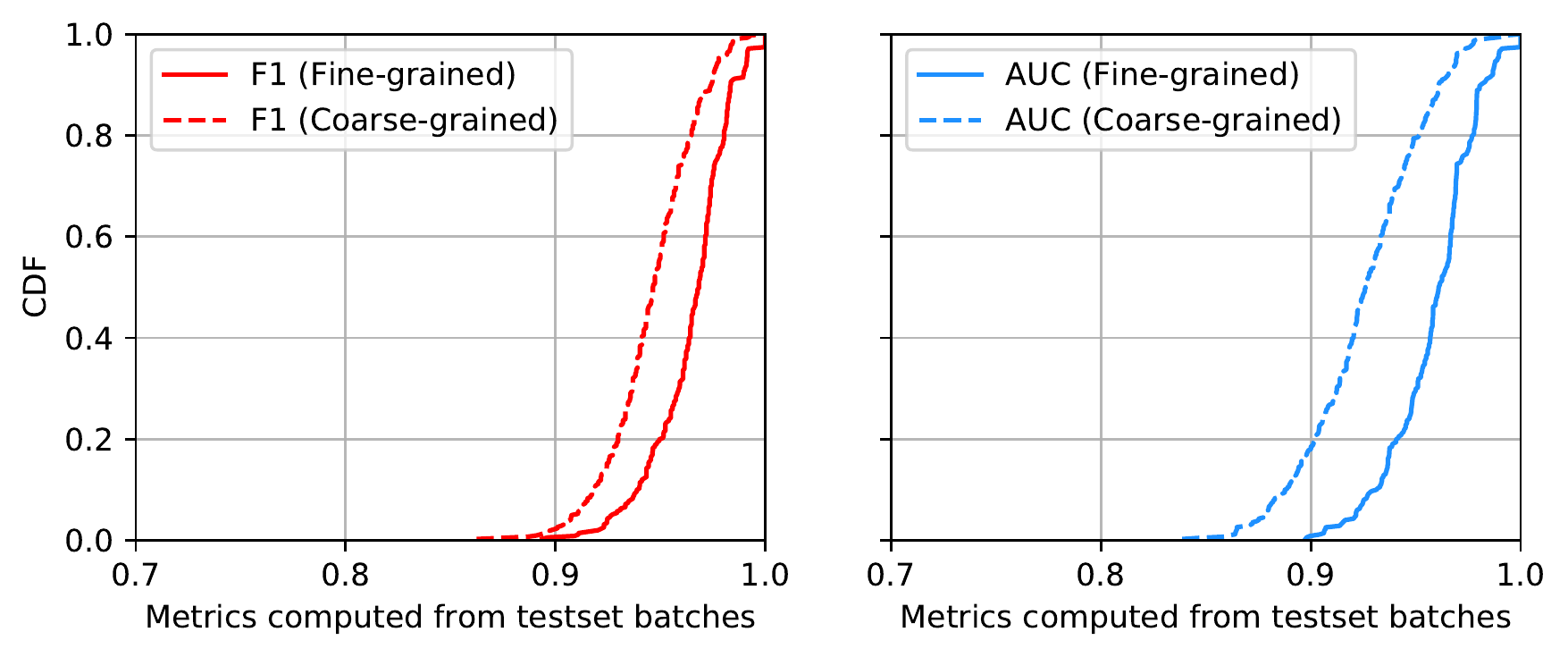}
	\caption{
		CDF comparison of F1 and AUC metrics
		between coarse-grained (dotted lines) and 
		well-balanced normalization (solid lines).
	}
	\label{fig:corase-balanced-grained}
\end{figure}

As a result,
we obtain higher F1 and AUC values
of $(0.965, 0.959)$ with well-balanced normalization
than those of $(0.946, 0.926)$ with coarse-grained 
normalization (See \autoref{t:fgn-bos-comp}).
\autoref{fig:corase-balanced-grained} shows
that the case applying well-balanced normalization has 
better F1 by $1.87\%$ and AUC by $3.30\%$.
Notably, well-balanced normalization decreases
the false positive rate by $5.25\%$.
Although the margins do not make surprisingly 
high improvement, we believe that
it may be fruitful for other downstream tasks
that are contextually sensitive.
%
%\subsubsection{Bag of Signature}
%\label{sss:bos}
%
% MOVED TO APPENDIX!
%

\boxbeg
\textbf{Answer to RQ3.}
%
%Feeding additional information when training
%a model (at a either pre-training or fine-tuning stage)
Well-balanced normalization
can enrich code representation
by feeding supplementary information.
We experimentally validate that 
the normalization process is 
essential for efficient learning 
toward semantic-aware code representation.
\boxend
\subsection{\syscopt Assessment}
\label{ss:eval-application}
This section introduces another useful downstream task
of \sys, that is, a classifier for 
a compiler or optimization level 
when an NF is given.
\begin{table}[!t]
	\centering
	\footnotesize
	\renewcommand{\arraystretch}{1.00}
	\scalebox{1.05} {
		\begin{tabular}{l|rrrr}
			\toprule
			\textbf{Metrics} & \textbf{Compiler} & \textbf{Optlevel} & 
\textbf{Optlevel (gcc)} & \textbf{Optlevel (clang)} \\
\hline
%\textbf{Accuracy} & 0.973 & 0.920 & 0.967 & 0.882 \\
\textbf{Precision} & 0.973 & 0.913 & 0.964 & 0.873 \\
\textbf{Recall} & 0.972 & 0.911 & 0.964 & 0.863 \\
\textbf{F1} & 0.962 & 0.910 & 0.963 & 0.862 \\
%\textbf{AUC} & 0.973 & 0.622 & 0.588 & 0.653 \\
			\bottomrule
		\end{tabular}
	}
	\caption{Metrics of 
		\syscopt classifiers: 
		compiler and optimization levels
		(\eg, gcc and clang). 
	}
	\label{t:copt-results}
\end{table}
Similar to previous experiments,
we generate a dataset that
contains all six combinations of two compilers and
three optimization levels with the same rate
(\eg, 106K functions per each), followed by
splitting it into (train, valid, test)=$(0.90,0.05,0.05)$.
We rule out \cc{-O2} based on our observation that
a large number of NFs
remain identical between 
\cc{-O2} and \cc{-O3} in \cc{clang}, and
\cc{-O1} and \cc{-O2} in \cc{gcc}.
This is partially because a function inlining optimization
drastically transforms its shape
~\footnote{\cc{clang} performs function inlining in 
	\cc{-O2} whereas \cc{gcc} in \cc{-O1} by default.},
and additional optimizations do not affect 
the function structure much.
According to our experiment, 
identical NF pairs with different labels
confused our classifier, resulting in a poor performance.

We generate two \syscopt models 
(\autoref{fig:copt-model}),
for compiler and optimization level prediction because
our model empirically shows a low performance when
attempting to categorize both compiler and optimization
simultaneously.
\autoref{t:copt-results} summarizes our results; an F1
of $0.96$ and $0.91$ for compiler and 
optimization classification.
We additionally carry out an experiment for optimization
prediction per each compiler,
resulting in a $8.5$\% (F1) better performance with \cc{gcc}.
%
%\XXX{Add the comparison with previous work if possible;
%	i think prior one has a quite limited (or easy) one}
%
As a similar setting~\footnote{O-glassesX~\cite{oglassesx}
 introduces a classifier with 19 labels using four
 compilers, two optimization levels (\ie, zero or max), and 
 two architectures(\ie, \texttt{x86} and \texttt{x86\_64}).}, 
O-glassesX~\cite{oglassesx} achieves
an accuracy of $0.956$ with an 16-instruction fixed input
for predicting compiler provenance, which shows
comparable performance.

\boxbeg
\textbf{Answer to RQ4.}
We showcase another downstream task, \syscopt,
which successfully classifies
a compiler and optimization level with a
high precision and recall.
\boxend
\subsection{Efficiency of \sys}
\label{ss:eval-performance}
In this section, we exhibit the efficiency of \sys
in terms of practicality.
\begin{table}[!t]
	\centering
	\footnotesize
	\renewcommand{\arraystretch}{1.1}
	\resizebox{\columnwidth}{!}{
		\begin{tabular}{l|rrr|rr}
			\toprule
			%\multirow{2}{*}{ \textbf{Measurement} } & \multicolumn{3}{c|}{ \textbf{Training} } & \multicolumn{2}{c}{ \textbf{Testing} } \\
\textbf{Metrics} & 
\makecell{\textbf{Pre-training} \\ \textbf{\syspre}} & 
\makecell{\textbf{Fine-tuning} \\ \textbf{\sysbs}} &
\makecell{\textbf{Fine-tuning} \\ \textbf{\syscopt}} &
\makecell{\textbf{Training} \\\textbf{\sysbs}} &
\makecell{\textbf{Training} \\\textbf{\syscopt}} 
\\
\hline
\textbf{Per Epoch} \\ \textbf{(seconds)} & 3553.11 & 2229.10 & 991.87 & 43.52 & 18.25 \\
\textbf{Per Batch} \\ \textbf{(seconds)} & 0.32 & 0.35 & 0.20 & 0.12 & 0.06 \\
\textbf{GPU (GB)} & 34.75 & 37.27 & 12.91 & 37.27 & 12.91 \\
			\bottomrule
		\end{tabular}
	}
	\caption{
		Comparison of computational
		resources across stages (\ie, training, testing) 
		and sub-tasks (\ie, \sysbs, \syscopt).
	}
	\label{t:performance}
\end{table}
\autoref{t:performance} concisely shows
a computational resource on average including 
i)~duration of pre-training, fine-tuning,
and testing a model per epoch and batch, and
ii)~GPU memory consumption for each job.
Creating an initial \syspre model
takes the longest amount of time 
while fine-tuning consumes relatively less expensive
resources.
%
%Note that a generic model (\syspre) can be 
%a one-time process, which can be re-purposed.
%
Once training is complete, testing can be
done much faster for processing 
(\eg, 194x faster than \syspre generation).
Generally, GPU memory consumption may vary
depending on different hyperparameter settings
(\eg, batch size, the number of 
attention/hidden layers,
 maximum length of input).

\boxbeg
\textbf{Answer to RQ5.}
\sys can be an efficient solution
to be applicable for subtasks
by carefully designing a fine-tuning model and
preparing a corresponding dataset in practice.
\boxend
\section{Discussion and Limitation}
\label{s:discussion}
%We have demonstrated that \sys effectively represents binary
%code with semantic-aware way.
This section discusses several cases to consider, 
feasible applications for future research, 
and limitations of our work.

\PP{Function Inlining and Splitting}
Optimization technique sometimes involves with function 
inlining (\ie, a function is part of
another for performance) during compilation.
This means that our labels in the binary similarity
classification problem may be slightly distorted
with the presence of inlining and splitting.
Say, a binary compiled with \texttt{-O3}
has Function A that includes Function B
where the one with \texttt{-O0} has both Function A and B.
The function name A has not been changed but
the actual content (\ie, instructions) could have
in the highly optimized binary.
In a similar vein, it may happen in case 
that a function has been separated at compilation.

\PP{Comparison Granularity}
Although most of functions consist of 
four basic blocks or 25 instructions on average
(\autoref{fig:fbi}), there are indeed functions
with a large size (as a long tail).
In such cases, comparison at the basic block level
would be more appropriate to seek similar blocks as
demonstrated in prior work~\cite{deepbindiff, ordermatters}.

\PP{Applicable Downstream Tasks}
\sys aims to support a wide range of
other downstream tasks that require semantic-aware
code representation by design, including (but not limited to)
a special type of vulnerability scanning,
software plagiarism detection,
malware behavior detection, malware family classification,
and bug patch detection.
One can also think of identifying a function or library
in a database like IDA FLIRT~\cite{flirt}.
%
%We showed two downstream tasks that utilizes binary code representation
%that \sys generates: predicting binary similarity and toolchain.
%%
%We believe that the semantic-aware vectorized binary function gained by \sys 
%can be used by diverse application such as vulnerabilities scanning, 
%plagiarism detection, and library identification and recognition for reverse engineering.
%%
%For instance, it is possible to store the function vectors that can be used
%for identifying function in database 
%for library recognition like IDA FLIRT~\cite{flirt}.
%%
%They are also useful for searching known vulnerabilities and recognizing 
%bug patch in large scales, which may require creating properly labeled
%dataset for purpose.

\PP{Normalization for Rarely Appeared Instructions}
As a rule of thumb, the embedding of an instruction
that has been rarely seen during pre-training
may have been hardly updated (\eg, close to an initial value).
Around 8.5\% of the whole 17,221 normalized instructions 
appear only once in the corpus, as illustrated 
in \autoref{fig:rankfreq}.
Namely, such instruction embeddings may not 
represent a meaningful context unless 
the number of occurrences is sufficient.

\PP{Room for Enhancement}
%
%\KK{Not sure if the following statement is required.}
We thoughtfully design well-balanced normalization
(\autoref{t:normalization-rules}), however, there
are different ways of instruction normalization 
such as tokenization of
separate opcode and operands. 
Besides, the BERT architecture can be replaced
with others like RoBERTa~\cite{roberta} or XLNet~\cite{xlnet}
that has been reported better performance in popular NLP tasks,
remaining as part of our future work.
%

%\PP{Proper Dataset Selection}
\PP{Limitations}
First, the current version of \sys
targets merely benign binaries.
Hence, it may not be possible to
directly apply \sys to a downstream task 
pertaining to malware that often 
ships with various packing, obfuscation or encryption.
Second, we have not yet tested \sys 
with cross-architecture,
which we leave as part of our future work.
Third, one may argue binary corpus representativeness
with a limited number of binaries due to the nature
of software diversity.
However, we carefully chose our corpus that
encompasses varying sizes, types (\eg, executable, library), 
functionalities (\eg, compiler, interpreter, compression, 
server, benchmark, AI-relevant code), and 
languages (\eg, C/C++/Fortran).
%
%As described in \autoref{t:dataset}, our dataset for pre-training is 
%composed of a set of amd64 binaries, which means our published model
%does not understand the semantics across architectures.
%
%We believe it can be resolved by training the model with dataset 
%including binaries from various architectures.
%
%Furthermore, \sys is not able to figure semantics out 
%from packed or encrypted
%binaries that are common attributes of malware 
%as packing and encryption break the normal structures of binary code
%that we assume by design.
%
Fourth, it is possible that
two different functions may have identical instructions
after a normalization process, albeit a rare case.
Such samples (\ie, same functions with different labels)
may confuse the current \sys when building a model.
Lastly, we also observe that a different set of 
corpora and hyperparameter settings during pre-training
may impact overall performance to a downstream task.
%
%In addition, different dataset may impact on the performance of downstream task 
%while fine-tuning, which leaves room to be improved by purpose of downstream task 
%in further research.

\section{Related work}
\label{s:relwk}
%
%It is a long-standing problem to penetrate
%\XXX{Reorganizing related work centered around
%	ML/DL techniques, and highlight Order Matters
%	that should be directly relevant to this paper.
%	Add more insights to the field.}
%
Penetrating the characteristics of a machine-interpretable
binary code can be applied to a wide range of
real-world applications including
i)~code clone (software plagiarism) or similarity
detection~\cite{binsimfse, esh-pldi16, esh2, blanket, 
	gemini, embedding-bar19, innereye, asm2vec, 
	bar-bbsimilarity, safe, deepbindiff, ordermatters}, 
ii)~malware family classification~\cite{mutantx},
    detection~\cite{polymorphic-worm, 
	self-mutating-malware, cfg-malware-detection},
	and analysis~\cite{bindnn, malware-aaai20, neurlux},
iii)~authorship prediction~\cite{malware-author-acsac12, 
	lineage-inference}, 
iv)~known bug discovery (code search)~\cite{bug-search-acsac14, 
	xarch-bug, discovre, acfg, bingo, fermadyine,
	vulseeker, struct2vec, binarm, alphadiff},
v)~patching analysis~\cite{exe-comp-dimva04,
	code-comp-saner16, spain-icse17}, and
vi)~toolchain provenance
~\cite{oglassesx, toolchain-recovery}, 
most of which pertain to binary similarity comparison.
Here we categorize such efforts into two approaches
based on the one with or without a machine learning technique.

\PP{Deterministic Approaches} 
%
%Prior to utilizing machine learning techniques,
Static and dynamic analyses are two mainstream techniques
to learn the underlying semantics of a binary code.
Luo et al.~\cite{binsimfse} introduce a means to detect
cloned code with the longest common subsequence of 
semantically equivalent basic blocks.
In a similar vein, Esh~\cite{esh-pldi16} leverages 
data-flow slices 
of basic blocks to detect binary similarity, and later
extending the idea through
re-optimization~\cite{esh2} to improve performance.
Blanket execution~\cite{blanket} takes an approach of
a dynamic equivalence testing primitive
by defining seven major features. 
Malware research has largely adopted deterministic approaches
including i)~malware classification by extracting static 
features~\cite{mutantx}, ii)~malware detection 
with structural similarities 
between multiple mutations~\cite{polymorphic-worm}
or control-flow graph matching~\cite{
	self-mutating-malware, cfg-malware-detection},
and iii)~malware authorship inference
~\cite{malware-author-acsac12, lineage-inference}.
Meanwhile, varying approaches have been proposed in
the field of known bug discovery by leveraging
i)~a tree edit distance
between the signature of a target basic block 
and that of other basic blocks~\cite{bug-search-acsac14},
ii)~the input/output behavior of basic blocks
from intermediate representation lifting~\cite{xarch-bug}, 
and iii)~dissimilar code filtering
with manual features such as
numeric and structural information~\cite{discovre}.
Further, Genius~\cite{acfg} and FERMADYNE~\cite{fermadyine}
devise another bug search engine 
for IoT firmware with both 
statistical and structural features where
BinGo~\cite{bingo}
supports both cross-architecture and cross-OS 
with a selective inlining technique to capture
function semantics.
%
%including values 
%read from/to the heap/stack, 
%library function calls, %via a \cc{.plt} section, 
%system calls, %made during execution, 
%and return values stored in the \cc{rax} register 
%upon function completion.
%
Oftentimes, static analysis suffers
from scalability  
(\eg, expensive graph matching algorithm, path explosion) 
or flexibility from a structural difference
(\eg, optimization), while
dynamic analysis struggles with 
incomplete code coverage (\eg, relying on inputs) and 
behavior undecidability.

\PP{Probabilistic Approaches} 
Recent advancements with machine learning techniques
have received much attention to be applicable
for binary analysis, fulfilling
both effectiveness and efficiency.
Malware analysis is one of popular applications:
i) BinDNN~\cite{bindnn} is one of the early works to
leverage deep neural networks (\eg, LSTM) 
to function matching for malware,
ii)~Zhang et al.~\cite{malware-aaai20} present
dynamic malware analysis with feature engineering
(API calls), and 
iii)~Neurlux~\cite{neurlux} proposes 
a system that learns features from a dynamic analysis
report (\ie, behavioral information of malware)
automatically.
Code similarity detection is another active
domain using a probabilistic approach.
Gemini~\cite{gemini} presents a cross-platform binary
code similarity detection with a graph embedding
network and Siamese network~\cite{siamese}.
Lately, varying efforts to deduce 
underlying semantics of a binary code 
have been made with \textit{deep learning}.
InnerEye~\cite{innereye} aims to
detect code similarity across different architectures,
borrowing the concept of neural machine translation (NMT)
from an NLP domain.
It generates vocabulary embeddings
with a Word2vec Skipgram~\cite{word2vec} model with
a coarse-grained normalization process.
Similarly, MIRROR~\cite{bar-bbsimilarity} presents 
an idea of basic block embedding across different ISAs
that utilizes an NMT model to establish the connection
between the two ISAs.
Asm2vec~\cite{asm2vec} takes a PV-DM model
for code clone search, demonstrating
that a code representation can be
robust over compiler optimization and 
even code obfuscation.
SAFE~\cite{safe} generates function embedding
based on a self-attentive neural network.
DeepBinDiff~\cite{deepbindiff} performs
a binary similarity task at the basic block granularity
with token embeddings for semantic information, 
feature vectors, and TADW algorithm for 
prorgram-wide contextual information.
Our experiment includes the two state-of-the-art
approaches from SAFE and DeepBinDiff for comparison.
Meanwhile, Redmond et al.~\cite{embedding-bar19} 
investigate the graph embedding network that
extracts relevant features automatically, resulting 
in similar performance compared to the architecture
without using any structural information, which
aligns our insights. 
Note that \sys use limited 
call graph information (\eg, \cc{libc} call).

\PP{Comparison with the Previous Study} 
%
%Word2vec CBOW for instruction (opcode/operands)
%FV(feature vector) = opcode * TF-IDF Weights || avg(operands)
%Graph merging (virtual node/edge) with TADW
%Block embedding with FV and Graph
%Unsupervised learning
%Preprocessing: IDA Pro CFGs
%
%\XXX{This part must be updated}
\textit{Order matters}~\cite{ordermatters} is 
one of the closest work with ours in terms of
adopting the BERT architecture.
The main difference is that \textit{Order matters} 
suggests the model that represents vectors for different binaries
(\ie, identical source but dissimilar binaries 
due to different platform and optimization)
as close as possible with a focus on seeking an identical binary. 
On the other hand, we propose the model that 
performs contextual similarity detection 
between binary functions. 
%
%It introduces a few combined techniques
%to assist code similarity detection including
%the MLM model for tokens, 
%ANP (Adjacency Node Prediction) for blocks, 
%BIG (Block Inside Graph) and GC (Graph Classification)
%for control flow graph.
%
For evaluation, \textit{Order matters} employs
rank-aware metrics such as Top1 (according to their
evaluation, they obtained 74\% accuracy), 
MRR (Mean Reciprocal Rank)
and NDCG (Normalized Discounted Cumulative Gain), whereas
ours employ a classification metric.
In this regard, the comparison through an experiment with 
\textit{Order matters} would be the best way to 
show the performance of \sys. 
However, we were unable to acquire the source code as
it is close-sourced~\footnote{The authors 
	refused to offer the source code.},
failing to construct an implementation 
based on the given conceptual 
information~\cite{ordermatters}
due to the absence of implementation details
(\eg, model hyperparameters).

%not only a direct comparison 
%with~\cite{ordermatters}
%but also the source code is unavailable
%~\footnote{Note that re-implementation based on 
%the given conceptual information~\cite{ordermatters} 
%was infeasible due to the absence of implementation details 
%(\eg, model hyperparameters).}; 
%we offered our full dataset (including all binaries
%and training/validation/test set) to the authors. 
%
%\KK{Need to check the following statement.}
% \SP{I think following paragraphs can be shown as aggressive comment}
%However, the responded results look unconvincing 
%because all values of P/R/F1 are consistently high
%(\eg, between 99.00\% and 99.92\%) regardless of compilers
%and optimization levels, which is against our
%intuition discussed in~\autoref{s:discussion}.
%
%We failed to debug on how the framework~\cite{ordermatters}
%reaches such outcomes (\eg, overfitting)
%because the authors refused to offer
%their source code.
%

\section{Conclusion}
\label{s:conclusion}

In this paper, we propose the \sys architecture
with BERT that allows for generating 
semantic-aware binary code representation.
We carefully design well-balanced normalization
for binary instructions to convey as much information
as possible during a learning process with BERT.
Moreover, our architecture leverages
the original BERT design to support
varying downstream tasks that require
the inference of code semantics
once a pre-trained model for generic code
representation is readily available.
Our experimental results from the two tasks 
demonstrate that \sysbs surpasses the performance of 
existing binary similarity comparison tools,
and \syscopt also shows acceptable outcomes.
We hope to aid further applications
with the idea of \sys.
%\input{ack}

%\balance

\footnotesize
\bibliographystyle{ACM-Reference-Format}
\bibliography{p,conf}


\begin{thebibliography}{00}

%%% the Web address.

\ifx \showCODEN    \undefined \def \showCODEN     #1{\unskip}     \fi
\ifx \showDOI      \undefined \def \showDOI       #1{#1}\fi
\ifx \showISBNx    \undefined \def \showISBNx     #1{\unskip}     \fi
\ifx \showISBNxiii \undefined \def \showISBNxiii  #1{\unskip}     \fi
\ifx \showISSN     \undefined \def \showISSN      #1{\unskip}     \fi
\ifx \showLCCN     \undefined \def \showLCCN      #1{\unskip}     \fi
\ifx \shownote     \undefined \def \shownote      #1{#1}          \fi
\ifx \showarticletitle \undefined \def \showarticletitle #1{#1}   \fi
\ifx \showURL      \undefined \def \showURL       {\relax}        \fi
% invisible to TeX
\providecommand\bibfield[2]{#2}
\providecommand\bibinfo[2]{#2}
\providecommand\natexlab[1]{#1}
\providecommand\showeprint[2][]{arXiv:#2}

\bibitem[\protect\citeauthoryear{??}{SEC}{2013}]%
        {SEC13}
 \bibinfo{year}{2013}\natexlab{}.
\newblock \bibinfo{booktitle}{{\em Proceedings of the 22nd USENIX Security
  Symposium (Security)}}. \bibinfo{address}{Washington, DC}.
\newblock


\bibitem[\protect\citeauthoryear{??}{NDS}{2016}]%
        {NDSS16}
 \bibinfo{year}{2016}\natexlab{}.
\newblock \bibinfo{booktitle}{{\em Proceedings of the 23rd Annual Network and
  Distributed System Security Symposium (NDSS)}}. \bibinfo{address}{San Diego,
  CA}.
\newblock


\bibitem[\protect\citeauthoryear{??}{ASE}{2018}]%
        {ASE18}
 \bibinfo{year}{2018}\natexlab{}.
\newblock \bibinfo{booktitle}{{\em Proceedings of the 33rd IEEE/ACM
  International Conference on Automated Software Engineering (ASE)}}.
  \bibinfo{address}{Montpellier, France}.
\newblock


\bibitem[\protect\citeauthoryear{??}{BAR}{2019}]%
        {BAR19}
 \bibinfo{year}{2019}\natexlab{}.
\newblock \bibinfo{booktitle}{{\em Proceedings of the 2nd Workshop on Binary
  Analysis Research (BAR)}}. \bibinfo{address}{San Diego, CA}.
\newblock


\bibitem[\protect\citeauthoryear{??}{AAA}{2020}]%
        {AAAI20}
 \bibinfo{year}{2020}\natexlab{}.
\newblock \bibinfo{booktitle}{{\em Proceedings of the 34st AAAI Conference on
  Artificial Intelligence (AAAI)}}. \bibinfo{address}{New York, NY}.
\newblock


\bibitem[\protect\citeauthoryear{??}{BAR}{2020}]%
        {BAR20}
 \bibinfo{year}{2020}\natexlab{}.
\newblock \bibinfo{booktitle}{{\em Proceedings of the 3rd Workshop on Binary
  Analysis Research (BAR)}}. \bibinfo{address}{San Diego, CA}.
\newblock


\bibitem[\protect\citeauthoryear{Alammar}{Alammar}{2018}]%
        {illustrated-bert}
\bibfield{author}{\bibinfo{person}{Jay Alammar}.}
  \bibinfo{year}{2018}\natexlab{}.
\newblock \bibinfo{title}{{The Illustrated BERT, ELMo, and co. (How NLP Cracked
  Transfer Learning)}}.
\newblock   (\bibinfo{year}{2018}).
\newblock
\newblock
\shownote{\url{https://jalammar.github.io/illustrated-bert/}.}


\bibitem[\protect\citeauthoryear{Angr}{Angr}{2016}]%
        {angr}
\bibfield{author}{\bibinfo{person}{Angr}.} \bibinfo{year}{2016}\natexlab{}.
\newblock \bibinfo{title}{{Python Framework for Analyzing Binaries}}.
\newblock   (\bibinfo{year}{2016}).
\newblock
\newblock
\shownote{\url{https://angr.io/}.}


\bibitem[\protect\citeauthoryear{Bahdanau, Cho, and Bengio}{Bahdanau
  et~al\mbox{.}}{2015}]%
        {attention}
\bibfield{author}{\bibinfo{person}{Dzmitry Bahdanau},
  \bibinfo{person}{Kyung~Hyun Cho}, {and} \bibinfo{person}{Yoshua Bengio}.}
  \bibinfo{year}{2015}\natexlab{}.
\newblock \showarticletitle{Neural Machine Translation by Jointly Learning to
  Align and Translate}. In \bibinfo{booktitle}{{\em Proceedings of the 3rd
  International Conference on Learning Representations (ICLR)}}.
\newblock


\bibitem[\protect\citeauthoryear{Bromley, Guyon, LeCun, Säckinger, and
  Shah}{Bromley et~al\mbox{.}}{1993}]%
        {siamese}
\bibfield{author}{\bibinfo{person}{Jane Bromley}, \bibinfo{person}{Isabelle~M
  Guyon}, \bibinfo{person}{Yann LeCun}, \bibinfo{person}{Eduard Säckinger},
  {and} \bibinfo{person}{Roopak Shah}.} \bibinfo{year}{1993}\natexlab{}.
\newblock \showarticletitle{Signature Verification using a Siamese Time Delay
  Neural Network}. In \bibinfo{booktitle}{{\em Proceedings of the 6th
  Conference on Neural Information Processing Systems (NeurIPS)}}.
\newblock


\bibitem[\protect\citeauthoryear{Bruschi, Martignoni, and Monga}{Bruschi
  et~al\mbox{.}}{2006}]%
        {self-mutating-malware}
\bibfield{author}{\bibinfo{person}{Danilo Bruschi}, \bibinfo{person}{Lorenzo
  Martignoni}, {and} \bibinfo{person}{Mattia Monga}.}
  \bibinfo{year}{2006}\natexlab{}.
\newblock \showarticletitle{Detecting Self-mutating Malware Using Control-flow
  Graph Matching}. In \bibinfo{booktitle}{{\em Proceedings of the 3rd
  Conference on Detection of Intrusions and Malware and Vulnerability
  Assessment (DIMVA)}}. \bibinfo{address}{Berlin, Germany}.
\newblock


\bibitem[\protect\citeauthoryear{Cesare, Xiang, and Zhou}{Cesare
  et~al\mbox{.}}{2014}]%
        {cfg-malware-detection}
\bibfield{author}{\bibinfo{person}{Silvio Cesare}, \bibinfo{person}{Yang
  Xiang}, {and} \bibinfo{person}{Wanlei Zhou}.}
  \bibinfo{year}{2014}\natexlab{}.
\newblock \showarticletitle{Control Flow-Based Malware Variant Detection}.
\newblock \bibinfo{journal}{{\em IEEE Transactions on Dependable and Secure
  Computing (TDSC)\/}} (\bibinfo{year}{2014}).
\newblock


\bibitem[\protect\citeauthoryear{Chandramohan, Xue, Xu, Liu, Cho, and
  Kuan}{Chandramohan et~al\mbox{.}}{2016}]%
        {bingo}
\bibfield{author}{\bibinfo{person}{Mahinthan Chandramohan},
  \bibinfo{person}{Yinxing Xue}, \bibinfo{person}{Zhengzi Xu},
  \bibinfo{person}{Yang Liu}, \bibinfo{person}{Chia~Yuan Cho}, {and}
  \bibinfo{person}{Tan Hee~Beng Kuan}.} \bibinfo{year}{2016}\natexlab{}.
\newblock \showarticletitle{BinGo: Cross-Architecture Cross-OS Binary Search}.
  In \bibinfo{booktitle}{{\em Proceedings of the 24th ACM SIGSOFT Symposium on
  the Foundations of Software Engineering (FSE)}}. \bibinfo{address}{Seattle,
  WA}.
\newblock


\bibitem[\protect\citeauthoryear{Chandramohan, Xue, Xu, Liu, Cho, and
  Kuan}{Chandramohan et~al\mbox{.}}{2018}]%
        {vulseeker}
\bibfield{author}{\bibinfo{person}{Mahinthan Chandramohan},
  \bibinfo{person}{Yinxing Xue}, \bibinfo{person}{Zhengzi Xu},
  \bibinfo{person}{Yang Liu}, \bibinfo{person}{Chia~Yuan Cho}, {and}
  \bibinfo{person}{Tan Hee~Beng Kuan}.} \bibinfo{year}{2018}\natexlab{}.
\newblock \showarticletitle{VulSeeker: a semantic learning based vulnerability
  seeker for cross-platform binary}, See \citeN{ASE18}.
\newblock


\bibitem[\protect\citeauthoryear{Chen, Egele, Woo, and Brumley}{Chen
  et~al\mbox{.}}{2016}]%
        {fermadyine}
\bibfield{author}{\bibinfo{person}{Daming~D. Chen}, \bibinfo{person}{Manuel
  Egele}, \bibinfo{person}{Maverick Woo}, {and} \bibinfo{person}{David
  Brumley}.} \bibinfo{year}{2016}\natexlab{}.
\newblock \showarticletitle{Towards Automated Dynamic Analysis for Linux-based
  Embedded Firmware}, See \citeN{NDSS16}.
\newblock


\bibitem[\protect\citeauthoryear{Cho, van Merriënboer, Gulcehre, Bahdanau,
  Bougares, Schwenk, and Bengio}{Cho et~al\mbox{.}}{2014}]%
        {gru}
\bibfield{author}{\bibinfo{person}{Kyunghyun Cho}, \bibinfo{person}{Bart van
  Merriënboer}, \bibinfo{person}{Caglar Gulcehre}, \bibinfo{person}{Dzmitry
  Bahdanau}, \bibinfo{person}{Fethi Bougares}, \bibinfo{person}{Holger
  Schwenk}, {and} \bibinfo{person}{Yoshua Bengio}.}
  \bibinfo{year}{2014}\natexlab{}.
\newblock \showarticletitle{Learning Phrase Representations using RNN
  Encoder-Decoder for Statistical Machine Translation}. In
  \bibinfo{booktitle}{{\em Proceedings of the 2014 Conference on Empirical
  Methods in Natural Language Processing (EMNLP)}}.
\newblock


\bibitem[\protect\citeauthoryear{Choi, Liu, Shang, Wang, Wang, Zhang, and
  Zhou}{Choi et~al\mbox{.}}{2020}]%
        {dl-based-binanalysis-survey}
\bibfield{author}{\bibinfo{person}{Yoon-Ho Choi}, \bibinfo{person}{Peng Liu},
  \bibinfo{person}{Zitong Shang}, \bibinfo{person}{Haizhou Wang},
  \bibinfo{person}{Zhilong Wang}, \bibinfo{person}{Lan Zhang}, {and}
  \bibinfo{person}{Junwei Zhou}.} \bibinfo{year}{2020}\natexlab{}.
\newblock \showarticletitle{Using Deep Learning to Solve Computer Security
  Challenges: a Survey}.
\newblock \bibinfo{journal}{{\em Cybersecurity\/}} (\bibinfo{year}{2020}).
\newblock


\bibitem[\protect\citeauthoryear{Dai and Le}{Dai and Le}{2015}]%
        {semi-supervised}
\bibfield{author}{\bibinfo{person}{Andrew~M. Dai} {and}
  \bibinfo{person}{Quoc~V. Le}.} \bibinfo{year}{2015}\natexlab{}.
\newblock \showarticletitle{Semi-supervised Sequence Learning}.
\newblock \bibinfo{journal}{{\em arXiv preprint arXiv:1511.01432\/}}
  (\bibinfo{year}{2015}).
\newblock
\showURL{%
\url{https://arxiv.org/abs/1511.01432}}


\bibitem[\protect\citeauthoryear{Dai and Le}{Dai and Le}{2019}]%
        {binsim-survey}
\bibfield{author}{\bibinfo{person}{Andrew~M. Dai} {and}
  \bibinfo{person}{Quoc~V. Le}.} \bibinfo{year}{2019}\natexlab{}.
\newblock \showarticletitle{A Survey of Binary Code Similarity}.
\newblock \bibinfo{journal}{{\em arXiv preprint arXiv:1909.11424\/}}
  (\bibinfo{year}{2019}).
\newblock
\showURL{%
\url{https://arxiv.org/pdf/1909.11424.pdf}}


\bibitem[\protect\citeauthoryear{David, Partush, and Yahav}{David
  et~al\mbox{.}}{2016}]%
        {esh-pldi16}
\bibfield{author}{\bibinfo{person}{Yaniv David}, \bibinfo{person}{Nimrod
  Partush}, {and} \bibinfo{person}{Eran Yahav}.}
  \bibinfo{year}{2016}\natexlab{}.
\newblock \showarticletitle{Statistical Similarity of Binaries}. In
  \bibinfo{booktitle}{{\em Proceedings of the 2016 ACM SIGPLAN Conference on
  Programming Language Design and Implementation (PLDI)}}.
  \bibinfo{address}{Santa Barbara, CA}.
\newblock


\bibitem[\protect\citeauthoryear{David, Partush, and Yahav}{David
  et~al\mbox{.}}{2017}]%
        {esh2}
\bibfield{author}{\bibinfo{person}{Yaniv David}, \bibinfo{person}{Nimrod
  Partush}, {and} \bibinfo{person}{Eran Yahav}.}
  \bibinfo{year}{2017}\natexlab{}.
\newblock \showarticletitle{Similarity of binaries through re-optimization}. In
  \bibinfo{booktitle}{{\em Proceedings of the 2017 ACM SIGPLAN Conference on
  Programming Language Design and Implementation (PLDI)}}.
  \bibinfo{address}{Barcelona, Spain}.
\newblock


\bibitem[\protect\citeauthoryear{Devlin, Chang, Lee, and Toutanova}{Devlin
  et~al\mbox{.}}{2019}]%
        {bert}
\bibfield{author}{\bibinfo{person}{Jacob Devlin}, \bibinfo{person}{Ming-Wei
  Chang}, \bibinfo{person}{Kenton Lee}, {and} \bibinfo{person}{Kristina
  Toutanova}.} \bibinfo{year}{2019}\natexlab{}.
\newblock \showarticletitle{{BERT}: Pre-training of Deep Bidirectional
  Transformers for Language Understanding}. In \bibinfo{booktitle}{{\em
  Proceedings of the 2019 Conference of the North {A}merican Chapter of the
  Association for Computational Linguistics: Human Language Technologies,
  Volume 1 (Long and Short Papers)}}. \bibinfo{address}{Minneapolis,
  Minnesota}, \bibinfo{pages}{4171--4186}.
\newblock


\bibitem[\protect\citeauthoryear{Dinga, Fung, and Charland}{Dinga
  et~al\mbox{.}}{2019}]%
        {asm2vec}
\bibfield{author}{\bibinfo{person}{Steven H.~H. Dinga},
  \bibinfo{person}{Benjamin C.~M. Fung}, {and} \bibinfo{person}{Philippe
  Charland}.} \bibinfo{year}{2019}\natexlab{}.
\newblock \showarticletitle{Asm2vec: Boosting static representation robustness
  for binary clone search against code obfuscation and compiler optimization}.
  In \bibinfo{booktitle}{{\em Proceedings of the 40th IEEE Symposium on
  Security and Privacy (Oakland)}}. \bibinfo{address}{San Francisco, CA}.
\newblock


\bibitem[\protect\citeauthoryear{Egele, Woo, Chapman, and Brumley}{Egele
  et~al\mbox{.}}{2014}]%
        {blanket}
\bibfield{author}{\bibinfo{person}{Manuel Egele}, \bibinfo{person}{Maverick
  Woo}, \bibinfo{person}{Peter Chapman}, {and} \bibinfo{person}{David
  Brumley}.} \bibinfo{year}{2014}\natexlab{}.
\newblock \showarticletitle{Blanket Execution: Dynamic Similarity Testing for
  Program Binaries and Components}. In \bibinfo{booktitle}{{\em Proceedings of
  the 23rd USENIX Security Symposium (Security)}}. \bibinfo{address}{San Diego,
  CA}.
\newblock


\bibitem[\protect\citeauthoryear{Eschweiler, Yakdan, and
  Gerhards-Padilla}{Eschweiler et~al\mbox{.}}{2016}]%
        {discovre}
\bibfield{author}{\bibinfo{person}{Sebastian Eschweiler},
  \bibinfo{person}{Khaled Yakdan}, {and} \bibinfo{person}{Elmar
  Gerhards-Padilla}.} \bibinfo{year}{2016}\natexlab{}.
\newblock \showarticletitle{discovRE: Efficient Cross-Architecture
  Identification of Bugs in Binary Code}, See \citeN{NDSS16}.
\newblock


\bibitem[\protect\citeauthoryear{Flake}{Flake}{2004}]%
        {exe-comp-dimva04}
\bibfield{author}{\bibinfo{person}{Halvar Flake}.}
  \bibinfo{year}{2004}\natexlab{}.
\newblock \showarticletitle{Structural Comparison of Executable Objects}. In
  \bibinfo{booktitle}{{\em Proceedings of the 1st Conference on Detection of
  Intrusions and Malware and Vulnerability Assessment (DIMVA)}}.
  \bibinfo{address}{Dortmund, Germany}.
\newblock


\bibitem[\protect\citeauthoryear{Ghaffarian and Shahriari}{Ghaffarian and
  Shahriari}{2017}]%
        {ml-dm-binanalysis-survey}
\bibfield{author}{\bibinfo{person}{Seyed~Mohammad Ghaffarian} {and}
  \bibinfo{person}{Hamid~Reza Shahriari}.} \bibinfo{year}{2017}\natexlab{}.
\newblock \showarticletitle{Software Vulnerability Analysis and Discovery Using
  Machine-Learning and Data-Mining Techniques: A Survey}.
\newblock \bibinfo{journal}{{\it Comput. Surveys}} (\bibinfo{year}{2017}).
\newblock


\bibitem[\protect\citeauthoryear{Google}{Google}{2020a}]%
        {tensorflow}
\bibfield{author}{\bibinfo{person}{Google}.} \bibinfo{year}{2020}\natexlab{a}.
\newblock \bibinfo{title}{{End-to-end open source machine learning platform}}.
\newblock   (\bibinfo{year}{2020}).
\newblock
\newblock
\shownote{\url{https://tensorflow.org}.}


\bibitem[\protect\citeauthoryear{Google}{Google}{2020b}]%
        {google-bertimpl}
\bibfield{author}{\bibinfo{person}{Google}.} \bibinfo{year}{2020}\natexlab{b}.
\newblock \bibinfo{title}{{Release of BERT Models}}.
\newblock   (\bibinfo{year}{2020}).
\newblock
\newblock
\shownote{\url{https://github.com/google-research/bert}.}


\bibitem[\protect\citeauthoryear{Hex-Rays}{Hex-Rays}{2005}]%
        {idapro}
\bibfield{author}{\bibinfo{person}{Hex-Rays}.} \bibinfo{year}{2005}\natexlab{}.
\newblock \bibinfo{title}{{IDA Pro Disassembler}}.
\newblock   (\bibinfo{year}{2005}).
\newblock
\newblock
\shownote{\url{https://www.hex-rays.com/products/ida/}.}


\bibitem[\protect\citeauthoryear{Hex-rays}{Hex-rays}{2019}]%
        {idapython}
\bibfield{author}{\bibinfo{person}{Hex-rays}.} \bibinfo{year}{2019}\natexlab{}.
\newblock \bibinfo{title}{{IDAPython Documentation}}.
\newblock   (\bibinfo{year}{2019}).
\newblock
\newblock
\shownote{\url{https://www.hex-rays.com/products/ida/support/idapython_docs/}.}


\bibitem[\protect\citeauthoryear{Hex-Rays}{Hex-Rays}{2020}]%
        {flirt}
\bibfield{author}{\bibinfo{person}{Hex-Rays}.} \bibinfo{year}{2020}\natexlab{}.
\newblock \bibinfo{title}{{IDA Pro Fast Library Identification and Recognition
  Technology}}.
\newblock   (\bibinfo{year}{2020}).
\newblock
\newblock
\shownote{\url{https://www.hex-rays.com/products/ida/tech/flirt/}.}


\bibitem[\protect\citeauthoryear{Hochreiter and Schmidhuber}{Hochreiter and
  Schmidhuber}{1997}]%
        {lstm}
\bibfield{author}{\bibinfo{person}{Sepp Hochreiter} {and}
  \bibinfo{person}{Jürgen Schmidhuber}.} \bibinfo{year}{1997}\natexlab{}.
\newblock \showarticletitle{Long Short-Term Memory}.
\newblock \bibinfo{journal}{{\em Neural Computation\/}} \bibinfo{volume}{9},
  \bibinfo{number}{8} (\bibinfo{year}{1997}), \bibinfo{pages}{1735--1780}.
\newblock


\bibitem[\protect\citeauthoryear{Hu, Bhatkar, Griffin, and Shin}{Hu
  et~al\mbox{.}}{2013}]%
        {mutantx}
\bibfield{author}{\bibinfo{person}{Xin Hu}, \bibinfo{person}{Sandeep Bhatkar},
  \bibinfo{person}{Kent Griffin}, {and} \bibinfo{person}{Kang~G. Shin}.}
  \bibinfo{year}{2013}\natexlab{}.
\newblock \showarticletitle{MutantX-S: Scalable Malware Clustering Based on
  Static Features}, See \citeN{SEC13}.
\newblock


\bibitem[\protect\citeauthoryear{Hu, Zhang, Li, and Gu}{Hu
  et~al\mbox{.}}{2016}]%
        {code-comp-saner16}
\bibfield{author}{\bibinfo{person}{Yikun Hu}, \bibinfo{person}{Yuanyuan Zhang},
  \bibinfo{person}{Juanru Li}, {and} \bibinfo{person}{Dawu Gu}.}
  \bibinfo{year}{2016}\natexlab{}.
\newblock \showarticletitle{Cross-architecture Binary Semantics Understanding
  via Similar Code Comparison}. In \bibinfo{booktitle}{{\em Proceedings of the
  2016 IEEE 23rd International Conference on Software Analysis, Evolution, and
  Reengineering (SANER)}}.
\newblock


\bibitem[\protect\citeauthoryear{huanghonggit}{huanghonggit}{2019}]%
        {bertimpl}
\bibfield{author}{\bibinfo{person}{huanghonggit}.}
  \bibinfo{year}{2019}\natexlab{}.
\newblock \bibinfo{title}{{BERT MLM with Pytorch}}.
\newblock   (\bibinfo{year}{2019}).
\newblock
\newblock
\shownote{\url{https://github.com/huanghonggit/Mask-Language-Model}.}


\bibitem[\protect\citeauthoryear{Jang, Woo, , and Brumley}{Jang
  et~al\mbox{.}}{2013}]%
        {lineage-inference}
\bibfield{author}{\bibinfo{person}{Jiyong Jang}, \bibinfo{person}{Maverick
  Woo}, \bibinfo{person}{}, {and} \bibinfo{person}{David Brumley}.}
  \bibinfo{year}{2013}\natexlab{}.
\newblock \showarticletitle{Towards Automatic Software Lineage Inference}, See
  \citeN{SEC13}.
\newblock


\bibitem[\protect\citeauthoryear{Jindal, Salls, Aghakhani, Long, Kruegel, and
  Vigna}{Jindal et~al\mbox{.}}{2019}]%
        {neurlux}
\bibfield{author}{\bibinfo{person}{Chani Jindal}, \bibinfo{person}{Christopher
  Salls}, \bibinfo{person}{Hojjat Aghakhani}, \bibinfo{person}{Keith Long},
  \bibinfo{person}{Christopher Kruegel}, {and} \bibinfo{person}{Giovanni
  Vigna}.} \bibinfo{year}{2019}\natexlab{}.
\newblock \showarticletitle{Neurlux: Dynamic Malware Analysis Without Feature
  Engineering}. In \bibinfo{booktitle}{{\em Proceedings of the Annual Computer
  Security Applications Conference (ACSAC)}}.
\newblock


\bibitem[\protect\citeauthoryear{Karpathy}{Karpathy}{2015}]%
        {linux-kernel-rnn}
\bibfield{author}{\bibinfo{person}{Andrej Karpathy}.}
  \bibinfo{year}{2015}\natexlab{}.
\newblock \bibinfo{title}{{The Unreasonable Effectiveness of Recurrent Neural
  Networks}}.
\newblock   (\bibinfo{year}{2015}).
\newblock
\newblock
\shownote{\url{http://karpathy.github.io/2015/05/21/rnn-effectiveness}.}


\bibitem[\protect\citeauthoryear{Kim, Lee, Kang, and Im}{Kim
  et~al\mbox{.}}{2019}]%
        {simcalc19}
\bibfield{author}{\bibinfo{person}{TaeGuen Kim}, \bibinfo{person}{Yeo~Reum
  Lee}, \bibinfo{person}{BooJoong Kang}, {and} \bibinfo{person}{Eul~Gyu Im}.}
  \bibinfo{year}{2019}\natexlab{}.
\newblock \showarticletitle{Binary Executable File Similarity Calculation using
  Function Matching}.
\newblock \bibinfo{journal}{{\em The Journal of Supercomputing\/}}
  (\bibinfo{year}{2019}).
\newblock


\bibitem[\protect\citeauthoryear{Kruegel, Kirda, Mutz, Robertson, and
  Vigna}{Kruegel et~al\mbox{.}}{2005}]%
        {polymorphic-worm}
\bibfield{author}{\bibinfo{person}{Christopher Kruegel}, \bibinfo{person}{Engin
  Kirda}, \bibinfo{person}{Darren Mutz}, \bibinfo{person}{William Robertson},
  {and} \bibinfo{person}{Giovanni Vigna}.} \bibinfo{year}{2005}\natexlab{}.
\newblock \showarticletitle{Polymorphic Worm Detection Using Structural
  Information of Executables}. In \bibinfo{booktitle}{{\em Proceedings of the
  8th International Symposium on Research in Attacks, Intrusions and Defenses
  (RAID)}}. \bibinfo{address}{Seattle, WA}.
\newblock


\bibitem[\protect\citeauthoryear{Lageman, Kilmer, Walls, and McDaniel}{Lageman
  et~al\mbox{.}}{2016}]%
        {bindnn}
\bibfield{author}{\bibinfo{person}{Nathaniel Lageman}, \bibinfo{person}{Eric~D.
  Kilmer}, \bibinfo{person}{Robert~J. Walls}, {and} \bibinfo{person}{Patrick~D.
  McDaniel}.} \bibinfo{year}{2016}\natexlab{}.
\newblock \showarticletitle{BinDNN: Resilient Function Matching Using Deep
  Learning}. In \bibinfo{booktitle}{{\em Proceedings of the 12th Security and
  Privacy in Communication Networks (SECOMM)}}. \bibinfo{address}{Guangzhou,
  China}.
\newblock


\bibitem[\protect\citeauthoryear{Lindorfer, Federico, Maggi, Comparetti, and
  Zanero}{Lindorfer et~al\mbox{.}}{2012}]%
        {malware-author-acsac12}
\bibfield{author}{\bibinfo{person}{Marina Lindorfer},
  \bibinfo{person}{Alessandro~Di Federico}, \bibinfo{person}{Federico Maggi},
  \bibinfo{person}{Paolo~Milani Comparetti}, {and} \bibinfo{person}{Stefano
  Zanero}.} \bibinfo{year}{2012}\natexlab{}.
\newblock \showarticletitle{Lines of Malicious Code: Insights Into the
  Malicious Software Industry}. In \bibinfo{booktitle}{{\em Proceedings of the
  Annual Computer Security Applications Conference (ACSAC)}}.
\newblock


\bibitem[\protect\citeauthoryear{Liu, Huo, Zhang, Li, Li, Piao, and Zou}{Liu
  et~al\mbox{.}}{2018}]%
        {alphadiff}
\bibfield{author}{\bibinfo{person}{Bingchang Liu}, \bibinfo{person}{Wei Huo},
  \bibinfo{person}{Chao Zhang}, \bibinfo{person}{Wenchao Li},
  \bibinfo{person}{Feng Li}, \bibinfo{person}{Aihua Piao}, {and}
  \bibinfo{person}{Wei Zou}.} \bibinfo{year}{2018}\natexlab{}.
\newblock \showarticletitle{{$\alpha$Diff}: Cross-version Binary Code
  Similarity Detection with DNN}, See \citeN{ASE18}.
\newblock


\bibitem[\protect\citeauthoryear{Liu, Ott, Goyal, Du, Joshi, Chen, Levy, Lewis,
  Zettlemoyer, and Stoyanov}{Liu et~al\mbox{.}}{2019}]%
        {roberta}
\bibfield{author}{\bibinfo{person}{Yinhan Liu}, \bibinfo{person}{Myle Ott},
  \bibinfo{person}{Naman Goyal}, \bibinfo{person}{Jingfei Du},
  \bibinfo{person}{Mandar Joshi}, \bibinfo{person}{Danqi Chen},
  \bibinfo{person}{Omer Levy}, \bibinfo{person}{Mike Lewis},
  \bibinfo{person}{Luke Zettlemoyer}, {and} \bibinfo{person}{Veselin
  Stoyanov}.} \bibinfo{year}{2019}\natexlab{}.
\newblock \showarticletitle{RoBERTa: {A} Robustly Optimized {BERT} Pretraining
  Approach}.
\newblock \bibinfo{journal}{{\em CoRR\/}}  \bibinfo{volume}{abs/1907.11692}
  (\bibinfo{year}{2019}).
\newblock
\showURL{%
\url{http://arxiv.org/abs/1907.11692}}


\bibitem[\protect\citeauthoryear{Luo, Ming, Wu, Liu, and Zhu}{Luo
  et~al\mbox{.}}{2014}]%
        {binsimfse}
\bibfield{author}{\bibinfo{person}{Lannan Luo}, \bibinfo{person}{Jiang Ming},
  \bibinfo{person}{Dinghao Wu}, \bibinfo{person}{Peng Liu}, {and}
  \bibinfo{person}{Sencun Zhu}.} \bibinfo{year}{2014}\natexlab{}.
\newblock \showarticletitle{Semantics-Based Obfuscation-Resilient Binary Code
  Similarity Comparison with Applications to Software Plagiarism Detection}. In
  \bibinfo{booktitle}{{\em Proceedings of the 22nd ACM SIGSOFT Symposium on the
  Foundations of Software Engineering (FSE)}}. \bibinfo{address}{Hong Kong}.
\newblock


\bibitem[\protect\citeauthoryear{Massarelli}{Massarelli}{2019}]%
        {safe-tool}
\bibfield{author}{\bibinfo{person}{Luca Massarelli}.}
  \bibinfo{year}{2019}\natexlab{}.
\newblock \bibinfo{title}{{Self Attentive Function Embedding Tool}}.
\newblock   (\bibinfo{year}{2019}).
\newblock
\newblock
\shownote{\url{https://github.com/gadiluna/SAFE}.}


\bibitem[\protect\citeauthoryear{Massarelli, Luna, , Petroni, Querzoni, and
  Baldoni}{Massarelli et~al\mbox{.}}{2019a}]%
        {struct2vec}
\bibfield{author}{\bibinfo{person}{Luca Massarelli}, \bibinfo{person}{Giuseppe
  A.~Di Luna}, \bibinfo{person}{}, \bibinfo{person}{Fabio Petroni},
  \bibinfo{person}{Leonardo Querzoni}, {and} \bibinfo{person}{Roberto
  Baldoni}.} \bibinfo{year}{2019}\natexlab{a}.
\newblock \showarticletitle{Investigating Graph Embedding Neural Networks with
  Unsupervised Features Extraction for Binary Analysis}, See \citeN{BAR19}.
\newblock


\bibitem[\protect\citeauthoryear{Massarelli, Luna, Petroni, Querzoni, and
  Baldoni}{Massarelli et~al\mbox{.}}{2019b}]%
        {safe}
\bibfield{author}{\bibinfo{person}{Luca Massarelli}, \bibinfo{person}{Giuseppe
  Antonio~Di Luna}, \bibinfo{person}{Fabio Petroni}, \bibinfo{person}{Leonardo
  Querzoni}, {and} \bibinfo{person}{Roberto Baldoni}.}
  \bibinfo{year}{2019}\natexlab{b}.
\newblock \showarticletitle{SAFE: Self-Attentive Function Embeddings for Binary
  Similarity}. In \bibinfo{booktitle}{{\em Proceedings of the 16th Conference
  on Detection of Intrusions and Malware and Vulnerability Assessment
  (DIMVA)}}. \bibinfo{address}{Göteborg, Sweden}.
\newblock


\bibitem[\protect\citeauthoryear{Mikolov, Sutskever, Chen, Corrado, and
  Dean}{Mikolov et~al\mbox{.}}{2013}]%
        {word2vec}
\bibfield{author}{\bibinfo{person}{Tomas Mikolov}, \bibinfo{person}{Ilya
  Sutskever}, \bibinfo{person}{Kai Chen}, \bibinfo{person}{Greg~S. Corrado},
  {and} \bibinfo{person}{Jeff Dean}.} \bibinfo{year}{2013}\natexlab{}.
\newblock \showarticletitle{Distributed Representations of Words and Phrases
  and their Compositionality}. In \bibinfo{booktitle}{{\em Proceedings of the
  26th Conference on Neural Information Processing Systems (NeurIPS)}}.
\newblock


\bibitem[\protect\citeauthoryear{Nginx}{Nginx}{2020}]%
        {nginx}
\bibfield{author}{\bibinfo{person}{Nginx}.} \bibinfo{year}{2020}\natexlab{}.
\newblock \bibinfo{title}{{High Performance Load-balancer and Web Server}}.
\newblock   (\bibinfo{year}{2020}).
\newblock
\newblock
\shownote{\url{https://nginx.com}.}


\bibitem[\protect\citeauthoryear{(NSA)}{(NSA)}{2019}]%
        {ghidra}
\bibfield{author}{\bibinfo{person}{National Security~Agency (NSA)}.}
  \bibinfo{year}{2019}\natexlab{}.
\newblock \bibinfo{title}{{Software Reverse Engineering (SRE) Suite of Tools}}.
\newblock   (\bibinfo{year}{2019}).
\newblock
\newblock
\shownote{\url{https://ghidra-sre.org/}.}


\bibitem[\protect\citeauthoryear{OpenSSL}{OpenSSL}{2020}]%
        {openssl}
\bibfield{author}{\bibinfo{person}{OpenSSL}.} \bibinfo{year}{2020}\natexlab{}.
\newblock \bibinfo{title}{{Cryptography and SSL/TLS Toolkit}}.
\newblock   (\bibinfo{year}{2020}).
\newblock
\newblock
\shownote{\url{https://www.openssl.org}.}


\bibitem[\protect\citeauthoryear{Otsubo, Otsuka, Mimura, Sakaki, and
  Ukegawa}{Otsubo et~al\mbox{.}}{2020}]%
        {oglassesx}
\bibfield{author}{\bibinfo{person}{Yuhei Otsubo}, \bibinfo{person}{Akira
  Otsuka}, \bibinfo{person}{Mamoru Mimura}, \bibinfo{person}{Takeshi Sakaki},
  {and} \bibinfo{person}{Hiroshi Ukegawa}.} \bibinfo{year}{2020}\natexlab{}.
\newblock \showarticletitle{o-glassesX: Compiler Provenance Recovery with
  Attention Mechanism from a Short Code Fragment}, See \citeN{BAR20}.
\newblock


\bibitem[\protect\citeauthoryear{Pang, Yu, Chen, Koskinen, Portokalidis, Mao,
  and Xu}{Pang et~al\mbox{.}}{2021}]%
        {sok-x86}
\bibfield{author}{\bibinfo{person}{Chengbin Pang}, \bibinfo{person}{Ruotong
  Yu}, \bibinfo{person}{Yaohui Chen}, \bibinfo{person}{Eric Koskinen},
  \bibinfo{person}{Georgios Portokalidis}, \bibinfo{person}{Bing Mao}, {and}
  \bibinfo{person}{Jun Xu}.} \bibinfo{year}{2021}\natexlab{}.
\newblock \showarticletitle{SoK: All You Ever Wanted to Know About x86/x64
  Binary Disassembly But Were Afraid to Ask}. In \bibinfo{booktitle}{{\em
  Proceedings of the 42nd IEEE Symposium on Security and Privacy (Oakland)}}.
  \bibinfo{address}{San Francisco, CA}.
\newblock


\bibitem[\protect\citeauthoryear{Peters, Neumann, Iyyer, Gardner, Clark, Lee,
  and Zettlemoyer}{Peters et~al\mbox{.}}{2018}]%
        {elmo}
\bibfield{author}{\bibinfo{person}{Matthew~E. Peters}, \bibinfo{person}{Mark
  Neumann}, \bibinfo{person}{Mohit Iyyer}, \bibinfo{person}{Matt Gardner},
  \bibinfo{person}{Christopher Clark}, \bibinfo{person}{Kenton Lee}, {and}
  \bibinfo{person}{Luke Zettlemoyer}.} \bibinfo{year}{2018}\natexlab{}.
\newblock \showarticletitle{Deep contextualized word representations}. In
  \bibinfo{booktitle}{{\em Proceedings of the 2018 Conference of the
  Association for Computational Linguistics (NAACL)}}.
\newblock


\bibitem[\protect\citeauthoryear{Pewny, Garmany, Gawlik, Rossow, and
  Holz}{Pewny et~al\mbox{.}}{2015}]%
        {xarch-bug}
\bibfield{author}{\bibinfo{person}{Jannik Pewny}, \bibinfo{person}{Behrad
  Garmany}, \bibinfo{person}{Robert Gawlik}, \bibinfo{person}{Christian
  Rossow}, {and} \bibinfo{person}{Thorsten Holz}.}
  \bibinfo{year}{2015}\natexlab{}.
\newblock \showarticletitle{Cross-Architecture Bug Search in Binary
  Executables}. In \bibinfo{booktitle}{{\em Proceedings of the 36th IEEE
  Symposium on Security and Privacy (Oakland)}}. \bibinfo{address}{San Jose,
  CA}.
\newblock


\bibitem[\protect\citeauthoryear{Pewny, Schuster, Rossow, Bernhard, and
  Holz}{Pewny et~al\mbox{.}}{2014}]%
        {bug-search-acsac14}
\bibfield{author}{\bibinfo{person}{Jannik Pewny}, \bibinfo{person}{Felix
  Schuster}, \bibinfo{person}{Christian Rossow}, \bibinfo{person}{Lukas
  Bernhard}, {and} \bibinfo{person}{Thorsten Holz}.}
  \bibinfo{year}{2014}\natexlab{}.
\newblock \showarticletitle{Leveraging Semantic Signatures for Bug Search in
  Binary Programs}. In \bibinfo{booktitle}{{\em Proceedings of the Annual
  Computer Security Applications Conference (ACSAC)}}.
\newblock


\bibitem[\protect\citeauthoryear{PyTorch}{PyTorch}{2019}]%
        {pytorch}
\bibfield{author}{\bibinfo{person}{PyTorch}.} \bibinfo{year}{2019}\natexlab{}.
\newblock \bibinfo{title}{{Open Source Machine Learning Framework}}.
\newblock   (\bibinfo{year}{2019}).
\newblock
\newblock
\shownote{\url{https://pytorch.org/}.}


\bibitem[\protect\citeauthoryear{Radare2}{Radare2}{2019}]%
        {radare2}
\bibfield{author}{\bibinfo{person}{Radare2}.} \bibinfo{year}{2019}\natexlab{}.
\newblock \bibinfo{title}{{Libre and Portable Reverse Engineering Framework}}.
\newblock   (\bibinfo{year}{2019}).
\newblock
\newblock
\shownote{\url{https://rada.re/n/}.}


\bibitem[\protect\citeauthoryear{Redmond, Luo, and Zeng}{Redmond
  et~al\mbox{.}}{2019}]%
        {embedding-bar19}
\bibfield{author}{\bibinfo{person}{Kimberly Redmond}, \bibinfo{person}{Lannan
  Luo}, {and} \bibinfo{person}{Qiang Zeng}.} \bibinfo{year}{2019}\natexlab{}.
\newblock \showarticletitle{A Cross-Architecture Instruction Embedding Model
  for Natural Language Processing-Inspired Binary Code Analysis}, See
  \citeN{BAR19}.
\newblock


\bibitem[\protect\citeauthoryear{Rosenblum, Miller, and Zhu}{Rosenblum
  et~al\mbox{.}}{2011}]%
        {toolchain-recovery}
\bibfield{author}{\bibinfo{person}{Nathan Rosenblum},
  \bibinfo{person}{Barton~P. Miller}, {and} \bibinfo{person}{Xiaojin Zhu}.}
  \bibinfo{year}{2011}\natexlab{}.
\newblock \showarticletitle{Recovering the toolchain provenance of binary
  code}. In \bibinfo{booktitle}{{\em Proceedings of the 2011 International
  Symposium on Software Testing and Analysis (ISSTA)}}.
\newblock


\bibitem[\protect\citeauthoryear{Shirani, Collard, Agba, Lebel, Debbabi, Wang,
  and Hanna}{Shirani et~al\mbox{.}}{2018}]%
        {binarm}
\bibfield{author}{\bibinfo{person}{Paria Shirani}, \bibinfo{person}{Leo
  Collard}, \bibinfo{person}{Basile~L. Agba}, \bibinfo{person}{Bernard Lebel},
  \bibinfo{person}{Mourad Debbabi}, \bibinfo{person}{Lingyu Wang}, {and}
  \bibinfo{person}{Aiman Hanna}.} \bibinfo{year}{2018}\natexlab{}.
\newblock \showarticletitle{BinArm: Scalable and Efficient Detection of
  Vulnerabilities in Firmware Images of Intelligent Electronic Device}. In
  \bibinfo{booktitle}{{\em Proceedings of the 15th Conference on Detection of
  Intrusions and Malware and Vulnerability Assessment (DIMVA)}}.
  \bibinfo{address}{Paris, France}.
\newblock


\bibitem[\protect\citeauthoryear{Stanford}{Stanford}{2017}]%
        {squad}
\bibfield{author}{\bibinfo{person}{Stanford}.} \bibinfo{year}{2017}\natexlab{}.
\newblock \bibinfo{title}{{The Stanford Question Answering Dataset}}.
\newblock   (\bibinfo{year}{2017}).
\newblock
\newblock
\shownote{\url{https://rajpurkar.github.io/SQuAD-explorer/}.}


\bibitem[\protect\citeauthoryear{Testa, Yin, Cheng, Feng, Zhou, and Xu}{Testa
  et~al\mbox{.}}{2016}]%
        {acfg}
\bibfield{author}{\bibinfo{person}{Brian Testa}, \bibinfo{person}{Heng Yin},
  \bibinfo{person}{Yao Cheng}, \bibinfo{person}{Qian Feng},
  \bibinfo{person}{Rundong Zhou}, {and} \bibinfo{person}{Chengcheng Xu}.}
  \bibinfo{year}{2016}\natexlab{}.
\newblock \showarticletitle{Scalable Graph-based Bug Search for Firmware
  Images}. In \bibinfo{booktitle}{{\em Proceedings of the 23rd ACM Conference
  on Computer and Communications Security (CCS)}}. \bibinfo{address}{Vienna,
  Austria}.
\newblock


\bibitem[\protect\citeauthoryear{Vaswani, Shazeer, Parmar, Uszkoreit, Jones,
  Gomez, Kaiser, and Polosukhin}{Vaswani et~al\mbox{.}}{2017}]%
        {transformer}
\bibfield{author}{\bibinfo{person}{Ashish Vaswani}, \bibinfo{person}{Noam
  Shazeer}, \bibinfo{person}{Niki Parmar}, \bibinfo{person}{Jakob Uszkoreit},
  \bibinfo{person}{Llion Jones}, \bibinfo{person}{Aidan~N. Gomez},
  \bibinfo{person}{Lukasz Kaiser}, {and} \bibinfo{person}{Illia Polosukhin}.}
  \bibinfo{year}{2017}\natexlab{}.
\newblock \showarticletitle{Attention Is All You Need}. In
  \bibinfo{booktitle}{{\em Proceedings of the 31th Conference on Neural
  Information Processing Systems (NeurIPS)}}.
\newblock


\bibitem[\protect\citeauthoryear{Vig}{Vig}{2019}]%
        {vig2019transformervis}
\bibfield{author}{\bibinfo{person}{Jesse Vig}.}
  \bibinfo{year}{2019}\natexlab{}.
\newblock \showarticletitle{A Multiscale Visualization of Attention in the
  Transformer Model}.
\newblock \bibinfo{journal}{{\em arXiv preprint arXiv:1906.05714\/}}
  (\bibinfo{year}{2019}).
\newblock
\showURL{%
\url{https://arxiv.org/abs/1906.05714}}


\bibitem[\protect\citeauthoryear{vsftpd}{vsftpd}{2020}]%
        {vsftpd}
\bibfield{author}{\bibinfo{person}{vsftpd}.} \bibinfo{year}{2020}\natexlab{}.
\newblock \bibinfo{title}{{Probably the Most Secure and Fastest FTP server}}.
\newblock   (\bibinfo{year}{2020}).
\newblock
\newblock
\shownote{\url{https://security.appspot.com/vsftpd.html}.}


\bibitem[\protect\citeauthoryear{Wang, Singh, Michael, Hill, and Bowman}{Wang
  et~al\mbox{.}}{2019}]%
        {glue}
\bibfield{author}{\bibinfo{person}{Alex Wang}, \bibinfo{person}{Amanpreet
  Singh}, \bibinfo{person}{Julian Michael}, \bibinfo{person}{Felix Hill}, {and}
  \bibinfo{person}{Omer Levy Samuel~R. Bowman}.}
  \bibinfo{year}{2019}\natexlab{}.
\newblock \showarticletitle{GLUE: A Multi-Task Benchmark and Analysis Platform
  for Natural Language Understanding}. In \bibinfo{booktitle}{{\em Proceedings
  of the 7th International Conference on Learning Representations (ICLR)}}.
\newblock


\bibitem[\protect\citeauthoryear{Wikipedia}{Wikipedia}{2020}]%
        {vanishing}
\bibfield{author}{\bibinfo{person}{Wikipedia}.}
  \bibinfo{year}{2020}\natexlab{}.
\newblock \bibinfo{title}{{Vanishing Gradient Problem}}.
\newblock   (\bibinfo{year}{2020}).
\newblock
\newblock
\shownote{\url{https://en.wikipedia.org/wiki/Vanishing_gradient_problem}.}


\bibitem[\protect\citeauthoryear{Williams, Nangia, and Bowman}{Williams
  et~al\mbox{.}}{2018}]%
        {multinli}
\bibfield{author}{\bibinfo{person}{Adina Williams}, \bibinfo{person}{Nikita
  Nangia}, {and} \bibinfo{person}{Samuel~R. Bowman}.}
  \bibinfo{year}{2018}\natexlab{}.
\newblock \showarticletitle{A Broad-Coverage Challenge Corpus for Sentence
  Understanding through Inference}.
\newblock \bibinfo{journal}{{\em arXiv preprint arXiv:1704.05426\/}}
  (\bibinfo{year}{2018}).
\newblock
\showURL{%
\url{https://arxiv.org/pdf/1704.05426.pdf}}


\bibitem[\protect\citeauthoryear{Wolf, Debut, Sanh, Chaumond, Delangue, Moi,
  Cistac, Rault, Louf, Funtowicz, Davison, Shleifer, von Platen, Ma, Jernite,
  Plu, Xu, Scao, Gugger, Drame, Lhoest, and Rush}{Wolf et~al\mbox{.}}{2019}]%
        {transformers}
\bibfield{author}{\bibinfo{person}{Thomas Wolf}, \bibinfo{person}{Lysandre
  Debut}, \bibinfo{person}{Victor Sanh}, \bibinfo{person}{Julien Chaumond},
  \bibinfo{person}{Clement Delangue}, \bibinfo{person}{Anthony Moi},
  \bibinfo{person}{Pierric Cistac}, \bibinfo{person}{Tim Rault},
  \bibinfo{person}{Rémi Louf}, \bibinfo{person}{Morgan Funtowicz},
  \bibinfo{person}{Joe Davison}, \bibinfo{person}{Sam Shleifer},
  \bibinfo{person}{Patrick von Platen}, \bibinfo{person}{Clara Ma},
  \bibinfo{person}{Yacine Jernite}, \bibinfo{person}{Julien Plu},
  \bibinfo{person}{Canwen Xu}, \bibinfo{person}{Teven~Le Scao},
  \bibinfo{person}{Sylvain Gugger}, \bibinfo{person}{Mariama Drame},
  \bibinfo{person}{Quentin Lhoest}, {and} \bibinfo{person}{Alexander~M. Rush}.}
  \bibinfo{year}{2019}\natexlab{}.
\newblock \showarticletitle{{HuggingFace\'s Transformers}: State-of-the-art
  Natural Language Processing}.
\newblock \bibinfo{journal}{{\em ArXiv\/}} (\bibinfo{year}{2019}).
\newblock


\bibitem[\protect\citeauthoryear{Xu, Chen, Chandramohan, Liu, and Song}{Xu
  et~al\mbox{.}}{2017}]%
        {spain-icse17}
\bibfield{author}{\bibinfo{person}{Zhengzi Xu}, \bibinfo{person}{Bihuan Chen},
  \bibinfo{person}{Mahinthan Chandramohan}, \bibinfo{person}{Yang Liu}, {and}
  \bibinfo{person}{Fu Song}.} \bibinfo{year}{2017}\natexlab{}.
\newblock \showarticletitle{SPAIN: Security Patch Analysis for Binaries towards
  Understanding the Pain and Pills}. In \bibinfo{booktitle}{{\em Proceedings of
  the 39th International Conference on Software Engineering (ICSE)}}.
  \bibinfo{address}{Buenos Aires, Argentina}.
\newblock


\bibitem[\protect\citeauthoryear{Xue, Sun, Venkataramani, and Lan}{Xue
  et~al\mbox{.}}{2019}]%
        {ml-based-binanalysis-survey}
\bibfield{author}{\bibinfo{person}{Hongfa Xue}, \bibinfo{person}{Shaowen Sun},
  \bibinfo{person}{Guru Venkataramani}, {and} \bibinfo{person}{Tian Lan}.}
  \bibinfo{year}{2019}\natexlab{}.
\newblock \showarticletitle{Machine Learning-Based Analysis of Program
  Binaries: A Comprehensive Study}.
\newblock \bibinfo{journal}{{\em IEEE Access\/}} (\bibinfo{year}{2019}).
\newblock


\bibitem[\protect\citeauthoryear{Yang, Dai, Yang, Carbonell, Salakhutdinov, and
  Le}{Yang et~al\mbox{.}}{2019}]%
        {xlnet}
\bibfield{author}{\bibinfo{person}{Zhilin Yang}, \bibinfo{person}{Zihang Dai},
  \bibinfo{person}{Yiming Yang}, \bibinfo{person}{Jaime Carbonell},
  \bibinfo{person}{Russ~R Salakhutdinov}, {and} \bibinfo{person}{Quoc~V Le}.}
  \bibinfo{year}{2019}\natexlab{}.
\newblock \showarticletitle{XLNet: Generalized Autoregressive Pretraining for
  Language Understanding}. In \bibinfo{booktitle}{{\em Advances in Neural
  Information Processing Systems}}, Vol.~\bibinfo{volume}{32}.
\newblock


\bibitem[\protect\citeauthoryear{Yu, Cao, Tang, Nie, Huang, and Wu}{Yu
  et~al\mbox{.}}{2017}]%
        {gemini}
\bibfield{author}{\bibinfo{person}{Zeping Yu}, \bibinfo{person}{Rui Cao},
  \bibinfo{person}{Qiyi Tang}, \bibinfo{person}{Sen Nie},
  \bibinfo{person}{Junzhou Huang}, {and} \bibinfo{person}{Shi Wu}.}
  \bibinfo{year}{2017}\natexlab{}.
\newblock \showarticletitle{Neural Network-based Graph Embedding for
  Cross-Platform Binary Code Similarity Detection}. In \bibinfo{booktitle}{{\em
  Proceedings of the 24th ACM Conference on Computer and Communications
  Security (CCS)}}. \bibinfo{address}{Dallas, TX}.
\newblock


\bibitem[\protect\citeauthoryear{Yu, Cao, Tang, Nie, Huang, and Wu}{Yu
  et~al\mbox{.}}{2020a}]%
        {deepbindiff}
\bibfield{author}{\bibinfo{person}{Zeping Yu}, \bibinfo{person}{Rui Cao},
  \bibinfo{person}{Qiyi Tang}, \bibinfo{person}{Sen Nie},
  \bibinfo{person}{Junzhou Huang}, {and} \bibinfo{person}{Shi Wu}.}
  \bibinfo{year}{2020}\natexlab{a}.
\newblock \showarticletitle{{DEEPBINDIFF}: Learning Program-Wide Code
  Representations for Binary Diffing}. In \bibinfo{booktitle}{{\em Proceedings
  of the 2020 Annual Network and Distributed System Security Symposium
  (NDSS)}}. \bibinfo{address}{San Diego, CA}.
\newblock


\bibitem[\protect\citeauthoryear{Yu, Cao, Tang, Nie, Huang, and Wu}{Yu
  et~al\mbox{.}}{2020b}]%
        {ordermatters}
\bibfield{author}{\bibinfo{person}{Zeping Yu}, \bibinfo{person}{Rui Cao},
  \bibinfo{person}{Qiyi Tang}, \bibinfo{person}{Sen Nie},
  \bibinfo{person}{Junzhou Huang}, {and} \bibinfo{person}{Shi Wu}.}
  \bibinfo{year}{2020}\natexlab{b}.
\newblock \showarticletitle{Order Matters: Semantic-Aware Neural Networks for
  Binary Code Similarity Detection}, See \citeN{AAAI20}.
\newblock


\bibitem[\protect\citeauthoryear{Yueduan}{Yueduan}{2020}]%
        {deepbindiff-tool}
\bibfield{author}{\bibinfo{person}{Yueduan}.} \bibinfo{year}{2020}\natexlab{}.
\newblock \bibinfo{title}{{Fine-grained Binary Diffing Tool for x86 Binaries}}.
\newblock   (\bibinfo{year}{2020}).
\newblock
\newblock
\shownote{\url{https://github.com/yueduan/DeepBinDiff}.}


\bibitem[\protect\citeauthoryear{Zhang, Sun, Pang, Liu, and Ma}{Zhang
  et~al\mbox{.}}{2020b}]%
        {bar-bbsimilarity}
\bibfield{author}{\bibinfo{person}{Xiaochuan Zhang}, \bibinfo{person}{Wenjie
  Sun}, \bibinfo{person}{Jianmin Pang}, \bibinfo{person}{Fudong Liu}, {and}
  \bibinfo{person}{Zhen Ma}.} \bibinfo{year}{2020}\natexlab{b}.
\newblock \showarticletitle{Similarity Metric Method for Binary Basic Blocks of
  Cross-Instruction Set Architecture}, See \citeN{BAR20}.
\newblock


\bibitem[\protect\citeauthoryear{Zhang, Qi, and Wang}{Zhang
  et~al\mbox{.}}{2020a}]%
        {malware-aaai20}
\bibfield{author}{\bibinfo{person}{Zhaoqi Zhang}, \bibinfo{person}{Panpan Qi},
  {and} \bibinfo{person}{Wei Wang}.} \bibinfo{year}{2020}\natexlab{a}.
\newblock \showarticletitle{Dynamic Malware Analysis with Feature Engineering
  and Feature Learning}, See \citeN{AAAI20}.
\newblock


\bibitem[\protect\citeauthoryear{Zipf}{Zipf}{1950}]%
        {zipfslaw}
\bibfield{author}{\bibinfo{person}{G.K. Zipf}.}
  \bibinfo{year}{1950}\natexlab{}.
\newblock \showarticletitle{Human behaviour and the principles of least
  effort}.
\newblock \bibinfo{journal}{{\em The Economic Journal\/}}
  (\bibinfo{year}{1950}).
\newblock


\bibitem[\protect\citeauthoryear{Zuo, Li, Young, Luo, and Zeng}{Zuo
  et~al\mbox{.}}{2019}]%
        {innereye}
\bibfield{author}{\bibinfo{person}{Fei Zuo}, \bibinfo{person}{Xiaopeng Li},
  \bibinfo{person}{Patrick Young}, \bibinfo{person}{Lannan Luo}, {and}
  \bibinfo{person}{Zhexin Zeng, Qiang~andZhang}.}
  \bibinfo{year}{2019}\natexlab{}.
\newblock \showarticletitle{Neural Machine Translation Inspired Binary Code
  Similarity Comparison Beyond Function Pairs}. In \bibinfo{booktitle}{{\em
  Proceedings of the 2019 Annual Network and Distributed System Security
  Symposium (NDSS)}}. \bibinfo{address}{San Diego, CA}.
\newblock


\end{thebibliography}

%\onecolumn
%
%\clearpage
\normalsize

\section*{Appendix}
\label{s:appendix}
\begin{figure}[!t]
	\centering
	\includegraphics[width=0.99\linewidth]{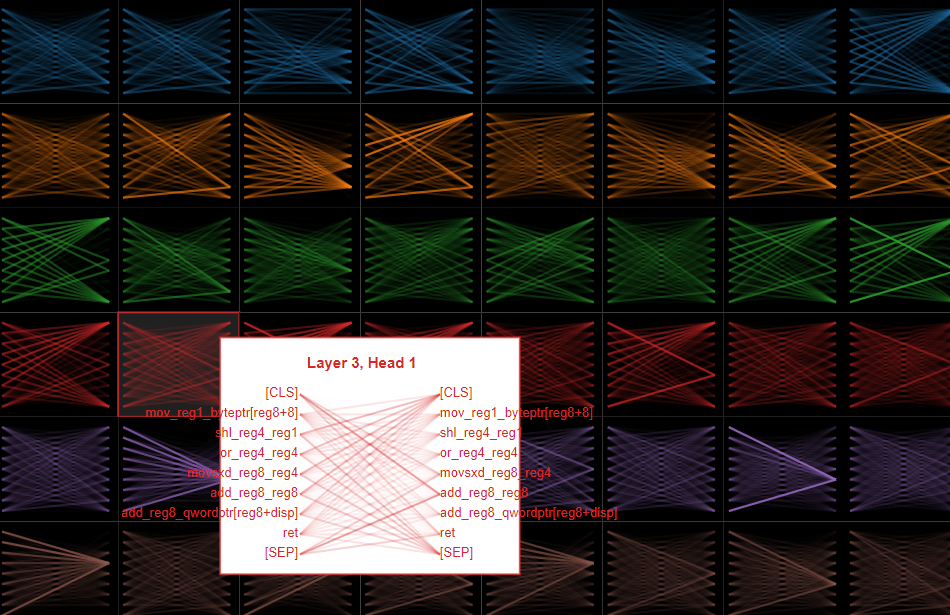}
	\caption{
		Visualization of a multi-head self-attention 
		architecture in Transformer with BERTViz~\cite{vig2019transformervis}. 
		Note that [CLS] and [SEP] correspond to [SOS] and [EOS] in \sys.
		%\KK{update the figure!}
		%
		% \XXX{explain and use. It shows CLS/SEP but you
		%	mentioned SOS/EOS.}
		%		the white box in layer3 head 1 looks very interesting,
		%		but needs interpretation (e.g., two thick lines from CLS to
		%		add_reg8_reg4 and to SEP; how should I understand this?).
		%		If possible, can we add one more such white box in other layer
		%		with distinctive patterns (blue right top or leftmost green).
	}
	\label{fig:bertviz}
\end{figure}

\PP{Visualization of Normalized Instructions in Transformer}
\autoref{fig:bertviz} illustrates an example of
how a different head at a different layer has
a different view in terms of the interactions
within input words using BERTViz~\cite{vig2019transformervis}.
This example has six layers with eight heads.
In this case, the first Head of the third Layer
pays an attention on two words
(\cc{[CLS]} and \cc{[SEP]}) as keys
for predicting \cc{add\_reg8\_reg4} as a query.
Note that \cc{[CLS]} and \cc{[SEP]} in
the original BERT correspond to
our \cc{[SOS]} and \cc{[EOS]}.

Word prediction, under the hood, can be achieved with 
keys ($k$; previous words), values ($v$; quantities
that represent the content of the words) and 
a query ($q$; words to predict), 
obtaining the attention distribution of previous words ($a_i$) 
and finally computing a hidden state by maximizing the distribution
as~\autoref{eq:kqv-attention} ($d_k$; dimension of key vectors):
\begin{equation}
\label{eq:kqv-attention}
\resizebox{.85\linewidth}{!}{$
	\displaystyle
	a_i = \text{softmax}(\frac{Q K^{T}}{\sqrt{d_k}}),
	\text{Attention}(Q,K,V) = \sum_{i}{a_i V_i}
	$}
\end{equation}
\begin{table*}[!htp]
	\centering
	\footnotesize
	\renewcommand{\arraystretch}{1.0}
	\scalebox{0.99} {
		\begin{tabular}{r|lrrr||r|lrrrr}
			\toprule
			\textbf{Rank} & \textbf{Normalized Instruction} & \textbf{Ratio} & \textbf{Cumulative} & \textbf{Group} & \textbf{Rank} & \textbf{Normalized Instruction} & \textbf{Ratio} & \textbf{Cumulative} & \textbf{Group} \\
\hline
1 & mov\_reg8\_reg8 & 8.243\% & 8.24\% & M & 73 & movzx\_reg4\_wordptr[reg8] & 0.163\% & 76.04\% & M \\
2 & call\_innerfunc & 5.735\% & 13.98\% & C & 74 & sub\_reg4\_immval & 0.163\% & 76.20\% & \\
3 & mov\_reg8\_qwordptr[bp8-disp] & 4.164\% & 18.14\% & M & 75 & mov\_dwordptr[bp8-disp]\_immval & 0.157\% & 76.36\% & M \\
4 & je\_jmpdst & 3.925\% & 22.07\% & J & 76 & mov\_qwordptr[ip8+disp]\_reg8 & 0.156\% & 76.51\% & M \\
5 & jmp\_jmpdst & 3.859\% & 25.93\% & J & 77 & mov\_qwordptr[sp8+disp]\_immval & 0.154\% & 76.67\% & M \\
6 & mov\_reg4\_immval & 3.486\% & 29.41\% & M & 78 & cmp\_qwordptr[bp8-disp]\_immval & 0.153\% & 76.82\% & \\
7 & jne\_jmpdst & 2.613\% & 32.03\% & J & 79 & mov\_qwordptr[reg8+8]\_reg8 & 0.151\% & 76.97\% & M \\
8 & mov\_reg8\_qwordptr[reg8+disp] & 2.267\% & 34.29\% & M & 80 & mov\_qwordptr[sp8+8]\_reg8 & 0.151\% & 77.12\% & M \\
9 & pop\_reg8 & 1.936\% & 36.23\% & & 81 & cmp\_dwordptr[bp8-disp]\_immval & 0.151\% & 77.27\% & \\
10 & mov\_reg4\_reg4 & 1.860\% & 38.09\% & M & 82 & mov\_reg4\_bp4 & 0.150\% & 77.42\% & M \\
11 & push\_reg8 & 1.844\% & 39.93\% & & 83 & cdqe & 0.146\% & 77.57\% & \\
12 & mov\_qwordptr[bp8-disp]\_reg8 & 1.839\% & 41.77\% & M & 84 & test\_reg1\_immval & 0.141\% & 77.71\% & \\
13 & xor\_reg4\_reg4 & 1.725\% & 43.50\% & & 85 & jg\_jmpdst & 0.140\% & 77.85\% & J \\
14 & ret & 1.477\% & 44.97\% & C & 86 & mov\_reg8\_sp8 & 0.139\% & 77.99\% & M \\
15 & test\_reg8\_reg8 & 1.387\% & 46.36\% & & 87 & mov\_reg4\_dispbss & 0.137\% & 78.13\% & M \\
16 & mov\_reg8\_qwordptr[reg8] & 1.370\% & 47.73\% & M & 88 & shr\_reg4\_immval & 0.136\% & 78.26\% & \\
17 & mov\_reg8\_qwordptr[sp8+disp] & 1.306\% & 49.04\% & M & 89 & mov\_reg8\_qwordptr[reg8+reg8*8] & 0.134\% & 78.40\% & M \\
18 & cmp\_reg4\_immval & 1.111\% & 50.15\% & & 90 & mov\_reg8\_qwordptr[sp8] & 0.134\% & 78.53\% & M \\
19 & add\_reg8\_immval & 1.107\% & 51.25\% & & 91 & call\_reg8 & 0.132\% & 78.66\% & C \\
20 & call\_externfunc & 1.061\% & 52.31\% & C & 92 & mulsd\_regxmm\_regxmm & 0.128\% & 78.79\% & \\
21 & mov\_reg4\_dispstr & 1.046\% & 53.36\% & M & 93 & movabs\_reg8\_immval & 0.127\% & 78.92\% & M \\
22 & test\_reg4\_reg4 & 0.936\% & 54.30\% & & 94 & mov\_dwordptr[reg8+disp]\_immval & 0.126\% & 79.04\% & M \\
23 & push\_bp8 & 0.891\% & 55.19\% & & 95 & shr\_reg8\_immval & 0.126\% & 79.17\% & \\
24 & mov\_reg4\_dwordptr[bp8-disp] & 0.872\% & 56.06\% & M & 96 & movzx\_reg4\_byteptr[reg8] & 0.126\% & 79.30\% & M \\
25 & sub\_sp8\_immval & 0.812\% & 56.87\% & & 97 & mov\_bp4\_immval & 0.124\% & 79.42\% & M \\
26 & mov\_qwordptr[sp8+disp]\_reg8 & 0.808\% & 57.68\% & M & 98 & sar\_reg8\_immval & 0.123\% & 79.54\% & \\
27 & mov\_reg8\_qwordptr[bp8-8] & 0.801\% & 58.48\% & M & 99 & cmp\_reg2\_immval & 0.117\% & 79.66\% & \\
28 & add\_sp8\_immval & 0.786\% & 59.26\% & & 100 & jl\_jmpdst & 0.116\% & 79.78\% & J \\
29 & pop\_bp8 & 0.757\% & 60.02\% & & 101 & lea\_reg8\_[reg8+reg8] & 0.115\% & 79.89\% & \\
30 & mov\_reg8\_qwordptr[ip8+disp] & 0.690\% & 60.71\% & M & 102 & sub\_reg4\_reg4 & 0.114\% & 80.01\% & \\
31 & lea\_reg8\_[bp8-disp] & 0.680\% & 61.39\% & & 103 & mov\_reg1\_immval & 0.113\% & 80.12\% & M \\
32 & mov\_reg8\_qwordptr[reg8+8] & 0.680\% & 62.07\% & M & 104 & mov\_reg8\_qwordptr[disp] & 0.112\% & 80.23\% & M \\
33 & lea\_reg8\_[sp8+disp] & 0.651\% & 62.72\% & & 105 & cmp\_qwordptr[reg8+disp]\_immval & 0.112\% & 80.34\% & \\
34 & mov\_reg8\_bp8 & 0.622\% & 63.35\% & M & 106 & cmp\_dwordptr[ip8+disp]\_immval & 0.111\% & 80.45\% & \\
35 & cmp\_reg8\_reg8 & 0.619\% & 63.96\% & & 107 & vmulsd\_regxmm\_regxmm\_regxmm & 0.109\% & 80.56\% & \\
36 & mov\_reg4\_dwordptr[reg8+disp] & 0.610\% & 64.57\% & M & 108 & lea\_reg8\_[reg8+reg8*8] & 0.107\% & 80.67\% & \\
37 & mov\_dwordptr[bp8-disp]\_reg4 & 0.584\% & 65.16\% & M & 109 & mov\_dwordptr[sp8+disp]\_immval & 0.106\% & 80.78\% & M \\
38 & mov\_bp8\_sp8 & 0.569\% & 65.73\% & M & 110 & movabs\_reg8\_dispstr & 0.106\% & 80.88\% & M \\
39 & add\_reg8\_reg8 & 0.550\% & 66.28\% & & 111 & sete\_reg1 & 0.105\% & 80.99\% & \\
40 & mov\_qwordptr[reg8+disp]\_reg8 & 0.512\% & 66.79\% & M & 112 & jge\_jmpdst & 0.103\% & 81.09\% & J \\
41 & mov\_reg4\_dispdata & 0.470\% & 67.26\% & M & 113 & cmp\_byteptr[reg8+disp]\_immval & 0.101\% & 81.19\% & \\
42 & and\_reg4\_immval & 0.466\% & 67.73\% & & 114 & js\_jmpdst & 0.100\% & 81.29\% & J \\
43 & mov\_qwordptr[bp8-8]\_reg8 & 0.459\% & 68.19\% & M & 115 & mov\_dwordptr[reg8]\_reg4 & 0.100\% & 81.39\% & M \\
44 & test\_reg1\_reg1 & 0.426\% & 68.61\% & & 116 & sub\_reg8\_immval & 0.099\% & 81.49\% & \\
45 & shl\_reg8\_immval & 0.422\% & 69.03\% & & 117 & shl\_reg4\_immval & 0.099\% & 81.59\% & \\
46 & add\_reg4\_immval & 0.397\% & 69.43\% & & 118 & test\_byteptr[reg8+disp]\_immval & 0.096\% & 81.68\% & \\
47 & lea\_reg8\_[reg8+disp] & 0.349\% & 69.78\% & & 119 & movapd\_regxmm\_regxmm & 0.096\% & 81.78\% & M \\
48 & movsxd\_reg8\_reg4 & 0.332\% & 70.11\% & M & 120 & mov\_qwordptr[sp8]\_reg8 & 0.093\% & 81.87\% & M \\
49 & mov\_bp8\_reg8 & 0.325\% & 70.44\% & M & 121 & mov\_byteptr[bp8-disp]\_reg1 & 0.093\% & 81.97\% & M \\
50 & mov\_reg4\_dwordptr[sp8+disp] & 0.322\% & 70.76\% & M & 122 & mov\_reg8\_qwordptrfs:[disp] & 0.093\% & 82.06\% & M \\
51 & mov\_qwordptr[reg8]\_reg8 & 0.299\% & 71.06\% & M & 123 & xor\_reg8\_qwordptrfs:[disp] & 0.093\% & 82.15\% & \\
52 & mov\_reg4\_dwordptr[reg8] & 0.284\% & 71.34\% & M & 124 & cmp\_dwordptr[reg8+disp]\_immval & 0.093\% & 82.24\% & \\
53 & call\_qwordptr[reg8+disp] & 0.280\% & 71.62\% & C & 125 & push\_immval & 0.092\% & 82.34\% & \\
54 & ja\_jmpdst & 0.278\% & 71.90\% & J & 126 & movsxd\_reg8\_dwordptr[bp8-disp] & 0.091\% & 82.43\% & M \\
55 & mov\_reg4\_dwordptr[ip8+disp] & 0.262\% & 72.16\% & M & 127 & setne\_reg1 & 0.091\% & 82.52\% & \\
56 & cmp\_reg4\_reg4 & 0.245\% & 72.41\% & & 128 & mov\_reg4\_dwordptr[reg8+8] & 0.091\% & 82.61\% & M \\
57 & jae\_jmpdst & 0.240\% & 72.65\% & J & 129 & test\_bp8\_bp8 & 0.091\% & 82.70\% & \\
58 & jle\_jmpdst & 0.239\% & 72.89\% & J & 130 & lea\_reg8\_[reg8+reg8*2] & 0.090\% & 82.79\% & \\
59 & sub\_reg8\_reg8 & 0.238\% & 73.12\% & & 131 & lea\_reg8\_[reg8+1] & 0.088\% & 82.88\% & \\
60 & cmp\_reg8\_immval & 0.236\% & 73.36\% & & 132 & movzx\_reg4\_reg2 & 0.086\% & 82.96\% & M \\
61 & jbe\_jmpdst & 0.232\% & 73.59\% & J & 133 & lea\_reg4\_[reg8+1] & 0.086\% & 83.05\% & \\
62 & mov\_dwordptr[sp8+disp]\_reg4 & 0.230\% & 73.82\% & M & 134 & mov\_qwordptr[reg8]\_dispdata & 0.086\% & 83.13\% & M \\
63 & mov\_qwordptr[reg8+disp]\_immval & 0.227\% & 74.05\% & M & 135 & mov\_bp4\_reg4 & 0.084\% & 83.22\% & M \\
64 & mov\_reg8\_qwordptr[bp8+disp] & 0.225\% & 74.27\% & M & 136 & vmovsd\_regxmm\_qwordptr[sp8+disp] & 0.084\% & 83.30\% & \\
65 & mov\_reg8\_qwordptr[sp8+8] & 0.224\% & 74.50\% & M & 137 & addsd\_regxmm\_regxmm & 0.083\% & 83.38\% & \\
66 & movzx\_reg4\_reg1 & 0.216\% & 74.71\% & M & 138 & mov\_qwordptr[bp8-disp]\_immval & 0.082\% & 83.47\% & M \\
67 & add\_reg4\_reg4 & 0.211\% & 74.92\% & & 139 & vmovsd\_qwordptr[sp8+disp]\_regxmm & 0.081\% & 83.55\% & \\
68 & jb\_jmpdst & 0.211\% & 75.14\% & J & 140 & and\_reg1\_immval & 0.080\% & 83.63\% & \\
69 & mov\_dwordptr[reg8+disp]\_reg4 & 0.195\% & 75.33\% & M & 141 & mov\_reg8\_qwordptr[bp8] & 0.080\% & 83.71\% & M \\
70 & cmp\_reg1\_immval & 0.193\% & 75.52\% & & 142 & pxor\_regxmm\_regxmm & 0.080\% & 83.79\% & \\
71 & leave & 0.186\% & 75.71\% & C & 143 & lea\_reg8\_[reg8+8] & 0.079\% & 83.87\% & \\
72 & movzx\_reg4\_byteptr[reg8+disp] & 0.167\% & 75.88\% & M & 144 & mov\_byteptr[reg8+disp]\_immval & 0.078\% & 83.94\% & M \\
			\bottomrule
		\end{tabular}
	}
	\caption{Top $144$ normalized instructions and their frequencies
		in our dataset. 
		The last $14\%$ of total instructions has rarely seen
		with a long tail of the distribution (\autoref{fig:rankfreq}).
		The group
	    field indicates (C)all, (J)ump and (M)ove instructions.}
	\label{t:freqrank}
\end{table*}

\PP{Bag of Signature (BoS) as Supplementary Information}
%
%\XXX{BoS does not improve too much but mention
%	the benefit slightly}
The idea behind BoS is that even a 
well-balanced normalization process
(\autoref{t:normalization-rules})
abandons a string or numeric constant
itself that may be fruitful to better understand
the context of a function,
as explained in~\autoref{ss:deepsemantic-overview}.
We devise the notion of BoS,
which assists a binary similarity prediction
classifier by feeding additional information
to our neural network.
Indeed, a number of prior approaches~\cite{acfg, fermadyine, blanket}
emphasize that such constants come into play
as important feature vectors to identify 
the behavior of a basic block or function.
We enumerate both string literals and numeric constants
via static analysis, combining them into a single bag
that can be used as a unique signature per each function.
We can simply compute a BoS similarity score
%an inner product over the multiplication
%of each square sum of two vectors 
with~\autoref{eq:bos}.
As a concrete example, a function ($v$) contains a list of
\cc{[1, 0x12, 8, 8, 8, 'Hello']}
(five numeric constants with three unique values,
and one string constant), whereas another function ($w$) 
holds \cc{[0x12, 8, 8, 'Hello']}.
Then, we can represent each vector based on 
counting those constants: $v = (1,1,3,1)$ and $w = (0,1,2,1)$
whose cosine similarity becomes $0.943$.
\begin{equation}
	\label{eq:bos}
	\resizebox{.85\linewidth}{!}{$
		\displaystyle
		BoS(\vec{v}, \vec{w}) = \
		\frac{\vec{v} \cdot \vec{w}}{|\vec{v}| |\vec{w}|} = \
		\frac{\sum_{i=1}^{n} v_i w_i}{\sqrt{\sum_{i=1}^{n} v_i^2} \sqrt{\sum_{i=1}^{n} w_i^2}}
		$}
\end{equation}
%
%Similar to \autoref{sss:balance-grained},
%we collect $513,727$ binary functions
%(out of $1,174,060$) that hold at least five
%BoS entries, followed by obtaining
%$1,801,684$ combination pairs for
%similar cases (\eg, label=True)
%and the same number of pairs
%for dissimilar cases (\eg, label=False)
%with negative sampling.
%%
%We exclude all cases where
%two normalized functions (NFs) are identical
%($135,358$ pairs)
%because those pairs would be too trivial
%(cosine similarity becomes always $1$) 
%for similarity decision.
%%
%We extract 100K cases as a final dataset,
%splitting them into the (train,valid,test) set.
%%with the ratio of $(0.9:0.05:0.05)$.
%
%\autoref{t:fgn-bos-comp} summarizes
%

For evaluation, we generate
another model by defining a new dataset 
%(including 6,502,405 trainable parameters)
that contains meaningful
BoS information 
(\eg, the number of the BoS list
is greater than or equal to five)
because functions in a small size
may not contain a numeric constant
or string reference, as illustrated in \autoref{t:dataset}
and \autoref{fig:fbi}.
Our results demonstrate a marginal $0.09\%$ enhancement in
both F1 and AUC.
We reason that
i)~approximately two-thirds of binary
functions in our corpus do not hold 
information such as immediate operands
or string literals and 
ii)~BoS has been
concatenated as a single dimensional vector 
(after computing cosine similarity)
with two hidden vectors (\eg, 256 dimensions)
from NFs, restricting a positive impact.
While the additional information has contributed to
our model with a slight margin, it indicates
the model would still be efficient even
when such knowledge is unavailable.

\end{document}